\documentclass[fleqn,usenatbib,usedcolumn]{mnras}
\pdfoutput=1

\usepackage{newtxtext,newtxmath}

\usepackage[T1]{fontenc}
\usepackage{ae,aecompl}

\usepackage{subfigure}
\usepackage{dblfloatfix}
\usepackage{multirow}
\usepackage{nccmath}
\usepackage{textcomp}

\usepackage{soul}
\soulregister\cite7
\soulregister\citet7
\soulregister\citep7
\soulregister\ref7
\soulregister\textit7
\soulregister\label7
\soulregister\subsection7

\usepackage{xcolor}

\usepackage{graphicx}	

\title[Bias in CMB T and SNe $z$ correlation]{Accidental deep field bias in CMB T and SNe $z$ correlation}

\author[T. Friday et al.]{Tracey Friday,$^{1}$\thanks{E-mail: tfriday@uclan.ac.uk}
Roger G. Clowes,$^{1}$
Srinivasan Raghunathan,$^{2}$
and \newauthor Gerard M. Williger$^{3,1}$
\\
$^{1}$Jeremiah Horrocks Institute, University of Central Lancashire, Preston PR1 2HE, UK\\
$^{2}$School of Physics, University of Melbourne, Parkville VIC 3010, Australia\\
$^{3}$Department of Physics and Astronomy, University of Louisville, Louisville KY 40292, USA
}

\date{Accepted 2018 May 21. Received 2018 May 14; in original form 2018 March 27}

\pubyear{2018}

\begin{document}
\label{firstpage}
\pagerange{\pageref{firstpage}--\pageref{lastpage}}
\maketitle

\begin{abstract}
Evidence presented by Yershov, Orlov and Raikov apparently showed that the \textit{WMAP}/\textit{Planck} cosmic microwave background (CMB) pixel-temperatures (T) at supernovae (SNe) locations tend to increase with increasing redshift ($z$). They suggest this correlation could be caused by the Integrated Sachs-Wolfe effect and/or by some unrelated foreground emission. Here, we assess this correlation independently using \textit{Planck} 2015 SMICA R2.01 data and, following Yershov et al., a sample of 2783 SNe from the Sternberg Astronomical Institute. Our analysis supports the prima facie existence of the correlation but attributes it to a composite selection bias (high CMB T $\times$ high SNe $z$) caused by the accidental alignment of seven deep survey fields with CMB hotspots. These seven fields contain 9.2 per cent of the SNe sample (256 SNe). Spearman's rank-order correlation coefficient indicates the correlation present in the whole sample ($\rho_s = 0.5$, p-value $= 6.7 \times 10^{-9}$) is insignificant for a sub-sample of the seven fields together ($\rho_s = 0.2$, p-value $= 0.2$) and entirely absent for the remainder of the SNe ($\rho_s = 0.1$, p-value $= 0.6$). We demonstrate the temperature and redshift biases of these seven deep fields, and estimate the likelihood of their falling on CMB hotspots by chance is at least $\sim$ 6.8 per cent (approximately 1 in 15). We show that a sample of 7880 SNe from the Open Supernova Catalogue exhibits the same effect and we conclude that the correlation is an accidental but not unlikely selection bias.
\end{abstract}

\begin{keywords}
cosmology: cosmic background radiation -- cosmology: observations -- supernovae: general -- methods: statistical -- surveys
\end{keywords}



\section{Introduction} \label{secIntro}

Observations of the cosmic microwave background (CMB) and Type Ia supernovae (SNIa) are exceptional probes of cosmological parameters. The measurements of the CMB by the \textit{WMAP} satellite \citep{Bennett2013} and SNIa by the high-$z$ supernova search team \citep{Riess1998} and the supernova cosmology project \citep{Perlmutter1999} have established the six parameter $\Lambda$CDM cosmological model. Precision measurements of the CMB temperature, polarisation, and lensing anisotropies from the South Pole Telescope, \textit{Planck} satellite, and Atacama Cosmology Telescope \citep[e.g.,][and references therein]{Story2013, Planck2016b, Louis2017} have strongly reinforced the preference for $\Lambda$CDM as the concordance model of cosmology.

Cross-correlation of CMB observations with the large-scale structure (LSS) of the Universe, revealed by surveys such as the \textit{Sloan Digital Sky Survey} \citep[SDSS,][]{Alam2015} and the \textit{Dark Energy Survey} \citep[DES,][]{DES2016}, provide powerful tests of $\Lambda$CDM \citep[e.g.,][]{Giannantonio2016, DES2017}. Increased efforts are currently being made for accurate and precise calibration of the distance-redshift relation using SNIa up to high redshifts \citep[e.g.,][]{LSST2009, Kessler2015} in order to understand the expansion history and late-time ($z \le 1$) accelerated expansion of the Universe attributed to dark energy.

\begin{figure} 
	\centering
	\includegraphics[width=0.8\columnwidth]{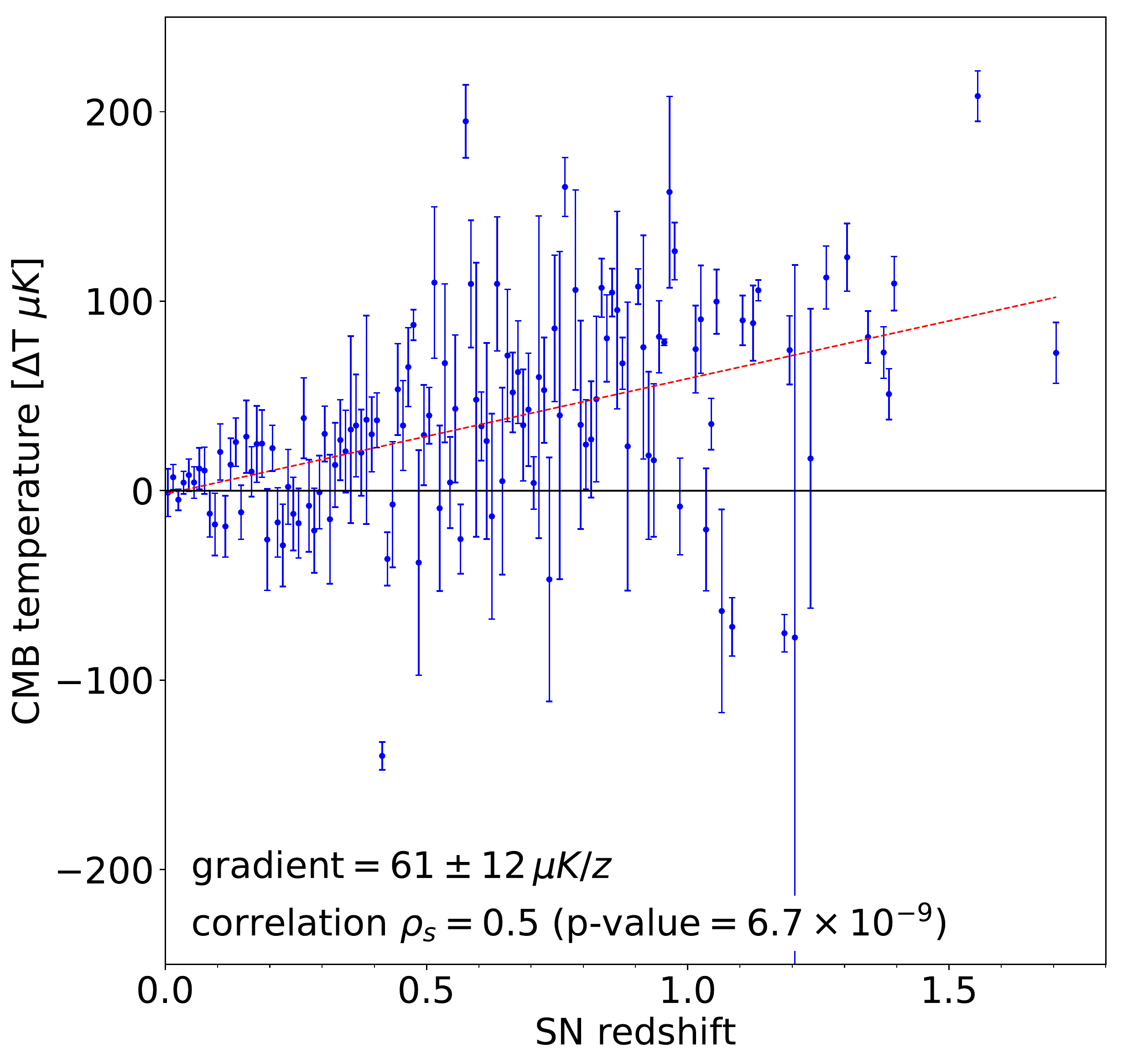}
	\caption{Plot of CMB temperature at SNe locations versus SNe redshift binned with bin sizes $\Delta z=0.01$. Data are restricted by Galactic latitude ($|b|>40^\circ$), redshift ($z>0.005$), and \textit{Planck} UT78 confidence mask. Error bars are the standard error on the bin mean. The dashed line indicates ordinary least squares linear regression, representing the correlation reported by \citet{Yershov2012, Yershov2014}.}
	\label{figBinnedScatter}
\end{figure}

\citet{Yershov2012, Yershov2014} combined CMB and supernovae (SNe) data and detected a correlation between the CMB temperature anisotropies (T) and the redshift (z) of the SNe (CMB T $\times$ SNe $z$). High $z$ SNe appear to be preferentially associated with hotter CMB temperatures (Fig.~\ref{figBinnedScatter}). This effect was particularly strong for SNIa. They concluded that the correlation is not caused by the Sunyaev-Zel\textquotesingle dovich (SZ) effect \citep{SunyaevZeldovich1970} and suggested it may instead be caused by the Integrated Sachs-Wolfe (ISW) effect \citep{SachsWolfe1967} or some remnant contamination in the CMB data, possibly from low redshift foreground \citep{Yershov2014}.

In this paper we re-analyse the SNe samples of \citet{Yershov2012, Yershov2014} and offer an alternative explanation for the correlation, namely that it is a composite selection bias caused by the chance alignment of certain deep survey fields with CMB hotspots. This bias (high CMB T $\times$ high SNe $z$) is the combined result of a selection bias (high $z$ SNe in deep fields) and the chance alignment of those deep fields with CMB hotspots.

The remainder of this paper describes our analyses of the reported correlation. In section~\ref{secDataMethod} we describe the data and summarise the variety of methods used to demonstrate the prima facie existence of the correlation in these data. We present the results from re-analysing the SNe sample of \citet{Yershov2012, Yershov2014} in section~\ref{secBias}. Specifically, we identify SNe fields with a high surface density of SNe (\ref{subsecFieldID}) and show that these cause the apparent correlation (\ref{subsecContribution}) due to their bias to hotter CMB temperature and higher redshift than the remainder of the SNe sample (\ref{subsecTempRedshift}). We quantify the likelihood of this bias occurring by chance: at least 6.8 per cent, or approximately 1 in 15 (\ref{subsecLikelihood}). We present corroborating results from analysing alternative data in section~\ref{secAlternativeData}. In section~\ref{secConclusions} we conclude that the correlation reported by \citet{Yershov2012, Yershov2014} is actually an accidental but not exceptionally unlikely composite selection bias and we briefly speculate on further potential implications for cosmology.

\section{Data and methods} \label{secDataMethod}

\subsection{SNe and CMB data} \label{subsecData}

We used CMB data from the \textit{Planck} 2015 \citep{Planck2016a} maps\footnote{\url{http://irsa.ipac.caltech.edu/data/Planck/release_2/all-sky-maps/matrix_cmb.html}}, specifically SMICA R2.01 with $N_{side}=2048$. These maps are provided and analysed using the Hierarchical Equal Area iso-Latitude Pixelisation scheme \citep[\textsc{HEALP}ix\footnote{\url{http://healpix.jpl.nasa.gov/}}, ][]{Gorski2005}. Our temperature distribution was consistent with that previously determined by \citet{Yershov2014} using the \textit{Planck} 2013 \citep{Planck2014a} SMICA R1.20 map. 

The \textit{Planck} component-separated CMB maps were produced using four techniques: Commander \citep{Eriksen2008}, NILC \citep{Delabrouille2009}, SEVEM \citep{FernandezCobos2012} and SMICA \citep{Cardoso2008}. \citet{Planck2016c} provide a critical analysis of the applicability of the resultant four 2015 maps, and confirm that, as in 2013 \citep{Planck2014b}, SMICA is preferred for high-resolution temperature analysis.

As recommended by \citet{Planck2016c}, for analysing component-separated CMB temperature maps, we used the \textit{Planck} UT78 common mask. This is the union of the Commander, SEVEM, and SMICA confidence masks. UT78 excludes point sources, some of the Galactic plane (note also our subsequent Galactic latitude restriction), and some other bright regions. It has a fraction of unmasked pixels of $f_{sky} = 77.6\%$. Note that our results were consistent using an alternative mask (UTA76), and without masking.

Our initial analysis used SNe data provided by \citet{Yershov2014} as supplementary data\footnote{\url{https://academic.oup.com/mnras/article-lookup/doi/10.1093/mnras/stu1932}}, derived from the Sternberg Astronomical Institute (SAI) Supernova Catalogue\footnote{\url{http://www.sai.msu.su/sn/sncat/}} \citep{Bartunov2007} as of October 2013. This provides a sample of 6359 SNe of all types. To avoid contamination from the Galactic plane we restricted this sample to high Galactic latitude $|b|>40^\circ$, the same conservative restriction used by \citet{Yershov2014}. We could not reproduce the identical sample for the redshift restriction apparently used by \citet{Yershov2014}, so we adopted a restriction of $z>0.005$, which yielded a similar sample size to theirs. We excluded SNe on masked (\textit{Planck} UT78) \textsc{HEALP}ix pixels. The resultant SAI sample contained 2783 SNe.

Above redshift $z\sim1.2$, SNe in the SAI sample become rather sparse and are predominantly associated with hotter than average CMB temperature. To analyse this high-redshift region further, we obtained $z>0.005$ data from the Open Supernova Catalogue\footnote{\url{https://sne.space/}} \citep[OSC,][]{Guillochon2017} as of June 2017. We removed SNe without co-ordinate information, those not yet confirmed as SNe ($Type=Candidate$) and gamma ray bursts ($Type=LGRB$). These selections provide a sample of 12879 SNe of all types. We also restricted this sample to high Galactic latitude $|b|>40^\circ$ and excluded SNe on masked (\textit{Planck} UT78) \textsc{HEALP}ix pixels. The resultant OSC sample contained 7880 SNe.

Unless specified otherwise, all analysis in this paper was performed on the SAI sample after the restrictions on Galactic latitude ($|b|>40^\circ$), redshift ($z>0.005$), and \textit{Planck} UT78 confidence mask. We repeated our analyses using the OSC sample (section~\ref{subsecOSC}) with the same restrictions to verify that our results are not specific to the SAI sample.

\subsection{Methods} \label{subsecMethod}

We constructed Fig.~\ref{figBinnedScatter} broadly following \citet{Yershov2014}. We determined the temperature\footnote{We follow the common practice of referring to temperature anisotropies ($\Delta T$) as temperature ($T$).} of the CMB map pixel at each SN location. The data were grouped into redshift bins of width $\Delta z = 0.01$ and the weighted mean CMB temperature of each bin ($\overline{T}$) was calculated as

\begin{ceqn}
\begin{equation}
\overline{T} = \sum_{i=1}^{n} w_iT_i / \sum_{i=1}^{n} w_i\ ,
\end{equation}
\end{ceqn}

\noindent where $T_i$ is the individual pixel temperature, $w_i$ is the weight of each pixel, and $n$ is the number of SNe per bin. Error bars are the standard error on the weighted bin mean ($\sigma_{\overline{T}}$), calculated from the weighted variances as

\begin{ceqn}
\begin{equation}
\sigma_{\overline{T}} = \sqrt{\sum_{i=1}^{n} w_i^2\sigma_i^2 / \left(\sum_{i=1}^{n} w_i\right)^2}\ ,
\end{equation}
\end{ceqn}

\noindent where $\sigma_i^2$ is our variance estimate for each pixel. Note that for bins with only one SN, error bars are $\sigma_{\overline{T}} = \sigma_i$.

Individual pixel variances ($\sigma_i^2$) were estimated\footnote{Following private communications with \textit{Planck} Legacy Archive and NASA/IPAC Infrared Science Archive.} by producing a squared, smoothed ($0.5^\circ$ FWHM) half-mission half-difference (HMHD) map from the two \textit{Planck} 2015 SMICA R2.01 half-mission maps . Weights ($w_i$) are thus

\begin{ceqn}
\begin{equation}
w_i = 1 / \sigma_i^2\ .
\end{equation}
\end{ceqn}

The choice of smoothing scale and the method of estimating pixel variance affects the resultant weights. We tested a number of these and our results are consistent. See Appendix~\ref{appVariances} for versions of Fig.~\ref{figBinnedScatter} produced using different smoothing scales (none, $5'$, 0.5$^\circ$, and 5$^\circ$ FWHM) and different estimates of individual pixel variance (\textit{Planck} 2013 R1.20 SMICA map noise, \textit{Planck} 2015 R2.01 SMICA HMHD and HRHD maps, and \textit{Planck} 2015 R2.02 143GHz and 217GHz frequency maps).

We fitted an ordinary least squares (OLS) linear regression to the binned data (dashed line) using Python's \texttt{statsmodels.api.OLS}, and calculated its slope plus the standard error on the gradient. We followed the same method to calculate the OLS linear regression gradient throughout this paper. Note that results using weighted least squares (WLS) linear regression (weighted by $\sigma_{\overline{T}}$ or $n$), and results using OLS and WLS linear regression of unbinned data, were consistent.

Several of our analyses compared various SNe sub-samples using the parametric independent 2-sample Welch's t-test \citep[or unequal variance t-test,][]{Welch1938} and the non-parametric 1-sided Mann-Whitney U (MWU) test. We used the unequal variance t-test to account for the different angular extent of the deep survey fields, and hence difference in variance of their CMB temperature. It also accommodates the wide variation in the number of SNe per redshift bin, and the resultant differing variance of both their CMB temperature and their redshift. The MWU test does not assume that the population follows any specific parameterised distribution, unlike the t-test which assumes a normal distribution, and it is less sensitive to outliers than the t-test.

The t-test gives the probability (p-value) of obtaining SNe sub-samples with differences in mean CMB temperature (or SNe redshift) at least as extreme as those observed, assuming the null hypothesis is true

\begin{ceqn}
\begin{equation}
H_{0}: \mu_{T1} = \mu_{T2} \ (H_{0}: \mu_{z1} = \mu_{z2})\ .
\end{equation}
\end{ceqn}

In other words, it tests whether the means of their populations differ. The 2-sample Welch's t-test was implemented using Python's \texttt{scipy.stats.ttest\_ind} with \texttt{equal\_var $=$ False}.

MWU combines the sub-samples of CMB temperature (or SNe redshift), ranks the combined sample, and determines the mean of the ranks ($\overline{R}$) for each sub-sample. It gives the probability (p-value) of obtaining SNe samples with differences in mean ranks at least as extreme as those observed, assuming the null hypothesis is true

\begin{ceqn}
\begin{equation}
H_{0}: \overline{R}_{T1} = \overline{R}_{T2} \ (H_{0}: \overline{R}_{z1} = \overline{R}_{z2})\ .
\end{equation}
\end{ceqn}

In practice, this is generally interpreted as whether the distributions of the sub-samples differ, since the ranks of the sub-samples will differ if so. We used the 1-sided MWU to test the alternative hypothesis that the CMB temperature (or SNe redshift) distribution of one sub-sample was greater than that of the other

\begin{ceqn}
\begin{equation}
H_{1}: \overline{R}_{T1} \ge \overline{R}_{T2} \ (H_{1}: \overline{R}_{z1} \ge \overline{R}_{z2})\ .
\end{equation}
\end{ceqn}

The 1-sided MWU test was implemented using Python's \texttt{scipy.stats.mannwhitneyu} with \texttt{alternative $=$ `greater'}.

To analyse which modes dominate the apparent correlation (section~\ref{subsubAngularScales}) we filtered the map to remove large angular scales. We used Python's \texttt{healpy.sphtfunc.almxfl} to apply a high pass filter to the spherical harmonic coefficients ($a_{\ell m}$) of the \textit{Planck} 2015 SMICA map. We then computed the filtered map from these filtered $a_{\ell m}$ values. We repeated the OLS linear regression and Spearman's rank-order correlation coefficient analyses for each filtered map.

In our analysis of likelihood (section~\ref{subsecLikelihood}) we performed randomisations of SNe locations within fields and of SNe field centres on the sky. We used Python's \texttt{random.uniform} to select random \textsc{HEALP}ix pixels, subject to the same Galactic latitude restriction ($|b|>40^\circ$) as the original sample. After each randomisation the masking (\textit{Planck} UT78) was re-applied, the CMB temperature and variance were re-sampled, and the weights were re-calculated before the OLS linear regression was re-fitted.

We also assessed the likelihood by creating simulations of the CMB. We used Python's \texttt{healpy.sphtfunc.anafast} to compute the power spectrum ($C_\ell$) of the original (unmasked) \textit{Planck} 2015 SMICA map. Note that this extracts $C_\ell$ values from the given map and does not assume a particular power spectrum or underlying cosmology. We then used \texttt{healpy.sphtfunc.synfast} to generate new synthetic maps from these $C_\ell$ values, at full resolution $5'$ FWHM, $N_{side} = 2048$ \citep{Planck2016c}, to match our fiducial \textit{Planck} 2015 SMICA map. After each simulation the CMB temperature was re-sampled. The SNe mask (\textit{Planck} UT78), variances, and weights were left unchanged (as the SNe had not moved) and the OLS linear regression was re-fitted.

We apply these methods to the data and various randomisations in the following sections.

\section{Results: deep field bias} \label{secBias}

\begin{table*}
	\centering
	\caption{SNe fields identified in the SAI sample. `No. SNe' is the number of SNe in each field after restrictions by Galactic latitude ($|b|>40^\circ$), redshift ($z>0.005$), and \textit{Planck} UT78 confidence mask. RA ($\alpha$, J2000) and Dec ($\delta$, J2000) are of the approximate field centres. Angular size is of a square (rectangle) in RA-Dec orientation encompassing each field 1-7 (Stripe 82). The equivalent surface density of SNe per deg$^2$ has been calculated. `No. pixels' is the number of CMB map pixels ($N_{side} = 2048$) whose centres are within the field and which are not masked by \textit{Planck} UT78. `Deep survey field(s)' lists examples coincident with the SNe fields. Values for the `remainder' sample (after removing fields 1-7 and Stripe 82), where applicable, have been shown for comparison.}
	\label{tabSNFields}
	\begin{tabular}{lrrrcrrl}
		\hline
		\multirow{2}{*}{Field} & No. & $\alpha$ (J2000) & $\delta$ (J2000) & Angular & No. SNe & No. & \multirow{2}{*}{Deep survey field(s)} \\
		& SNe & h m s & $^\circ\ '\ ''$ & size & deg$^{-2}$ & pixels & \\
		\hline
		Field 1 & 50 & 14:19:28 & 52:40:28 & $1.1^\circ \times 1.1^\circ$ & 41.3 & 1599 & SNLS D3, EGS \\
		Field 2 & 29 & 12:36:55 & 62:16:40 & $0.4^\circ \times 0.4^\circ$ & 181.3 & 221 & HDF-N, GOODS-N \\
		Field 3 & 54 & 02:31:41 & -08:24:43 & $2.0^\circ \times 2.0^\circ$ & 13.5 & 4944 & ESSENCE wdd \\
		Field 4 & 22 & 02:25:55 & -04:30:58 & $1.1^\circ \times 1.1^\circ$ & 18.2 & 1596 & SNLS D1 \\
		Field 5 & 64 & 02:07:53 & -04:19:04 & $2.0^\circ \times 2.0^\circ$ & 16.0 & 5000 & ESSENCE wcc, NDWFS \\
		Field 6 & 21 & 22:15:36 & -17:42:17 & $1.1^\circ \times 1.1^\circ$ & 17.4 & 1593 & SNLS D4 \\
		Field 7 & 16 & 03:32:26 & -27:38:25 & $0.4^\circ \times 0.4^\circ$ & 100.0 & 219 & CDF-S, GOODS-S \\
		Stripe 82 & 665 & 00:55:00 & 00:00:00 & $90^\circ \times 2.8^\circ$ & 2.8 & 352871 & SDSS Stripe 82 \\
		&&&&&&&\\
		Remainder & 1862 & n/a & n/a & 14474 deg$^2$ & 0.2 & 16971444 & n/a \\
		\hline
	\end{tabular}
\end{table*}

\begin{figure} 
	\includegraphics[width=\columnwidth]{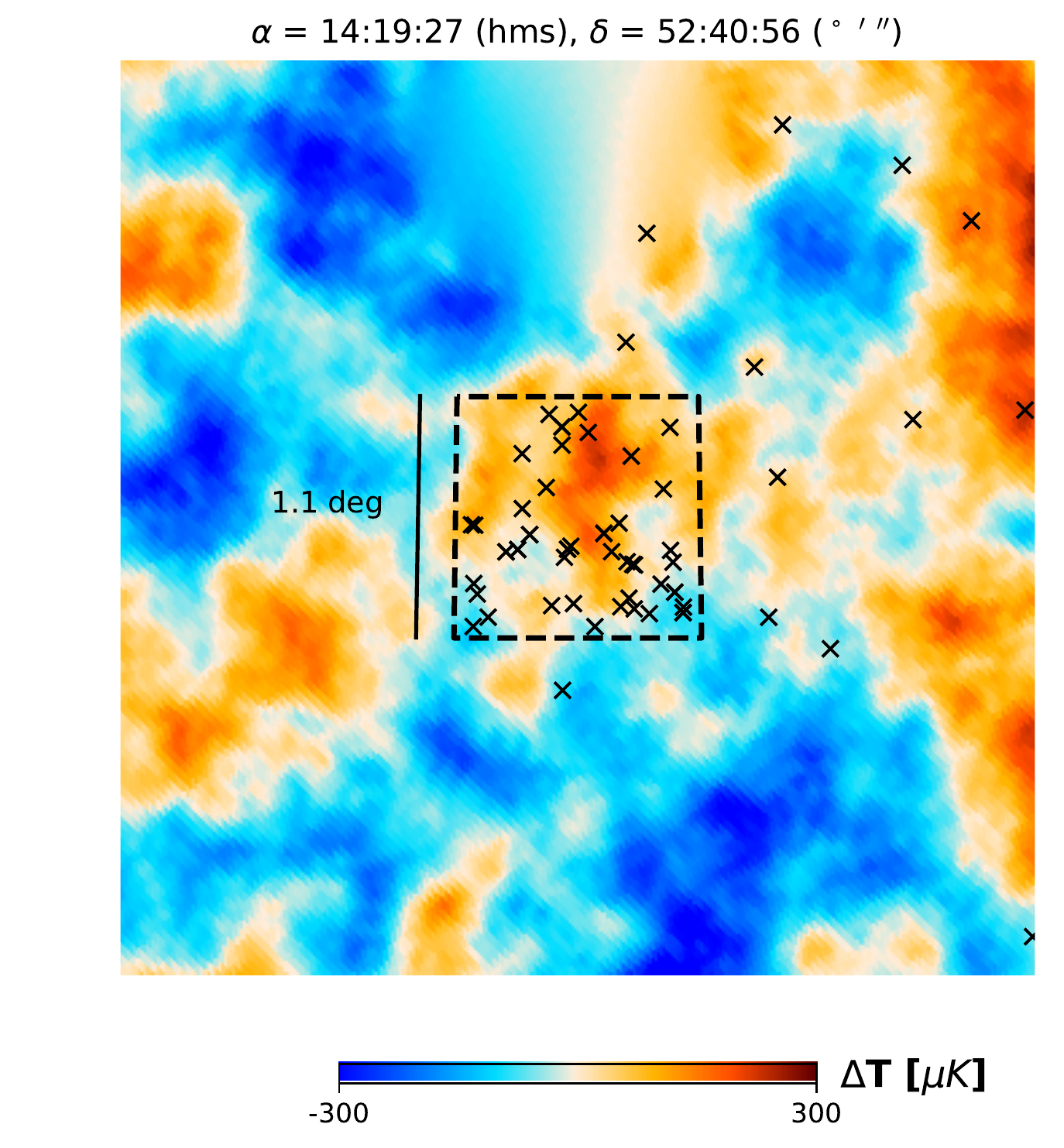}
	\caption{Gnomic projection of the \textit{Planck} 2015 SMICA CMB map in the vicinity of field 1 (colour on-line). The location of SAI sample SNe are plotted with crosses ($\times$). The dashed square is the boundary of field 1, centred on the SNLS D3 deep survey field at the specified coordinates ($\alpha$ and $\delta$, J2000). See Appendix~\ref{appFields1to7} for all fields 1-7.}
	\label{figField1}
\end{figure}

\citet{Yershov2012, Yershov2014} detected a correlation between SNe redshifts and CMB temperature using OLS linear regression and Pearson's correlation coefficient. We verified this correlation using the independent 2-sample Welch's t-test, the 1-sided MWU test, and Spearman's rank-order correlation coefficient, so we do not dispute that a correlation is found.

However, we do not believe that there is any astrophysical origin of the correlation and conclude that it is a composite selection bias caused by the chance alignment of certain deep survey fields with CMB hotspots. In this section we describe our identification of the deep survey fields in question and determine the significance of their contribution to the correlation. We show how their temperature and redshift biases cause this composite selection bias and we quantify the likelihood of it occurring by chance.

\subsection{Identification of fields} \label{subsecFieldID}

SNe are transient objects, historically detected both by chance and by repeated observations of specific fields, galaxy clusters etc. Reliably detecting and following the lightcurves of SNe at high redshift requires particularly targeted approaches \citep[e.g.,][]{Filippenko1998, Dawson2009}. As a result SNe are not evenly detected across the sky and most SNe datasets are not spatially uniform. This situation is changing with wide and time-domain surveys such as DES \citep{DES2016}.

We analysed the SAI sample to identify regions with a high surface density of SNe. Visual inspection of a \textit{Topcat} Sky Plot suggested that defining these as regions containing $\geqslant 20$ SNe with an average surface density of $\geqslant 20$ SNe per square degree would be appropriate and productive. We placed no constraint on the overall angular size of the region. Our algorithm identified 7 SNe fields meeting these criteria, plus 2 additional fields within SDSS Stripe 82.

Table~\ref{tabSNFields} lists the 7 SNe fields plus Stripe 82. These fields contain a total of 921 SNe (33.1 per cent), with Stripe 82 containing 665 SNe (23.9 per cent) and fields 1-7 containing 256 SNe (9.2 per cent) of the sample. For each field 1-7 we identified corresponding deep survey fields coincident with the SNe field. There were 3 fields from the Supernova Legacy Survey \citep[SNLS,][]{Astier2006}, 2 from the ESSENCE supernova survey \citep{Miknaitis2007}, the Hubble Deep Field North \citep[HDF-N,][]{Williams1996}, and the Chandra Deep Field South \citep[CDF-S,][]{Giacconi2001}.

For each field 1-7 we defined a square in RA and Dec orientation consistent with the deep survey field footprint and encompassing the bulk of the SNe identified by our algorithm. For the majority of the fields the best fit was to increase the deep survey field edge lengths by 10 per cent. For field 2 and field 7, coincident with HDF-N and CDF-S respectively, the best fit was to rotate (to RA and Dec orientation) a square enclosing the deep survey field and increase the deep survey field edge lengths by 20 per cent. Note that localising our SNe fields to the corresponding deep survey fields in this way generally reduced their angular size and the number of SNe they contained, which in some cases reduced the number of SNe and/or their surface density below the initial detection thresholds used.

Fig.~\ref{figField1} shows the resultant boundary of field 1 (dashed box), with SNe positions plotted with crosses ($\times$) and CMB temperature shown by the colour-bar (red indicating hotter than average, blue indicating colder than average). For all fields 1-7 see Appendix.~\ref{appFields1to7}.

\subsection{Contribution to correlation} \label{subsecContribution}

To test whether these fields contribute to the correlation we compared the OLS linear regression both with and without them in the sample. We created `remainder' samples containing SNe from the SAI sample minus those in fields 1-7 combined, minus those in Stripe 82, and minus those in fields 1-7 and Stripe 82 together. We calculated the OLS linear regression gradient and Spearman's rank-order correlation coefficient of these remainders and compared them with those of the whole sample. Note that results using WLS linear regression, and values of Pearson's correlation coefficient, were consistent.

\setlength{\tabcolsep}{1pt} 
\begin{table} 
	\centering
	\caption{OLS gradient (with uncertainty of standard error on the gradient) and Spearman's rank-order correlation coefficient for the SAI sample after removing subsets of SNe fields. `No. SNe' is the number of SNe in each `remainder' sample.}
	\label{tabGradients}
	\begin{tabular}{llcrrrclcrll}
		\hline
		\multirow{2}{*}{Fields removed} && No. && \multicolumn{4}{c}{Gradient} && \multicolumn{3}{c}{Corr. Coeff.}\\
		&& SNe && \multicolumn{4}{c}{($\mu K / z$)} && $\rho_s$ && p-value\\
		\hline
		None &~~~& 2783 &~~~&~~~& 61 & $\pm$ & 12 &~~~& 0.5 &~~~& $6.7 \times 10^{-9}$\\
		Fields 1-7 && 2527 &&& -2 & $\pm$ & 22 && 0.1 && 0.6\\
		Stripe 82 && 2118 &&& 54 & $\pm$ & 14 && 0.4 && $4.9 \times 10^{-7}$\\
		Fields 1-7 \& Stripe 82 && 1862 &&& -2 & $\pm$ & 25 && 0.0 && 0.7\\
		\hline
	\end{tabular}
\end{table}
\setlength{\tabcolsep}{6pt}

Table~\ref{tabGradients} shows the gradient of the OLS linear regression slope for each remainder sample in units of $\mu K$ per unit redshift, plus the standard error on the gradient. In these units the gradient of the whole sample is $61 \pm 12 \,\mu K / z$, which is significantly above zero. Spearman's rank-order correlation coefficient for the whole sample shows a moderate correlation ($\rho_s = 0.5$) which is statistically significant (p-value $= 6.7 \times 10^{-9}$).

SDSS Stripe 82 is the largest field we identified, both in terms of the number of SNe (665) and angular size. Therefore the SNe in Stripe 82 cover a wider variety of CMB pixels and any statistical contribution from them should be much less prone to selection bias. Indeed, removing Stripe 82 from the sample does not significantly affect the OLS linear regression slope ($54 \pm 14 \,\mu K / z$) or Spearman's rank-order correlation coefficient ($\rho_s = 0.4$, p-value $= 4.9 \times 10^{-7}$). However, removing fields 1-7 (256 SNe) reduces the gradient dramatically to $-2 \pm 22 \,\mu K / z$, consistent with zero, and there is no correlation evident ($\rho_s = 0.1$, p-value $= 0.6$) in the remainder. Removing both fields 1-7 and Stripe 82 together has a similar effect.

\begin{figure} 
	\centering
	\subfigure[Whole sample (2783 SNe)]{\includegraphics[width=0.8\columnwidth]{binned_scatter_fields_1to7_all_square.pdf} \label{subBinnedScatterFields1to7All}}
	\\
	\subfigure[Fields 1-7 sample (256 SNe)]{\includegraphics[width=0.8\columnwidth]{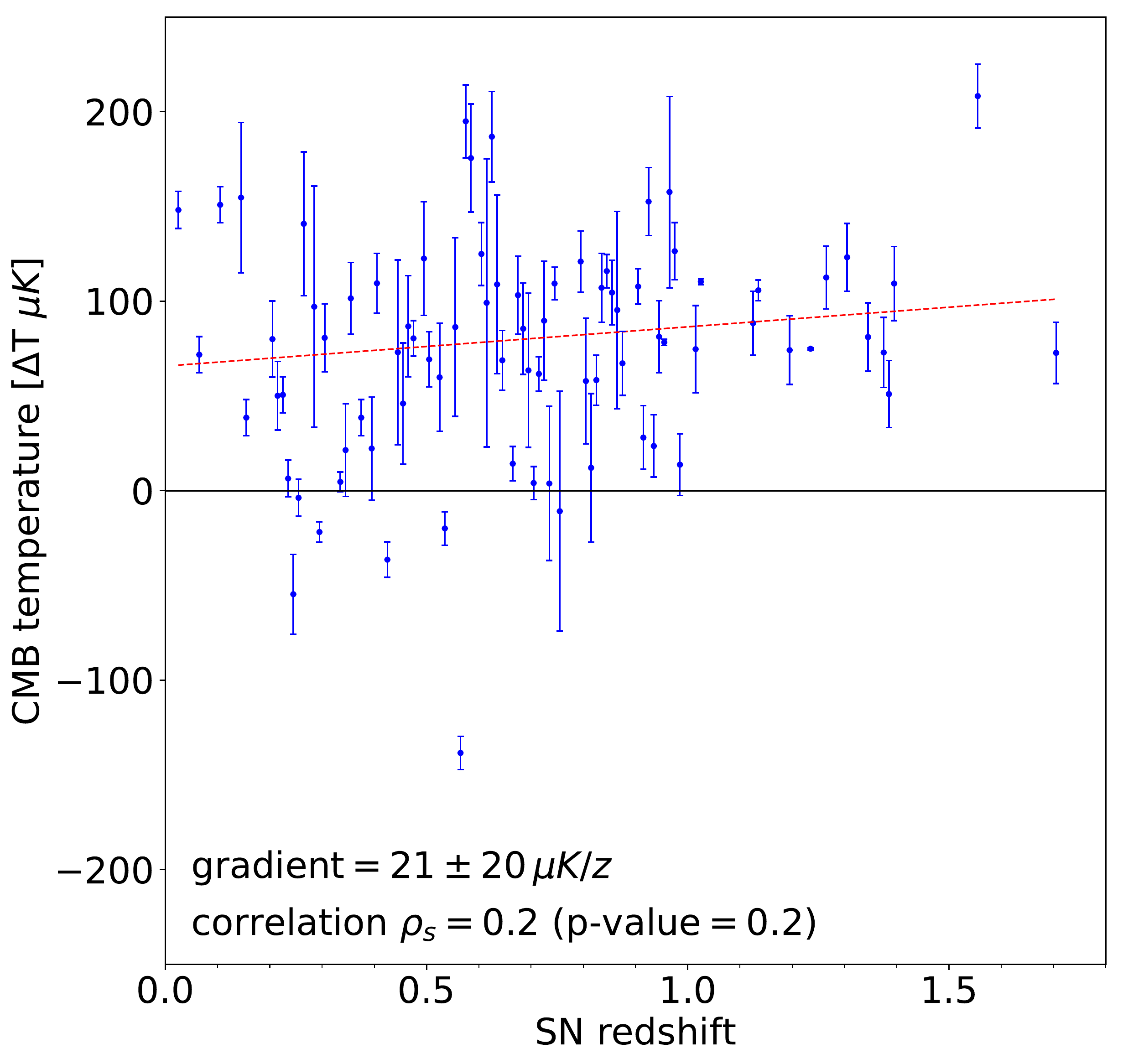} \label{subBinnedScatterFields1to7Fields}}
	\\
	\subfigure[Remainder sample (2527 SNe)]{\includegraphics[width=0.8\columnwidth]{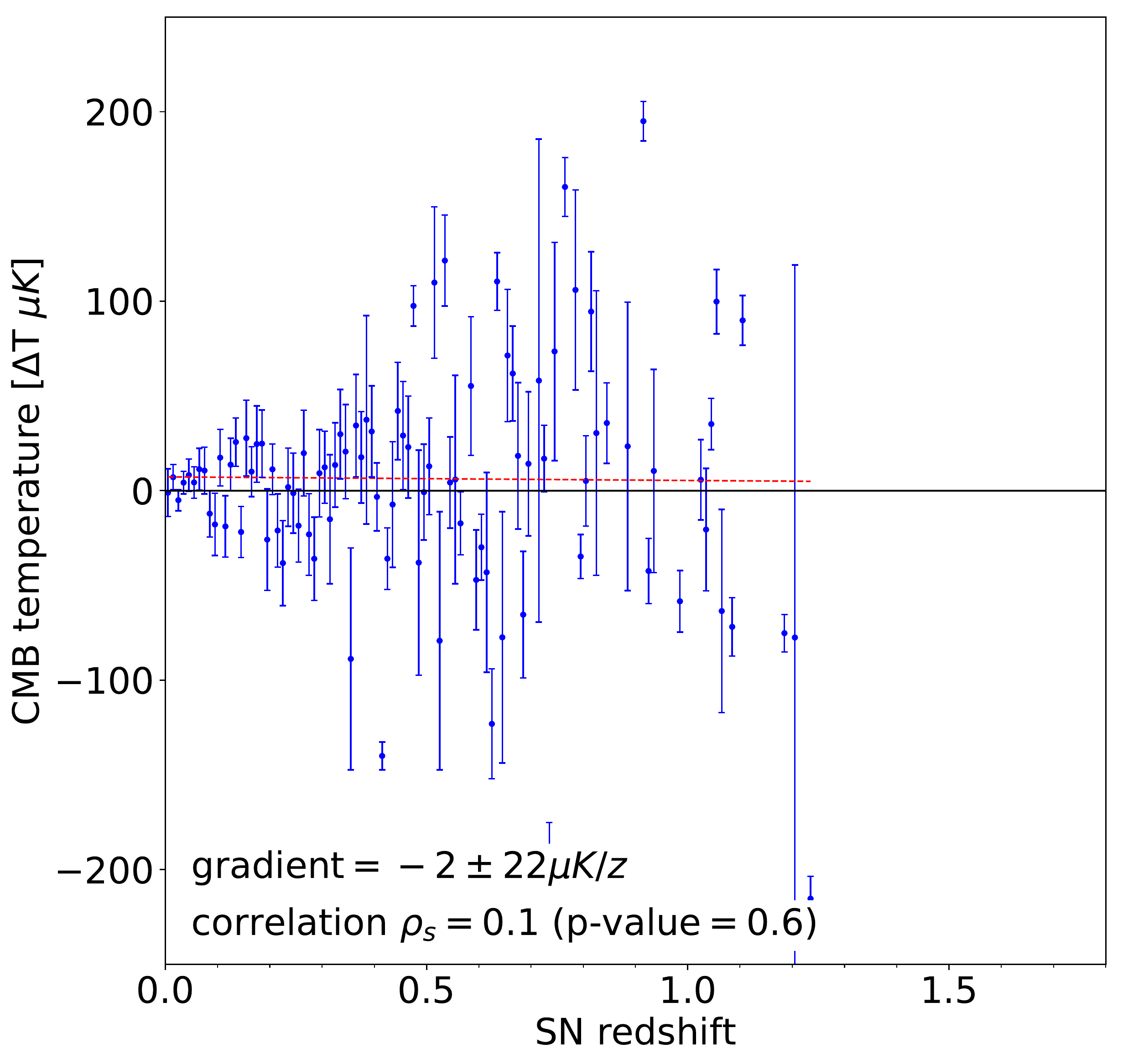} \label{subBinnedScatterFields1to7Remainder}}
	\caption{Plot of CMB temperature at SAI sample SNe locations versus SNe redshift binned with bin sizes $\Delta z=0.01$. Data are restricted by Galactic latitude ($|b|>40^\circ$), redshift ($z>0.005$), and \textit{Planck} UT78 confidence mask. Sub-figures are \protect\subref{subBinnedScatterFields1to7All} the whole sample, \protect\subref{subBinnedScatterFields1to7Fields} fields 1-7 sample, and \protect\subref{subBinnedScatterFields1to7Remainder} remainder (fields 1-7 removed) sample. Error bars are the standard error on the bin mean. The dashed line indicates ordinary least squares linear regression.}
	\label{figBinnedScatterFields1to7}
\end{figure}

The result of removing fields 1-7 from the SAI sample is illustrated in Fig.~\ref{figBinnedScatterFields1to7}, which plots the weighted mean CMB temperature at SNe locations in redshift bins of $\Delta z = 0.01$. This plot is repeated for the whole sample (\ref{subBinnedScatterFields1to7All}), fields 1-7 only (\ref{subBinnedScatterFields1to7Fields}) and the remainder of the sample after fields 1-7 are removed (\ref{subBinnedScatterFields1to7Remainder}). Note that OLS linear regression gradients of unbinned data were consistent.

The OLS gradient and correlation present in the whole sample (gradient $= 61 \pm 12 \,\mu K / z$, $\rho_s = 0.5$, p-value $= 6.7 \times 10^{-9}$) are entirely absent in the remainder (gradient $= -2 \pm 22 \,\mu K / z$, $\rho_s = 0.1$, p-value $= 0.6$). The OLS gradient of the fields 1-7 sample is slightly positive ($21 \pm 20 \,\mu K / z$) but Spearman's rank-order correlation coefficient indicates that there is no significant correlation evident ($\rho_s = 0.2$, p-value $ = 0.2$).

The data clearly indicate that the correlation is caused by fields 1-7 and that SDSS Stripe 82 does not contribute significantly.

\subsubsection{Angular scales} \label{subsubAngularScales}

We checked whether large-scale hot/cold spots, or anisotropies on scales of $\sim 1^\circ$ ($\ell \sim 100$), dominate the apparent correlation. We filtered the \textit{Planck} 2015 SMICA map to remove large angular scales ($\ell < 10$, $\ell < 50$, and $\ell < 100$) and repeated the OLS linear regression and Spearman's rank-order correlation coefficient analyses. In all cases there was no significant correlation evident (e.g., $\ell < 50$, $\rho_s = 0.1$, p-value $= 0.1$). The large angular scales are dominating, as expected, indicating that the CMB map pixels at SNe locations contribute no more than any other pixels within these scales.

This supports our likelihood results (section~\ref{subsecLikelihood}, Fig.~\ref{subShuffleInFields}), which indicate that the CMB map pixels at SNe locations are no more relevant than any other pixels within fields 1-7 (angular size from $0.4^\circ \times 0.4^\circ$ to $2.0^\circ \times 2.0^\circ$).

\subsection{Temperature and redshift} \label{subsecTempRedshift}

We investigated whether the CMB temperature at SNe locations and/or the redshift of SNe within fields 1-7 and Stripe 82 differ from those in the rest of the sample. We calculated the mean CMB temperature ($\overline{T} \pm \sigma_{\overline{T}}$) and mean SNe redshift ($\overline{z} \pm \sigma_{\overline{z}}$) for each sample compared with the remainder. We also analysed the CMB temperature and SNe redshift distributions using the independent 2-sample Welch's t-test and 1-sided MWU test. Note that results using the median CMB temperature at SNe locations, and median SNe redshift, were consistent.

\subsubsection{Mean CMB T and mean SNe $z$} \label{subsubMeanTempRedshift}

\begin{figure*} 
	\includegraphics[width=0.6\textwidth]{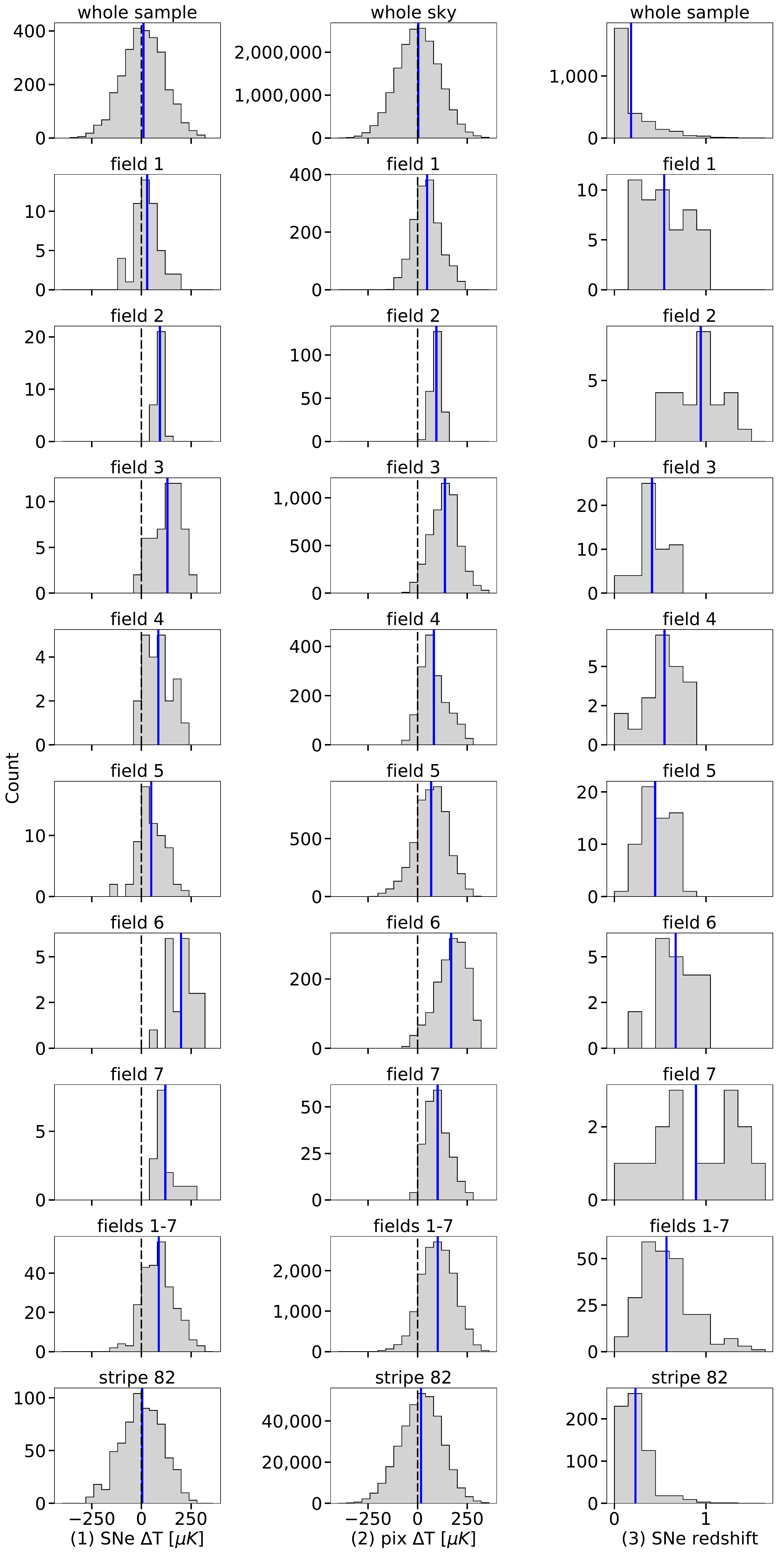}
	\caption{Histograms of SAI sample CMB temperature and SNe redshift distributions for fields 1-7, Stripe 82, and the whole sample or sky. Column (1) shows the CMB temperature distribution at SNe locations within each field sample binned with bin sizes $\Delta T = 40 \,\mu K$. Column (2) shows the CMB temperature distribution of all \textsc{HEALP}ix pixels within each field sample area binned with bin sizes $\Delta T = 40 \,\mu K$. Column (3) shows the redshift distribution of SNe within each field sample binned with bin sizes $\Delta z = 0.15$. The temperature and redshift scales are indicated on the bottom plot of each column. SNe are restricted by Galactic latitude ($|b|>40^\circ$), redshift ($z>0.005$), and \textit{Planck} UT78 confidence mask. Pixels are restricted by Galactic latitude ($|b|>40^\circ$) and \textit{Planck} UT78 confidence mask. Solid vertical lines are arithmetic (unweighted) mean values for each distribution. Dashed vertical lines in columns (1) and (2) are at $\Delta T = 0$.}
	\label{figHistograms}
\end{figure*}

\setlength{\tabcolsep}{1pt} 
\begin{table} 
	\centering
	\caption{Arithmetic (unweighted) mean CMB temperature, at SNe locations within each field sample and of all \textsc{HEALP}ix pixels within each field sample area, and mean SNe redshift for each field sample and the whole sample or sky. SNe are restricted by Galactic latitude ($|b|>40^\circ$), redshift ($z>0.005$), and \textit{Planck} UT78 confidence mask. Pixels are restricted by Galactic latitude ($|b|>40^\circ$) and \textit{Planck} UT78 confidence mask. The uncertainty is the standard error of the mean.}
	\label{tabHistograms}
	\begin{tabular}{lrclcrclcrcl}
		\hline
		\multirow{2}{*}{Field} & \multicolumn{7}{c}{CMB temperature ($\mu K$)} && \multicolumn{3}{c}{Redshift} \\
		& \multicolumn{3}{c}{SNe} && \multicolumn{3}{c}{pixels} && \multicolumn{3}{c}{SNe} \\
		\hline
		Field 1 & 29.1 & $\pm$ & 9.1 &~~~& 47.5 & $\pm$ & 1.7 &~~~& 0.55 & $\pm$ & 0.03  \\
		Field 2 & 92.9 & $\pm$ & 3.3 && 94.3 & $\pm$ & 1.7 && 0.94 & $\pm$ & 0.06 \\
		Field 3 & 130.8 & $\pm$ & 9.2 && 137.5 & $\pm$ & 1.0 && 0.41 & $\pm$ & 0.02 \\
		Field 4 & 85.3 & $\pm$ & 14.3 && 81.7 & $\pm$ & 1.7 && 0.55 & $\pm$ & 0.05 \\
		Field 5 & 49.7 & $\pm$ & 8.5 && 68.2 & $\pm$ & 1.2 && 0.45 & $\pm$ & 0.02 \\
		Field 6 & 199.3 & $\pm$ &13.6 && 168.9 & $\pm$ & 1.9 && 0.67 & $\pm$ & 0.04 \\
		Field 7 & 120.5 & $\pm$ & 12.0 && 100.8 & $\pm$ & 4.0 && 0.89 & $\pm$ & 0.11 \\
		Fields 1-7 & 87.4 & $\pm$ & 5.0 && 101.5 & $\pm$ & 0.7 && 0.57 & $\pm$ & 0.02 \\
		Stripe 82 & 3.8 & $\pm$ & 4.0 && 17.2 & $\pm$ & 0.2 && 0.23 & $\pm$ & 0.01 \\
		Whole sample & 10.3 & $\pm$ & 2.0 && 3.1 & $\pm$ & 0.02 && 0.18 & $\pm$ & 0.00 \\
		\hline
	\end{tabular}
\end{table}
\setlength{\tabcolsep}{6pt}

The mean CMB temperature ($\overline{T} \pm \sigma_{\overline{T}}$) at SNe locations, and of all CMB map \textsc{HEALP}ix pixels within each sample, and mean SNe redshift ($\overline{z} \pm \sigma_{\overline{z}}$) are specified in Table~\ref{tabHistograms}. The samples are fields 1-7 individually, fields 1-7 combined, Stripe 82, and the whole sample. Fig.~\ref{figHistograms} illustrates these distributions, namely CMB temperature at SNe locations (1), CMB temperature of all CMB map \textsc{HEALP}ix pixels within each sample (2), and SNe redshift (3) in these samples. In both Table~\ref{tabHistograms} and Fig.~\ref{figHistograms} SNe are restricted by Galactic latitude ($|b|>40^\circ$), redshift ($z>0.005$), and \textit{Planck} UT78 confidence mask and CMB map \textsc{HEALP}ix pixels are restricted by Galactic latitude ($|b|>40^\circ$) and \textit{Planck} UT78 confidence mask.

SNe in all fields 1-7 are biased to CMB temperatures hotter than the mean of the whole sample ($10.3 \pm 2.0 \,\mu K$). Fields 3, 6 and 7 are particularly extreme, with mean CMB temperatures at SNe locations of $130.8 \pm 9.2 \,\mu K$, $199.3 \pm 13.6 \,\mu K$, and $120.5 \pm 12.0 \,\mu K$ respectively. SNe in SDSS Stripe 82 are not biased to CMB temperatures hotter than the mean of the whole sample. For all the fields, fields 1-7 and Stripe 82, the CMB temperature  distribution (and mean) at SNe locations is generally representative of the CMB map \textsc{HEALP}ix pixel temperature distribution (and mean) to within $\pm \sim 30 \,\mu K$.

SNe in all fields 1-7 are also biased to higher redshift than the mean of the whole sample ($z = 0.18 \pm 0.00$). Fields 2, 6 and 7 are particularly extreme, with mean SNe redshifts of $z = 0.94 \pm 0.06$, $z = 0.67 \pm 0.04$ and $z = 0.89 \pm 0.11$ respectively. SNe in SDSS Stripe 82 are biased to slightly higher redshift of $z = 0.23 \pm 0.01$ than the whole sample. However, although Stripe 82 is deeper it is not hotter, which explains why it does not significantly contribute to the correlation.

\begin{table} 
	\centering
	\caption{Results (p-values) from independent 2-sample Welch's t-tests and 1-sided (greater) MWU tests of CMB temperature and SNe redshift for each field sample. All tests are against a constant remainder sample after removing fields 1-7 and Stripe 82.}
	\label{tabFieldMWU}
	\begin{tabular}{lllll}
		\hline
		\multirow{3}{*}{Field} & \multicolumn{4}{c}{p-values} \\
		& \multicolumn{2}{c}{CMB temperature} & \multicolumn{2}{c}{SN redshift} \\
		& t-test & MWU & t-test & MWU \\
		\hline
		Field 1 & $5.8 \times 10^{-3}$ & $1.8 \times 10^{-2}$ & $5.8 \times 10^{-17}$ & $3.0 \times 10^{-26}$ \\
		Field 2 & $2.9 \times 10^{-31}$ & $9.9 \times 10^{-9}$ & $9.0 \times 10^{-15}$ & $1.7 \times 10^{-19}$ \\
		Field 3 & $9.6 \times 10^{-20}$ & $1.8 \times 10^{-18}$ & $2.3 \times 10^{-20}$ & $3.9 \times 10^{-25}$ \\
		Field 4 & $1.2 \times 10^{-5}$ & $3.5 \times 10^{-5}$ & $9.6 \times 10^{-9}$ & $1.4 \times 10^{-12}$ \\
		Field 5 & $1.0 \times 10^{-6}$ & $1.7 \times 10^{-5}$ & $1.5 \times 10^{-23}$ & $1.9 \times 10^{-29}$ \\
		Field 6 & $3.8 \times 10^{-12}$ & $1.0 \times 10^{-12}$ & $2.9 \times 10^{-11}$ & $1.2 \times 10^{-13}$ \\
		Field 7 & $6.1 \times 10^{-8}$ & $7.6 \times 10^{-7}$ & $6.0 \times 10^{-6}$ & $8.2 \times 10^{-11}$ \\
		Fields 1-7 & $3.8 \times 10^{-42}$ & $1.6 \times 10^{-36}$ & $3.7 \times 10^{-71}$ & $1.2 \times 10^{-112}$ \\
		Stripe 82 & 0.7 & 0.2 & $8.3 \times 10^{-49}$ & $1.8 \times 10^{-118}$ \\
		\hline
	\end{tabular}
\end{table}

\subsubsection{MWU and t-test}

We analysed whether the CMB temperature distribution at SNe locations and/or SNe redshift distribution within fields 1-7 and Stripe 82 differ from the rest of the sample using the independent 2-sample Welch's t-test and 1-sided MWU test. To recap from section~\ref{subsecMethod}, Welch's t-test (or unequal variance t-test) accommodates both the different angular extent of the deep survey fields, and hence difference in variance of the CMB temperature, and the variation in the number of SNe per redshift bin. The MWU test makes fewer assumptions (in particular, the t-test assumes a normal distribution) and is less sensitive to outliers than the t-test. Clearly not all the samples we tested are normally distributed (see Fig.~\ref{figHistograms}) but we have included all the results for completeness.

We performed all the analysis using a constant `remainder' sample created by removing fields 1-7 and Stripe 82 from the sample. This was tested against samples containing SNe from fields 1-7 individually, fields 1-7 combined, and Stripe 82. See Table~\ref{tabFieldMWU} for the results (p-values). Note that comparison between the very small p-values is unlikely to be meaningful.

For CMB temperature both p-values for SDSS Stripe 82 and the MWU p-value for field~1 are above the $\alpha = 1$ per cent significance level. Therefore we cannot reject the null hypotheses that Stripe 82 has the same mean temperature and same temperature distribution as the remainder of the sample, nor the null hypothesis that field 1 has the same temperature distribution as the remainder of the sample.

However, for all the other tests the p-values indicate that the individual field samples do not have the same mean temperature as the remainder, and that the temperature distribution of the fields is significantly hotter than that of the remainder.

For SNe redshift all the p-values of all the samples indicate that the individual fields do not have the same mean redshift as the remainder, and that the redshift distribution of the fields is significantly higher than that of the remainder.

\bigskip

\noindent We have demonstrated that fields 1-7 are biased to hotter CMB temperatures, specifically at SNe locations but also at all CMB map \textsc{HEALP}ix pixels within the fields. We believe this is the result of the chance alignment of those fields with CMB hotspots. This would not on its own be sufficient to lead to the correlation reported by \citet{Yershov2012, Yershov2014}. However, fields 1-7 are also biased to higher redshifts because they are the result of deep survey fields. The remainder of the SNe are generally lower redshift and are spread more uniformly across the sky, so they have a mean CMB temperature closer to the mean of the whole CMB map.

The composite effect is to introduce enough high-redshift SNe at locations of sufficiently high CMB temperature to skew all the analyses we have performed to demonstrate the presence of the correlation, namely OLS linear regression, Welch's t-test, MWU test, and Spearman's rank-order correlation coefficient. This effect was caused by 256 SNe, comprising 9.2 per cent of the restricted SAI sample of 2783 SNe.

\subsection{Likelihood} \label{subsecLikelihood}

We quantified the likelihood of this selection bias happening by chance by analysing the effect on the OLS gradient of moving SNe to random positions within fields 1-7, and by moving fields 1-7 to random positions on the sky. In both analyses the SNe were not moved between fields. We also analysed the effect on the OLS gradient of simulating the CMB sky, without moving the SNe at all.

Within each field 1-7 we moved SNe to 1000 random positions within the field boundaries defined in section~\ref{subsecFieldID}. We also moved each field 1-7 to 10000 random positions on the sky, compliant with the Galactic latitude restriction ($|b|>40^\circ$), whilst keeping the field size and shape constant and the SNe in approximately the same position within each field (within small angle approximation). In both cases all other SNe outside fields 1-7 were left in their original positions. After each move (within each field or of each field on the sky) the masking (\textit{Planck} UT78) was re-applied, the CMB temperature and variance were re-sampled, and the weights were re-calculated.

We created 10000 simulations of the CMB sky from the power spectrum of our fiducial \textit{Planck} 2015 SMICA map, as described in section~\ref{subsecMethod}. All SNe were left in their original positions. After each simulation the CMB temperature was re-sampled. The masking (\textit{Planck} UT78) and variances were left unchanged as the SNe remained on their original CMB map \textsc{HEALP}ix pixel.

Following each move or simulation and subsequent derivations/calculations we binned the data, re-calculated the weighted mean CMB temperature of each bin, fitted an OLS linear regression, and determined the gradient of the slope as previously described (section~\ref{subsecMethod}).

\begin{figure} 
	\subfigure[Move SNe within fields 1-7]{\includegraphics[width=\columnwidth]{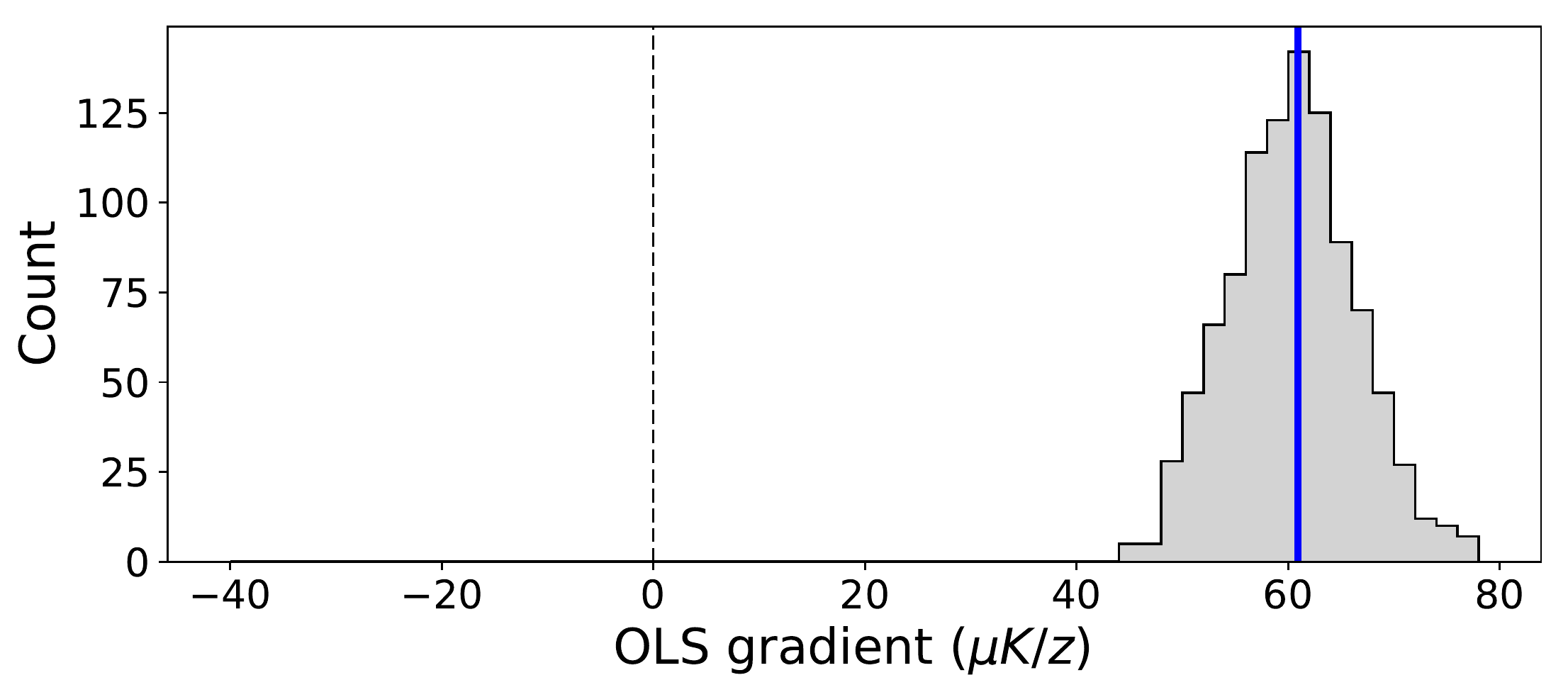} \label{subShuffleInFields}}
	\\
	\subfigure[Move fields 1-7 on sky]{\includegraphics[width=\columnwidth]{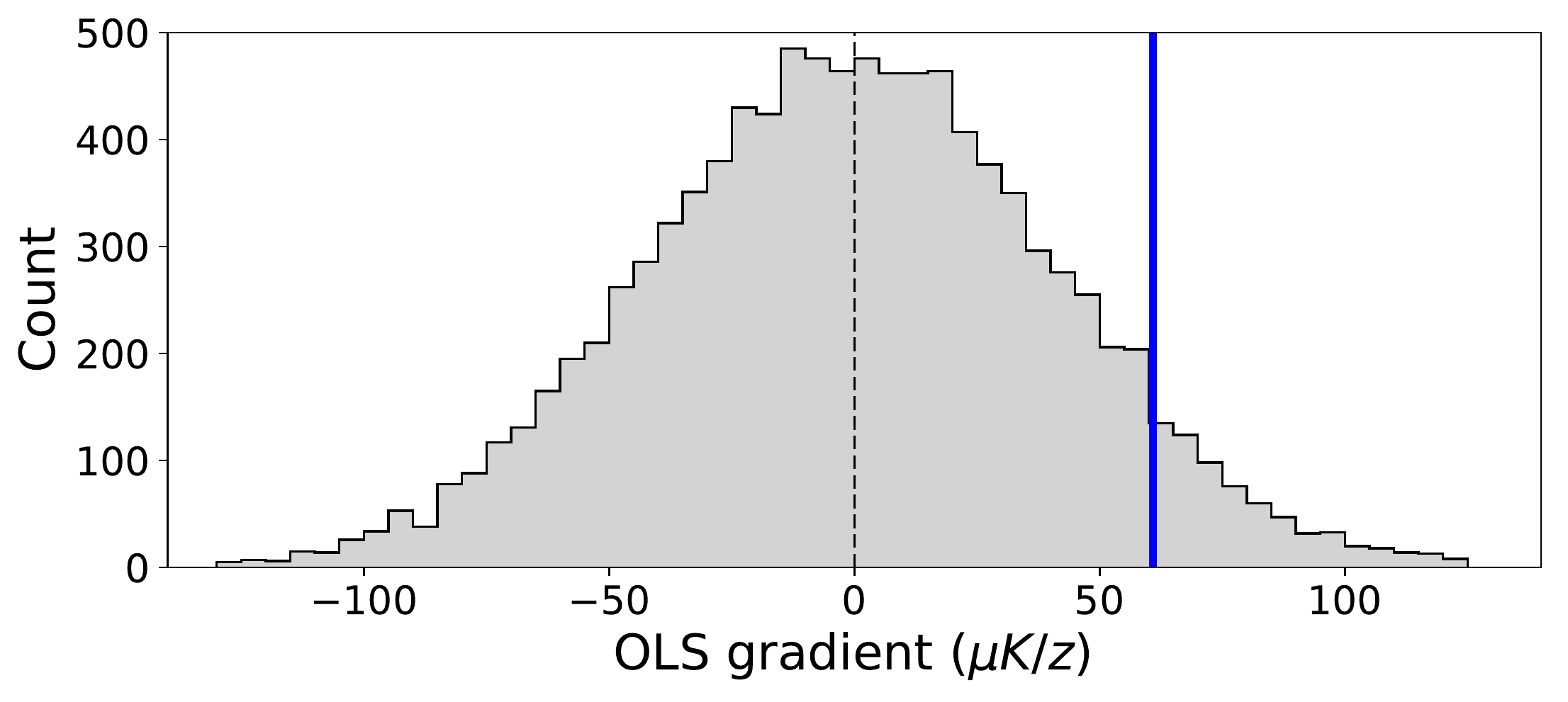} \label{subShuffleOnSky}}
	\\
	\subfigure[Simulate CMB]{\includegraphics[width=\columnwidth]{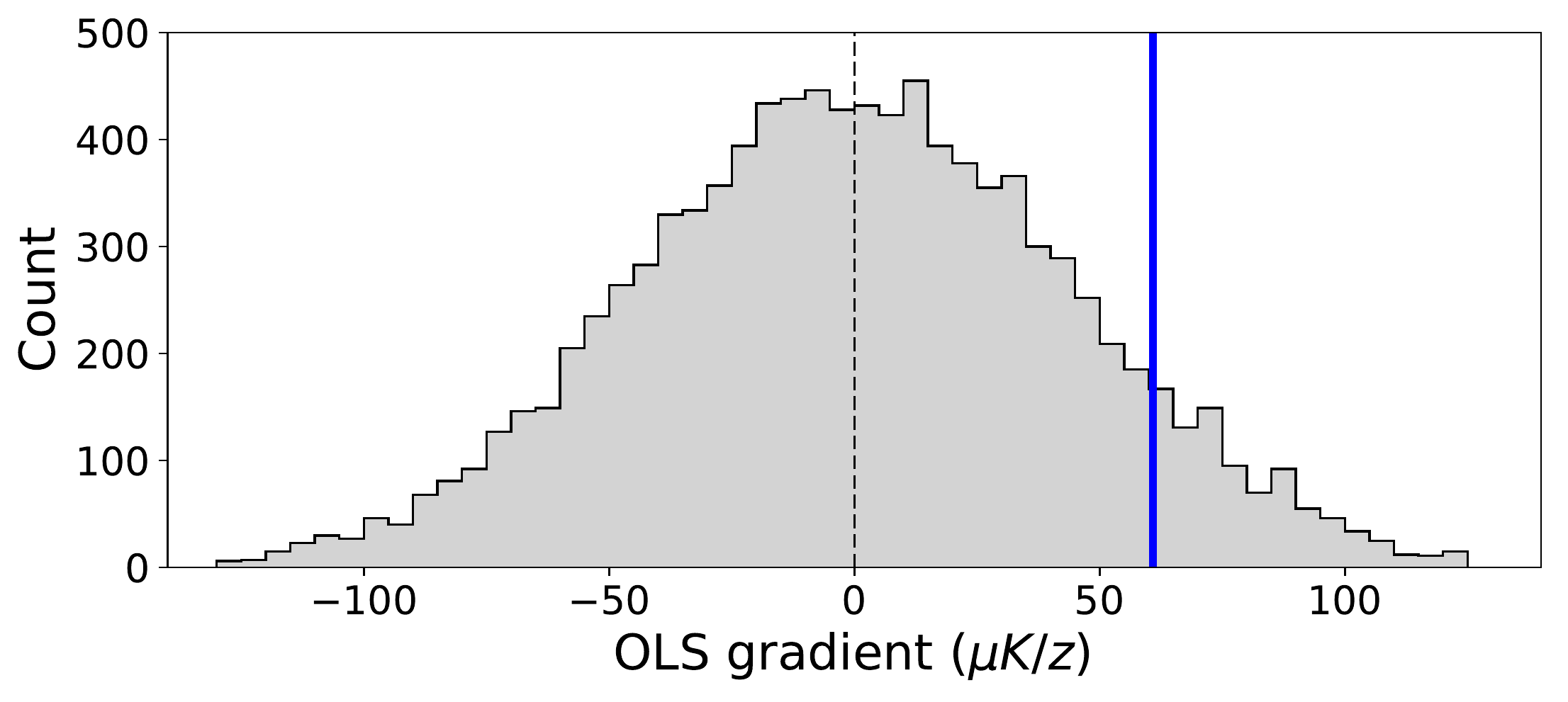} \label{subSimCMB}}
	\caption{Histogram of OLS gradient for the SAI sample after \protect\subref{subShuffleInFields} moving SNe to 1000 random positions within each field 1-7, \protect\subref{subShuffleOnSky} moving fields 1-7 to 10000 random positions on the sky ($|b|>40^\circ$), and \protect\subref{subSimCMB} 10000 simulations of the CMB sky. SNe positions are restricted by Galactic latitude ($|b|>40^\circ$) and \textit{Planck} UT78 confidence mask. Dashed vertical lines are at zero gradient. Solid vertical lines are original value of the gradient.}
	\label{figShuffle}
\end{figure}

Fig.~\ref{figShuffle} shows the distribution of OLS gradients after these random moves and simulations. For comparison, the original gradient ($61 \pm 12 \,\mu K / z$) is shown as a solid vertical line. Note that the uncertainty in the original gradient is the standard error on the gradient as calculated by the OLS linear regression, whereas the uncertainties in the means of the distributions, described below, are the standard errors of the means.

After moving SNe within each field 1-7 to 1000 random positions within the fields, the mean of the OLS gradient distribution (\ref{subShuffleInFields}) is $60 \pm 0 \,\mu K / z$. The distribution is narrow and consistent with the original gradient, indicating that the position of SNe within fields 1-7 does not significantly affect the correlation.

After moving each field 1-7 to 10000 random positions on the sky, the mean of the OLS gradient distribution (\ref{subShuffleOnSky}) is $-1 \pm 0 \,\mu K / z$. The distribution is wide, centred near zero, and inconsistent with the original gradient, unsurprisingly indicating that the position of fields 1-7 on the sky is responsible for the correlation.

After 10000 simulations of the CMB sky, the mean of the OLS gradient distribution (\ref{subSimCMB}) is $-0 \pm 1 \,\mu K / z$. The distribution is consistent with the results from moving each field 1-7 to 10000 random positions on the sky.

Assuming a standard normal distribution we calculated the z-score (standard score) of the original OLS gradient (X) as

\begin{ceqn}
\begin{equation}
z = (X - \mu) / \sigma \ ,
\end{equation}
\end{ceqn}

\noindent where $\mu$ is the mean of the gradient distribution and $\sigma$ is its standard deviation. We then used the standard normal distribution table to provide the probability of observing a gradient at least as extreme as X within our gradient distributions. For moving fields 1-7 on the sky (\ref{subShuffleOnSky}) the probability is 6.8 per cent (approximately 1 in 15) and for simulating the CMB (\ref{subSimCMB}) it is 8.9 per cent (approximately 1 in 11). Therefore the chance alignment of fields 1-7 with CMB hotspots is not an exceptionally unlikely event.

\section{Results: alternative data} \label{secAlternativeData}

\subsection{SNe types} \label{subsecSNTypes}

\citet{Yershov2014} demonstrated that the correlation between SNe redshifts and CMB temperature was particularly strong for the SNIa sub-sample, whereas for the rest of the SNe it vanished. Is this consistent with our assertion that the correlation is the result of a composite selection bias caused by the chance alignment of certain deep survey fields (fields 1-7) with CMB hotspots?

\begin{table}
	\centering
	\caption{Number and proportion (per cent) of SNIa within the whole SAI sample, fields 1-7 sample, and remainder (fields 1-7 removed) sample.}
	\label{tabSNIa}
	\begin{tabular}{lrrr}
		\hline
		\multirow{2}{*}{Sub-sample} & No. & No. & \% \\
		& SNe & SNIa & SNIa \\
		\hline
		Whole sample & 2783 & 1,749 & 62.8\% \\
		Fields 1-7 & 256 & 235 & 91.8\% \\
		Remainder & 2527 & 1,514 & 59.9\% \\
		\hline
	\end{tabular}
\end{table}

Table~\ref{tabSNIa} shows the number and proportion of Type Ia SNe in our SNe samples. Supernova surveys such as SNLS and ESSENCE primarily targeted SNIa \citep{Pritchet2005, Miknaitis2007}, so it is unsurprising that fields 1-7 contain predominantly SNIa (91.8 per cent). As expected in the whole sample, a little over half the SNe are SNIa. Removing fields 1-7 from the sample does not significantly decrease the proportion of SNIa, which drops from 62.8 per cent in the whole sample to 59.9 per cent in the remainder. However, we have shown that the correlation present in the whole sample (Fig.~\ref{subBinnedScatterFields1to7All}) is entirely absent in this remainder (Fig.~\ref{subBinnedScatterFields1to7Remainder})

Fields 1-7 together comprise 9.2 per cent of the whole sample, but when the sample is restricted to SNIa only this increases to 13.4 per cent. Thus, restricting the sample to SNIa increases the influence of fields 1-7. We suggest that the correlation is not caused by SNIa themselves, but that it is inadvertently enhanced by restricting the sample to SNIa due to the dominance of SNIa in fields 1-7.

\subsection{SNe catalogues} \label{subsecOSC}

We have demonstrated that the correlation reported by \citet{Yershov2012, Yershov2014} is a composite selection bias caused by the chance alignment of certain deep survey fields with CMB hotspots. \citet{Yershov2012, Yershov2014} analysed the Sternberg Astronomical Institute \citep[SAI,][]{Bartunov2007} SNe catalogue, but it seems reasonable that other SNe catalogues could show a similar effect.

We repeated our analyses from section~\ref{secBias} using the Open Supernova Catalogue \citep[OSC,][]{Guillochon2017}. Data were obtained, restricted and weighted as described in section~\ref{secDataMethod}, yielding a sample of 7880 SNe. We found that the OSC sample does indeed exhibit a similar apparent correlation (with a gradient of $42 \pm 7 \,\mu K / z$) to the SAI sample.

We applied the same SNe field detection algorithm with the same detection thresholds described in section~\ref{subsecFieldID} to the OSC sample.

\begin{table}
	\centering
	\caption{SNe fields identified in the SAI and OSC samples and the number of SNe within each. See Table~\ref{tabSNFields} for field positions, sizes and further information.}
	\label{tabSNFieldsOSC}
	\begin{tabular}{lrr}
		\hline
		\multirow{2}{*}{Field} & \multicolumn{2}{c}{No. SNe} \\
		& SAI & OSC \\
		\hline
		Field 1 & 50 & 152 \\
		Field 2 & 29 & 91 \\
		Field 3 & 54 & 79 \\
		Field 4 & 22 & 116 \\
		Field 5 & 64 & 91 \\
		Field 6 & 21 & 92 \\
		Field 7 & 16 & 55 \\
		Stripe 82 & 665 & 2445 \\
		\hline
	\end{tabular}
\end{table}

Our algorithm identified the same 7 SNe fields with the same boundaries, but with somewhat increased SNe membership, plus 13 fields within Stripe 82. These fields (Table~\ref{tabSNFieldsOSC}) contain a total of 3,121 SNe (39.6 per cent), with Stripe 82 containing 2,445 SNe (31.0 per cent) and fields 1-7 containing 676 SNe (8.6 per cent) of the OSC sample. We compared the OLS linear regression both with and without these fields in the sample and calculated Spearman's rank-order correlation coefficient of these samples, as described in section~\ref{subsecContribution}.

Table~\ref{tabGradientsOSC} shows the gradient of the OLS linear regression slope for each OSC remainder sample in units of $\mu K$ per unit redshift, plus the standard error on the gradient. In these units the gradient of the whole sample is $42 \pm 7 \,\mu K / z$, which as for the SAI sample is significantly above zero. Spearman's rank-order correlation coefficient for the whole sample shows a moderate correlation ($\rho_s = 0.6$) which is statistically significant (p-value $= 2.0 \times 10^{-15}$). These results are consistent with those for the whole SAI sample (gradient $= 61 \pm 12 \,\mu K / z$, $\rho_s = 0.5$ and p-value $= 6.7 \times 10^{-9}$).

SDSS Stripe 82 is again the largest field we identified, both in terms of the number of SNe (2,445) and angular size. Removing Stripe 82 from the OSC sample does not significantly affect the OLS linear regression slope ($38 \pm 7 \,\mu K / z$) or Spearman's rank-order correlation coefficient ($\rho_s = 0.5$, p-value $= 1.1 \times 10^{-11}$). However, removing fields 1-7 (676 SNe) reduces the gradient dramatically to $10 \pm 11 \,\mu K / z$, and there is no correlation evident in the remainder ($\rho_s = 0.1$, p-value $= 0.5$). Removing both fields 1-7 and Stripe 82 together has a similar effect.

\setlength{\tabcolsep}{1pt} 
\begin{table} 
	\centering
	\caption{OLS gradient (with uncertainty of standard error on the gradient) and Spearman's rank-order correlation coefficient for the OSC sample after removing subsets of SNe fields. `No. SNe' is the number of SNe in each `remainder' sample.}
	\label{tabGradientsOSC}
	\begin{tabular}{llcrrrclcrll}
		\hline
		\multirow{2}{*}{Fields removed} && No. && \multicolumn{4}{c}{Gradient} && \multicolumn{3}{c}{Corr. Coeff.}\\
		&& SNe && \multicolumn{4}{c}{($\mu K / z$)} && $\rho_s$ && p-value\\
		\hline
		None &~~~& 7880 &~~~&~~~& 42 & $\pm$ & 7 &~~~& 0.6 &~~~& $2.0 \times 10^{-15}$\\
		Fields 1-7 && 7204 &&& 10 & $\pm$ & 11 && 0.1 && 0.5\\
		Stripe 82 && 5435 &&& 38 & $\pm$ & 7 && 0.5 && $1.1 \times 10^{-11}$\\
		Fields 1-7 \& Stripe 82 && 4759 &&& 11 & $\pm$ & 13 && -0.0 && 0.1\\
		\hline
	\end{tabular}
\end{table}
\setlength{\tabcolsep}{6pt}

\begin{figure} 
	\centering
	\subfigure[Whole sample (7880 SNe)]{\includegraphics[width=0.8\columnwidth]{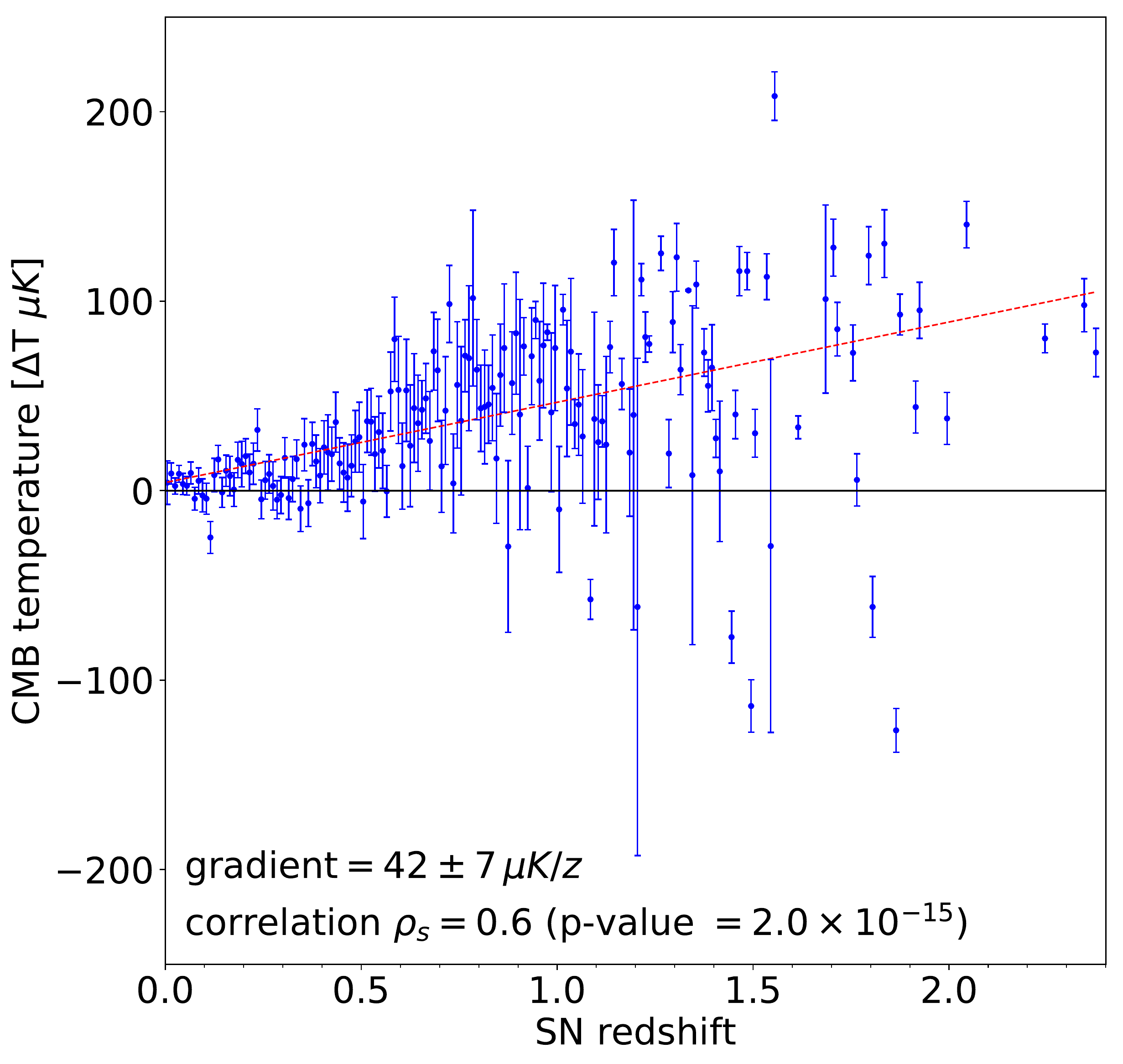} \label{subBinnedScatterFields1to7AllOSC}}
	\\
	\subfigure[Fields 1-7 sample (676 SNe)]{\includegraphics[width=0.8\columnwidth]{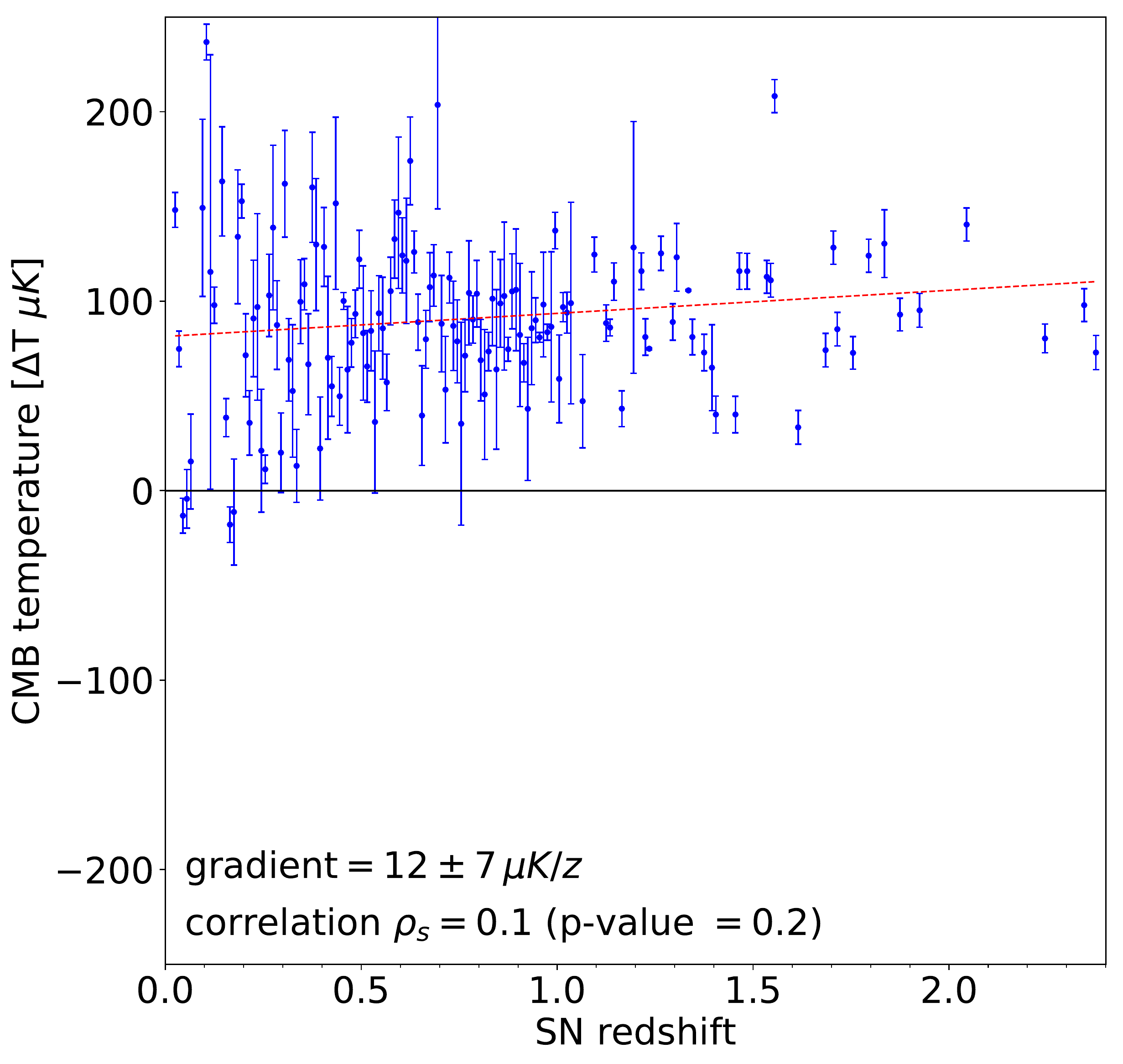} \label{subBinnedScatterFields1to7FieldsOSC}}
	\\
	\subfigure[Remainder sample (7204 SNe)]{\includegraphics[width=0.8\columnwidth]{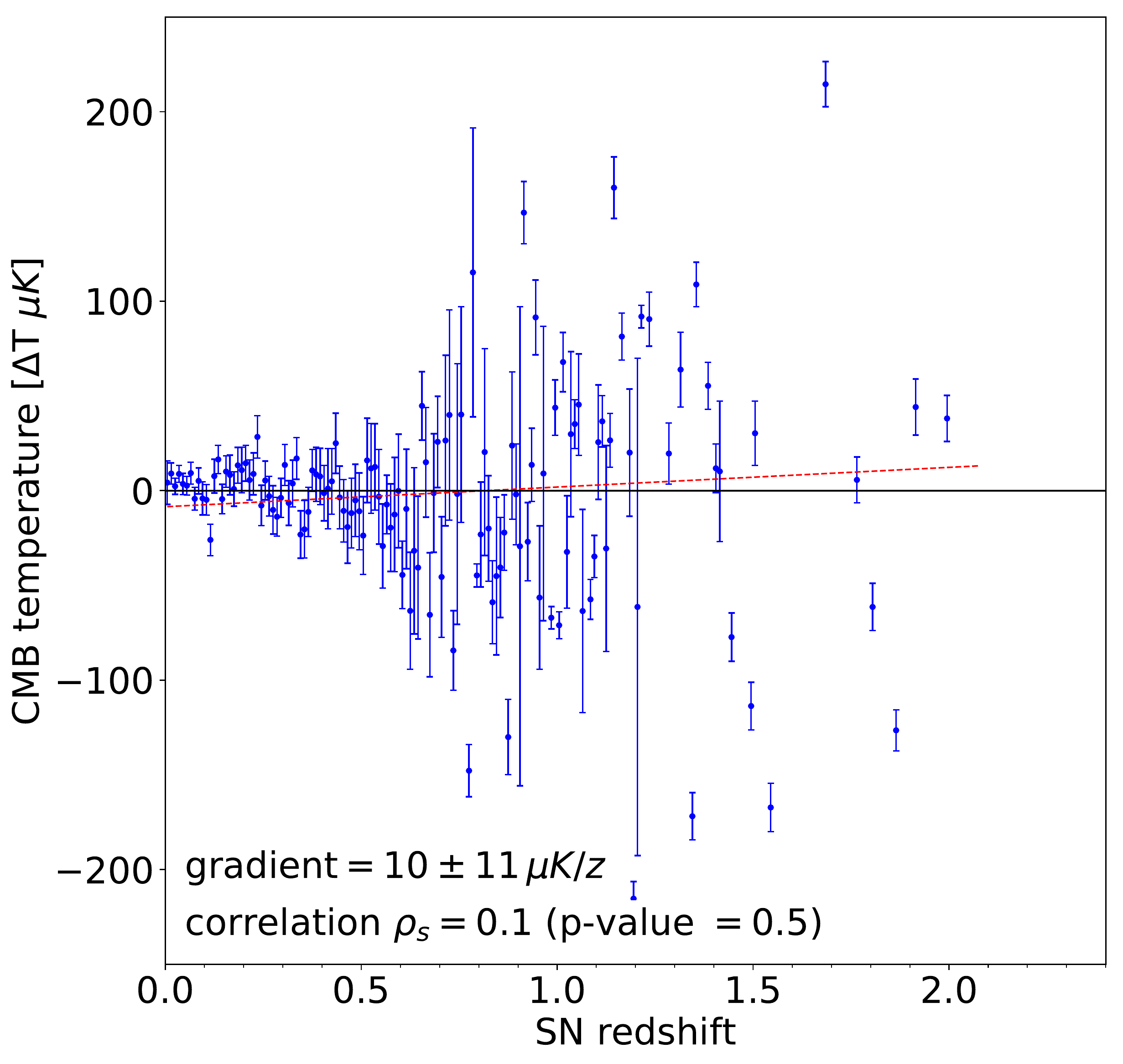} \label{subBinnedScatterFields1to7RemainderOSC}}
	\caption{Same as Fig.~\ref{figBinnedScatterFields1to7} but for the OSC sample.}
	\label{figBinnedScatterFields1to7OSC}
\end{figure}

The result of removing fields 1-7 from the OSC sample is illustrated in Fig.~\ref{figBinnedScatterFields1to7OSC}, which plots the weighted mean CMB temperature at SNe locations in redshift bins of $\Delta z = 0.01$. This plot is repeated for the whole sample (\ref{subBinnedScatterFields1to7AllOSC}), fields 1-7 only (\ref{subBinnedScatterFields1to7FieldsOSC}) and the remainder of the sample after fields 1-7 are removed (\ref{subBinnedScatterFields1to7RemainderOSC}).

The results for the OSC sample indicate that the correlation is caused by fields 1-7 and that SDSS Stripe 82 does not contribute significantly, which is consistent with those for the SAI sample.

The results of our final two OSC sample analyses, namely determining the temperature and redshift biases of fields 1-7 and quantifying the likelihood of the selection bias happening by chance, are entirely consistent with those for the SAI sample (sections~\ref{subsecTempRedshift} and~\ref{subsecLikelihood} respectively).

\subsection{\textit{Planck} CMB maps} \label{subsecMaps}

Both our analysis and that of \citet{Yershov2014} used maps produced by the \textit{Planck} SMICA component separation pipeline (\textit{Planck} 2015 SMICA R2.01 and \textit{Planck} 2013 SMICA R1.20 respectively). To check our results are consistent across all four of the \textit{Planck} component separation pipelines we repeated selected analyses from section~\ref{secBias} using the \textit{Planck} 2015 Commander, NILC, and SEVEM CMB maps. In all cases the pixel variance estimates were calculated from the corresponding HMHD maps as described in section~\ref{subsecMethod}.

We repeated the OLS linear regression gradient and Spearman's rank-order correlation coefficient analyses from section~\ref{subsecContribution}. For all four maps (Commander, NILC, SEVEM, and SMICA) the OLS gradient and correlation present in the whole sample are entirely absent in the remainder once fields 1-7 are removed. For SMICA the contribution of fields 1-7 to the correlation was illustrated in Fig.~\ref{figBinnedScatterFields1to7}. For Commander, NILC, and SEVEM see Appendix~\ref{appAltPlanckMaps} Figs.~\ref{figBinnedScatterFields1to7Commander}, \ref{figBinnedScatterFields1to7nilc}, and \ref{figBinnedScatterFields1to7sevem} respectively.

We repeated the mean CMB temperature analysis from section~\ref{subsubMeanTempRedshift}. For all four maps SNe in fields 1-7, and all \textsc{HEALP}ix pixels within each sample, are biased to CMB temperatures hotter than the mean of the whole sample. In all cases fields 3, 6, and 7 are particularly extreme(Appendix~\ref{appAltPlanckMaps} Table~\ref{tabHistogramsAltPlanckMaps}).

Our results are entirely consistent across all four \textit{Planck} maps.

\section{Discussion and conclusions} \label{secConclusions}

We have shown that the apparent correlation of CMB temperature and SNe redshift reported by \citet{Yershov2012, Yershov2014} using OLS linear regression, Pearson's correlation coefficient, and an SAI SNe sample, is also evident using Spearman's rank-order correlation coefficient, Welch's t-test, and MWU test, and it is discernible in at least one other SNe sample (OSC).

Whilst our analysis supports the prima facie existence of the apparent correlation, the data indicate that it is actually a composite selection bias (high CMB T $\times$ high SNe $z$) caused by the accidental alignment of seven deep survey fields (fields 1-7) with CMB hotspots. These fields include 3 from the Supernova Legacy Survey, 2 from the ESSENCE supernova survey, HDF-N and CDF-S. These comprise 9.2 per cent of the SAI sample and 8.6 per cent of the OSC sample. These deep fields by their very nature contain SNe at higher redshift than the remainder of the samples. We have shown that the SNe within fields 1-7 are also biased to hotter CMB temperature than the remainder of the samples. Our results are consistent across all four of the \textit{Planck} maps.

We have quantified the likelihood of fields 1-7 falling on CMB hotspots by chance and have found this to be at least 6.8 per cent, or approximately 1 in 15. We conclude that the correlation reported by \citet{Yershov2012, Yershov2014} is a composite selection bias caused by the chance alignment of certain deep survey fields with CMB hotspots. This bias (high CMB T $\times$ high SNe $z$) is the combined result of both a selection bias (high $z$ SNe in deep fields) and the chance alignment of those deep fields with CMB hotspots.

This selection bias results in heteroscedastic data, where the variance of CMB temperature at SNe locations is unequal across the range of redshifts. We have shown that high redshift SNe tend to be in deep survey fields which, given the chance alignments, generally give hot \textit{Planck} pixel temperatures. Low redshift SNe are more uniformly scattered across the sky and thus have much wider variance of hot and cold \textit{Planck} pixel temperatures. This heteroscedasticity was hidden by binning the data.

This paper shows that deep survey fields have biased SNe cross-correlation with CMB temperature, but the implications could extend further. Deep fields could potentially bias any cross-correlation between astronomical objects (e.g., SNe, galaxies, GRBs, quasars) and the CMB. It is conceivable that deep fields could, by chance, also be aligned with distant large-scale structures, voids, cosmic bulk flows, or even regions of anisotropic cosmic expansion (should they exist).

Furthermore, perhaps the spatial non-uniformity of SNe datasets could help explain some of the tensions that have been reported between and within them \citep[e.g.,][]{Choudhury2005, Nesseris2007, BuenoSanchez2009, Karpenka2015}.

\section*{Acknowledgements}

We are indebted to the anonymous referee for helpful and constructive comments. We are grateful for advice on calculating individual pixel variances received from the NASA/IPAC Infrared Science Archive (IRSA) Help Desk and the \textit{Planck} Legacy Archive (PLA) Helpdesk. We thank Christian Reichardt for reading and commenting on the draft manuscript. TF is in receipt of a STFC PhD studentship. SR acknowledges support from the Australian Research Council's Discovery Projects scheme (DP150103208).

This research has made use of the PLA maintained by the European Space Agency, the IRSA, which is operated by the Jet Propulsion Laboratory, California Institute of Technology, under contract with the National Aeronautics and Space Administration (NASA), and NASA's Astrophysics Data System. It has used data from the supernova catalogue maintained by the Sternberg Astronomical Institute at Moscow State University, and the Open Supernova Catalogue maintained by James Guillochon and Jerod Parrent at Harvard University. The analysis used the \textsc{HEALP}ix pixelisation scheme and software packages maintained by NASA's Jet Propulsion Laboratory at the California Institute of Technology.




\bibliographystyle{mnras}
\bibliography{bibliography} 

\begin{thebibliography}{}
\makeatletter
\relax
\def\mn@urlcharsother{\let\do\@makeother \do\$\do\&\do\#\do\^\do\_\do\%\do\~}
\def\mn@doi{\begingroup\mn@urlcharsother \@ifnextchar [ {\mn@doi@}
  {\mn@doi@[]}}
\def\mn@doi@[#1]#2{\def\@tempa{#1}\ifx\@tempa\@empty \href
  {http://dx.doi.org/#2} {doi:#2}\else \href {http://dx.doi.org/#2} {#1}\fi
  \endgroup}
\def\mn@eprint#1#2{\mn@eprint@#1:#2::\@nil}
\def\mn@eprint@arXiv#1{\href {http://arxiv.org/abs/#1} {{\tt arXiv:#1}}}
\def\mn@eprint@dblp#1{\href {http://dblp.uni-trier.de/rec/bibtex/#1.xml}
  {dblp:#1}}
\def\mn@eprint@#1:#2:#3:#4\@nil{\def\@tempa {#1}\def\@tempb {#2}\def\@tempc
  {#3}\ifx \@tempc \@empty \let \@tempc \@tempb \let \@tempb \@tempa \fi \ifx
  \@tempb \@empty \def\@tempb {arXiv}\fi \@ifundefined
  {mn@eprint@\@tempb}{\@tempb:\@tempc}{\expandafter \expandafter \csname
  mn@eprint@\@tempb\endcsname \expandafter{\@tempc}}}

\bibitem[\protect\citeauthoryear{{Alam} et~al.,}{{Alam}
  et~al.}{2015}]{Alam2015}
{Alam} S.,  et~al., 2015, \mn@doi [\apjs] {10.1088/0067-0049/219/1/12}, 219, 12

\bibitem[\protect\citeauthoryear{{Astier} et~al.,}{{Astier}
  et~al.}{2006}]{Astier2006}
{Astier} P.,  et~al., 2006, \mn@doi [\aap] {10.1051/0004-6361:20054185}, 447,
  31

\bibitem[\protect\citeauthoryear{{Bartunov}, {Tsvetkov}  \&
  {Pavlyuk}}{{Bartunov} et~al.}{2007}]{Bartunov2007}
{Bartunov} O.~S.,  {Tsvetkov} D.~Y.,   {Pavlyuk} N.~N.,  2007, \mn@doi
  [Highlights Astron.] {10.1017/S1743921307010812}, 14, 316

\bibitem[\protect\citeauthoryear{{Bennett} et~al.,}{{Bennett}
  et~al.}{2013}]{Bennett2013}
{Bennett} C.~L.,  et~al., 2013, \mn@doi [\apjs] {10.1088/0067-0049/208/2/20},
  208, 20

\bibitem[\protect\citeauthoryear{{Bueno Sanchez}, {Nesseris}  \&
  {Perivolaropoulos}}{{Bueno Sanchez} et~al.}{2009}]{BuenoSanchez2009}
{Bueno Sanchez} J.~C.,  {Nesseris} S.,   {Perivolaropoulos} L.,  2009, \mn@doi
  [\jcap] {10.1088/1475-7516/2009/11/029}, 11, 029

\bibitem[\protect\citeauthoryear{{Cardoso}, {Le Jeune}, {Delabrouille},
  {Betoule}  \& {Patanchon}}{{Cardoso} et~al.}{2008}]{Cardoso2008}
{Cardoso} J.-F.,  {Le Jeune} M.,  {Delabrouille} J.,  {Betoule} M.,
  {Patanchon} G.,  2008, \mn@doi [IEEE Journal of Selected Topics in Signal
  Processing] {10.1109/JSTSP.2008.2005346}, 2, 735

\bibitem[\protect\citeauthoryear{{Choudhury} \& {Padmanabhan}}{{Choudhury} \&
  {Padmanabhan}}{2005}]{Choudhury2005}
{Choudhury} T.~R.,  {Padmanabhan} T.,  2005, \mn@doi [\aap]
  {10.1051/0004-6361:20041168}, 429, 807

\bibitem[\protect\citeauthoryear{{DES Collaboration} et~al.,}{{DES
  Collaboration} et~al.}{2016}]{DES2016}
{DES Collaboration} et~al., 2016, \mn@doi [\mnras] {10.1093/mnras/stw641}, 460,
  1270

\bibitem[\protect\citeauthoryear{{DES Collaboration} et~al.,}{{DES
  Collaboration} et~al.}{2017}]{DES2017}
{DES Collaboration} et~al., 2017, preprint (\mn@eprint {arXiv} {1708.01530})

\bibitem[\protect\citeauthoryear{{Dawson} et~al.,}{{Dawson}
  et~al.}{2009}]{Dawson2009}
{Dawson} K.~S.,  et~al., 2009, \mn@doi [\aj] {10.1088/0004-6256/138/5/1271},
  138, 1271

\bibitem[\protect\citeauthoryear{{Delabrouille}, {Cardoso}, {Le Jeune},
  {Betoule}, {Fay}  \& {Guilloux}}{{Delabrouille}
  et~al.}{2009}]{Delabrouille2009}
{Delabrouille} J.,  {Cardoso} J.-F.,  {Le Jeune} M.,  {Betoule} M.,  {Fay} G.,
   {Guilloux} F.,  2009, \mn@doi [\aap] {10.1051/0004-6361:200810514}, 493, 835

\bibitem[\protect\citeauthoryear{{Eriksen}, {Jewell}, {Dickinson}, {Banday},
  {G{\'o}rski}  \& {Lawrence}}{{Eriksen} et~al.}{2008}]{Eriksen2008}
{Eriksen} H.~K.,  {Jewell} J.~B.,  {Dickinson} C.,  {Banday} A.~J.,
  {G{\'o}rski} K.~M.,   {Lawrence} C.~R.,  2008, \mn@doi [\apj]
  {10.1086/525277}, 676, 10

\bibitem[\protect\citeauthoryear{{Fern{\'a}ndez-Cobos}, {Vielva}, {Barreiro}
  \& {Mart{\'{\i}}nez-Gonz{\'a}lez}}{{Fern{\'a}ndez-Cobos}
  et~al.}{2012}]{FernandezCobos2012}
{Fern{\'a}ndez-Cobos} R.,  {Vielva} P.,  {Barreiro} R.~B.,
  {Mart{\'{\i}}nez-Gonz{\'a}lez} E.,  2012, \mn@doi [\mnras]
  {10.1111/j.1365-2966.2011.20182.x}, 420, 2162

\bibitem[\protect\citeauthoryear{{Filippenko} \& {Riess}}{{Filippenko} \&
  {Riess}}{1998}]{Filippenko1998}
{Filippenko} A.~V.,  {Riess} A.~G.,  1998, \mn@doi [\physrep]
  {10.1016/S0370-1573(98)00052-0}, 307, 31

\bibitem[\protect\citeauthoryear{{Giacconi} et~al.,}{{Giacconi}
  et~al.}{2001}]{Giacconi2001}
{Giacconi} R.,  et~al., 2001, preprint (\mn@eprint {arXiv} {astro-ph/0112184})

\bibitem[\protect\citeauthoryear{{Giannantonio} et~al.,}{{Giannantonio}
  et~al.}{2016}]{Giannantonio2016}
{Giannantonio} T.,  et~al., 2016, \mn@doi [\mnras] {10.1093/mnras/stv2678},
  456, 3213

\bibitem[\protect\citeauthoryear{{G{\'o}rski}, {Hivon}, {Banday}, {Wandelt},
  {Hansen}, {Reinecke}  \& {Bartelmann}}{{G{\'o}rski}
  et~al.}{2005}]{Gorski2005}
{G{\'o}rski} K.~M.,  {Hivon} E.,  {Banday} A.~J.,  {Wandelt} B.~D.,  {Hansen}
  F.~K.,  {Reinecke} M.,   {Bartelmann} M.,  2005, \mn@doi [\apj]
  {10.1086/427976}, 622, 759

\bibitem[\protect\citeauthoryear{{Guillochon}, {Parrent}, {Kelley}  \&
  {Margutti}}{{Guillochon} et~al.}{2017}]{Guillochon2017}
{Guillochon} J.,  {Parrent} J.,  {Kelley} L.~Z.,   {Margutti} R.,  2017,
  \mn@doi [\apj] {10.3847/1538-4357/835/1/64}, 835, 64

\bibitem[\protect\citeauthoryear{{Karpenka}, {Feroz}  \& {Hobson}}{{Karpenka}
  et~al.}{2015}]{Karpenka2015}
{Karpenka} N.~V.,  {Feroz} F.,   {Hobson} M.~P.,  2015, \mn@doi [\mnras]
  {10.1093/mnras/stv415}, 449, 2405

\bibitem[\protect\citeauthoryear{{Kessler} et~al.,}{{Kessler}
  et~al.}{2015}]{Kessler2015}
{Kessler} R.,  et~al., 2015, \mn@doi [\aj] {10.1088/0004-6256/150/6/172}, 150,
  172

\bibitem[\protect\citeauthoryear{{LSST Science Collaboration} et~al.,}{{LSST
  Science Collaboration} et~al.}{2009}]{LSST2009}
{LSST Science Collaboration} et~al., 2009, preprint (\mn@eprint {arXiv}
  {0912.0201})

\bibitem[\protect\citeauthoryear{{Louis} et~al.,}{{Louis}
  et~al.}{2017}]{Louis2017}
{Louis} T.,  et~al., 2017, \mn@doi [\jcap] {10.1088/1475-7516/2017/06/031}, 6,
  031

\bibitem[\protect\citeauthoryear{{Miknaitis} et~al.,}{{Miknaitis}
  et~al.}{2007}]{Miknaitis2007}
{Miknaitis} G.,  et~al., 2007, \mn@doi [\apj] {10.1086/519986}, 666, 674

\bibitem[\protect\citeauthoryear{{Nesseris} \& {Perivolaropoulos}}{{Nesseris}
  \& {Perivolaropoulos}}{2007}]{Nesseris2007}
{Nesseris} S.,  {Perivolaropoulos} L.,  2007, \mn@doi [\jcap]
  {10.1088/1475-7516/2007/02/025}, 2, 025

\bibitem[\protect\citeauthoryear{{Perlmutter} et~al.,}{{Perlmutter}
  et~al.}{1999}]{Perlmutter1999}
{Perlmutter} S.,  et~al., 1999, \mn@doi [\apj] {10.1086/307221}, 517, 565

\bibitem[\protect\citeauthoryear{{Planck Collaboration} et~al.,}{{Planck
  Collaboration} et~al.}{2014a}]{Planck2014a}
{Planck Collaboration} et~al., 2014a, \mn@doi [\aap]
  {10.1051/0004-6361/201321529}, 571, A1

\bibitem[\protect\citeauthoryear{{Planck Collaboration} et~al.,}{{Planck
  Collaboration} et~al.}{2014b}]{Planck2014b}
{Planck Collaboration} et~al., 2014b, \mn@doi [\aap]
  {10.1051/0004-6361/201321580}, 571, A12

\bibitem[\protect\citeauthoryear{{Planck Collaboration} et~al.,}{{Planck
  Collaboration} et~al.}{2016a}]{Planck2016a}
{Planck Collaboration} et~al., 2016a, \mn@doi [\aap]
  {10.1051/0004-6361/201527101}, 594, A1

\bibitem[\protect\citeauthoryear{{Planck Collaboration} et~al.,}{{Planck
  Collaboration} et~al.}{2016b}]{Planck2016c}
{Planck Collaboration} et~al., 2016b, \mn@doi [\aap]
  {10.1051/0004-6361/201525936}, 594, A9

\bibitem[\protect\citeauthoryear{{Planck Collaboration} et~al.,}{{Planck
  Collaboration} et~al.}{2016c}]{Planck2016b}
{Planck Collaboration} et~al., 2016c, \mn@doi [\aap]
  {10.1051/0004-6361/201525830}, 594, A13

\bibitem[\protect\citeauthoryear{{Pritchet} \& {SNLS Collaboration}}{{Pritchet}
  \& {SNLS Collaboration}}{2005}]{Pritchet2005}
{Pritchet} C.~J.,  {SNLS Collaboration} 2005, in {Wolff} S.~C.,  {Lauer} T.~R.,
   eds,  Astronomical Society of the Pacific Conference Series Vol. 339,
  Observing Dark Energy. p.~60 (\mn@eprint {} {astro-ph/0406242})

\bibitem[\protect\citeauthoryear{{Riess} et~al.,}{{Riess}
  et~al.}{1998}]{Riess1998}
{Riess} A.~G.,  et~al., 1998, \mn@doi [\aj] {10.1086/300499}, 116, 1009

\bibitem[\protect\citeauthoryear{{Sachs} \& {Wolfe}}{{Sachs} \&
  {Wolfe}}{1967}]{SachsWolfe1967}
{Sachs} R.~K.,  {Wolfe} A.~M.,  1967, \mn@doi [\apj] {10.1086/148982}, 147, 73

\bibitem[\protect\citeauthoryear{{Story} et~al.,}{{Story}
  et~al.}{2013}]{Story2013}
{Story} K.~T.,  et~al., 2013, \mn@doi [\apj] {10.1088/0004-637X/779/1/86}, 779,
  86

\bibitem[\protect\citeauthoryear{{Sunyaev} \& {Zel\textquotesingle
  dovich}}{{Sunyaev} \& {Zel\textquotesingle
  dovich}}{1970}]{SunyaevZeldovich1970}
{Sunyaev} R.~A.,  {Zel\textquotesingle dovich} Y.~B.,  1970, \mn@doi [\apss]
  {10.1007/BF00653471}, 7, 3

\bibitem[\protect\citeauthoryear{Welch}{Welch}{1938}]{Welch1938}
Welch B.~L.,  1938, Biometrika, 29, 350

\bibitem[\protect\citeauthoryear{{Williams} et~al.,}{{Williams}
  et~al.}{1996}]{Williams1996}
{Williams} R.~E.,  et~al., 1996, \mn@doi [\aj] {10.1086/118105}, 112, 1335

\bibitem[\protect\citeauthoryear{{Yershov}, {Orlov}  \& {Raikov}}{{Yershov}
  et~al.}{2012}]{Yershov2012}
{Yershov} V.~N.,  {Orlov} V.~V.,   {Raikov} A.~A.,  2012, \mn@doi [\mnras]
  {10.1111/j.1365-2966.2012.21026.x}, 423, 2147

\bibitem[\protect\citeauthoryear{{Yershov}, {Orlov}  \& {Raikov}}{{Yershov}
  et~al.}{2014}]{Yershov2014}
{Yershov} V.~N.,  {Orlov} V.~V.,   {Raikov} A.~A.,  2014, \mn@doi [\mnras]
  {10.1093/mnras/stu1932}, 445, 2440

\makeatother
\end{thebibliography}




\appendix

\section{Fields 1-7} \label{appFields1to7}

\begin{figure*} 
	\centering
	\subfigure[Field 1, centred on SNLS D3]{\includegraphics[width=0.3\textwidth]{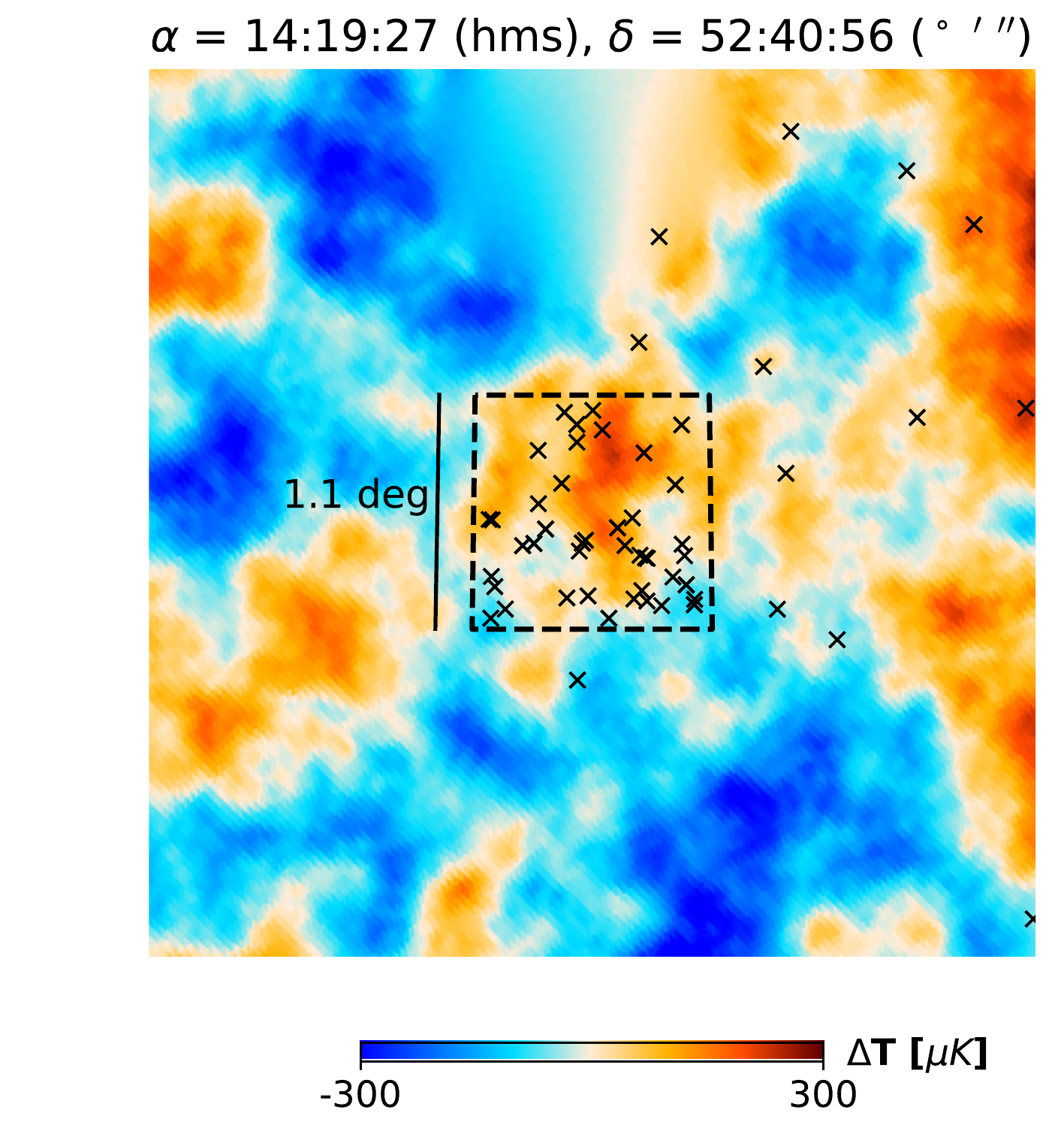} \label{subField1}}
	\subfigure[Field 2, centred on HDF-N]{\includegraphics[width=0.3\textwidth]{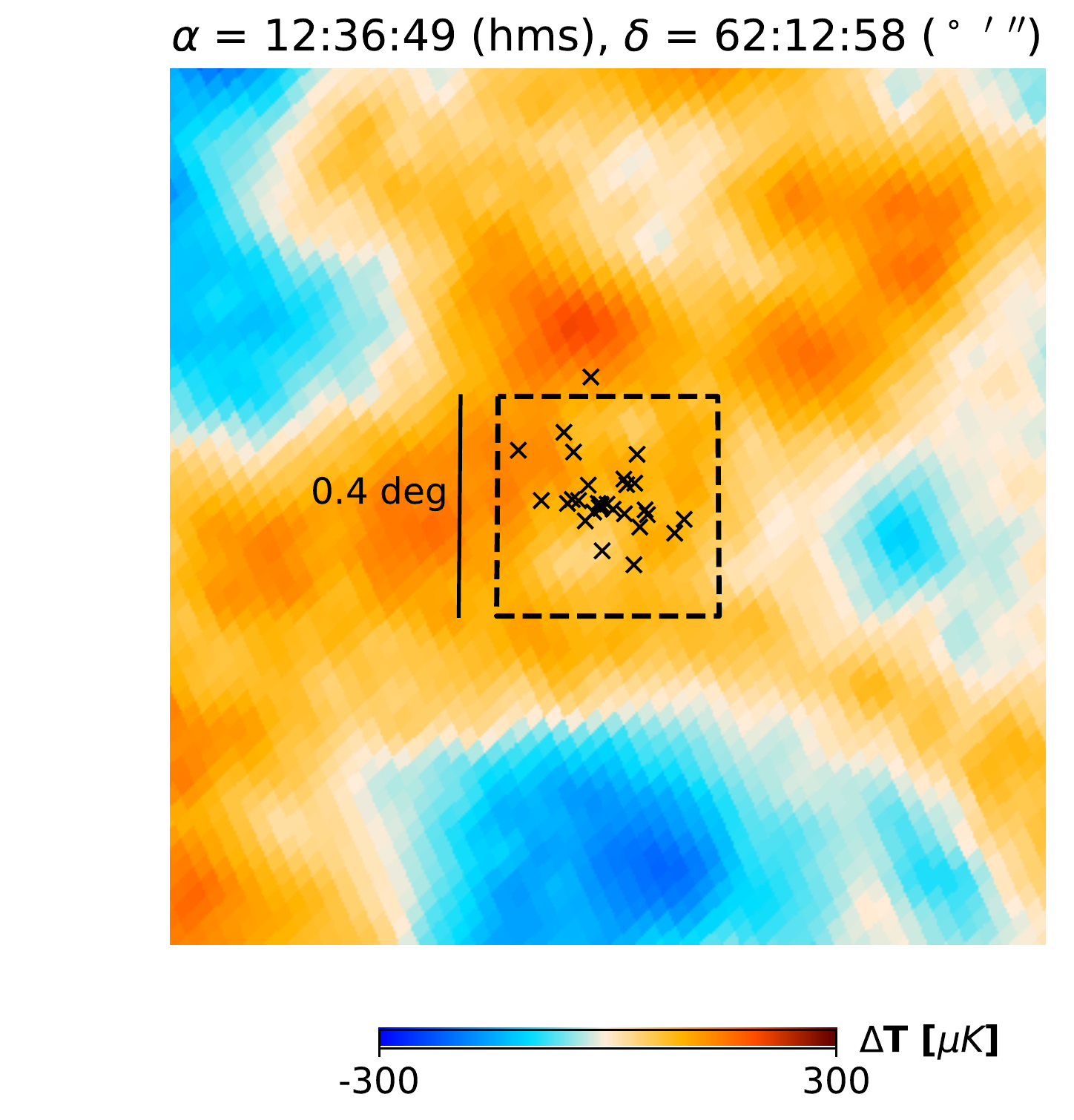} \label{subField2}}
	\subfigure[Field 3, centred on ESSENCE wdd]{\includegraphics[width=0.3\textwidth]{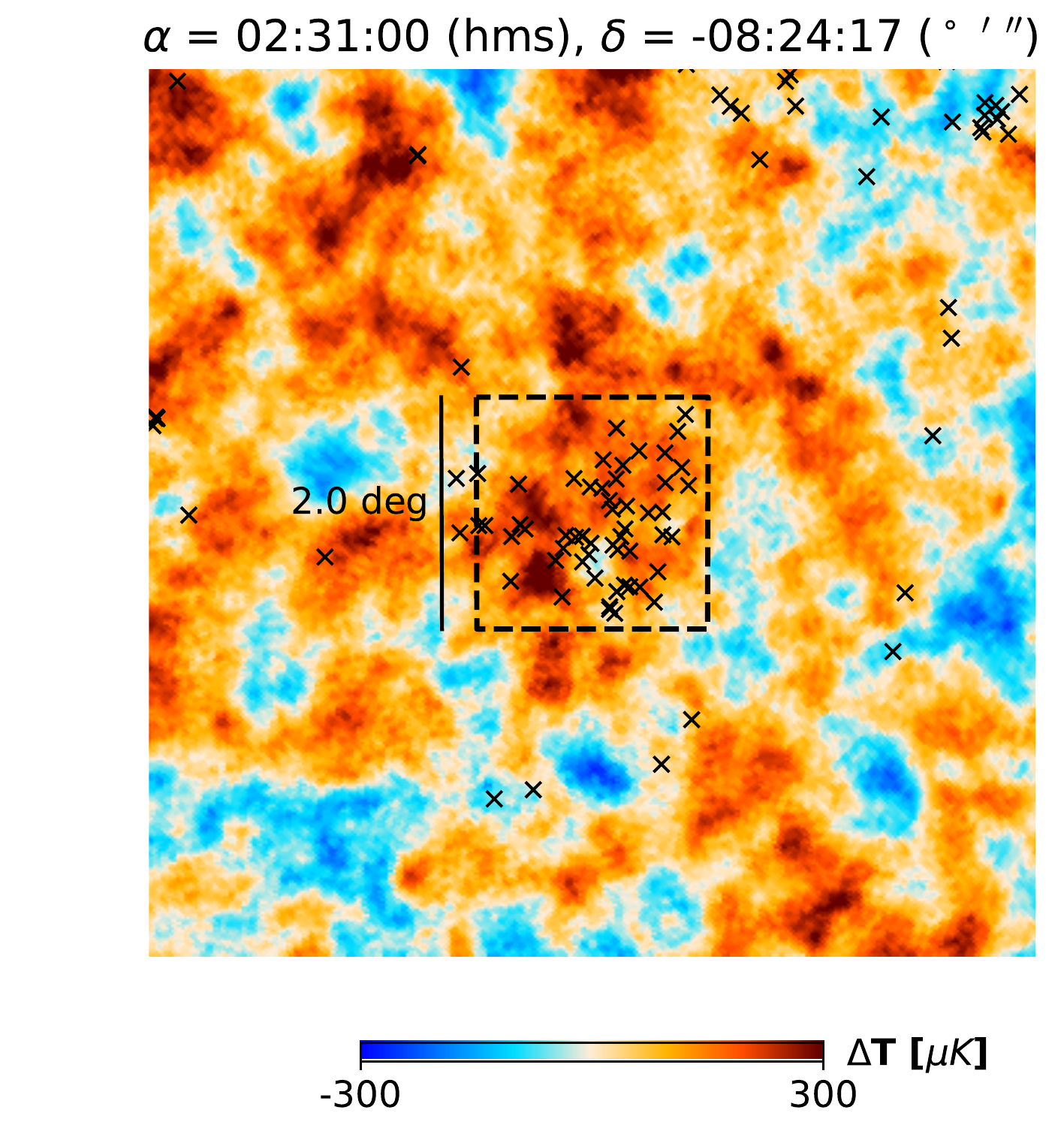} \label{subField3}}
	\\
	\subfigure[Field 4, centred on SNLS D1]{\includegraphics[width=0.3\textwidth]{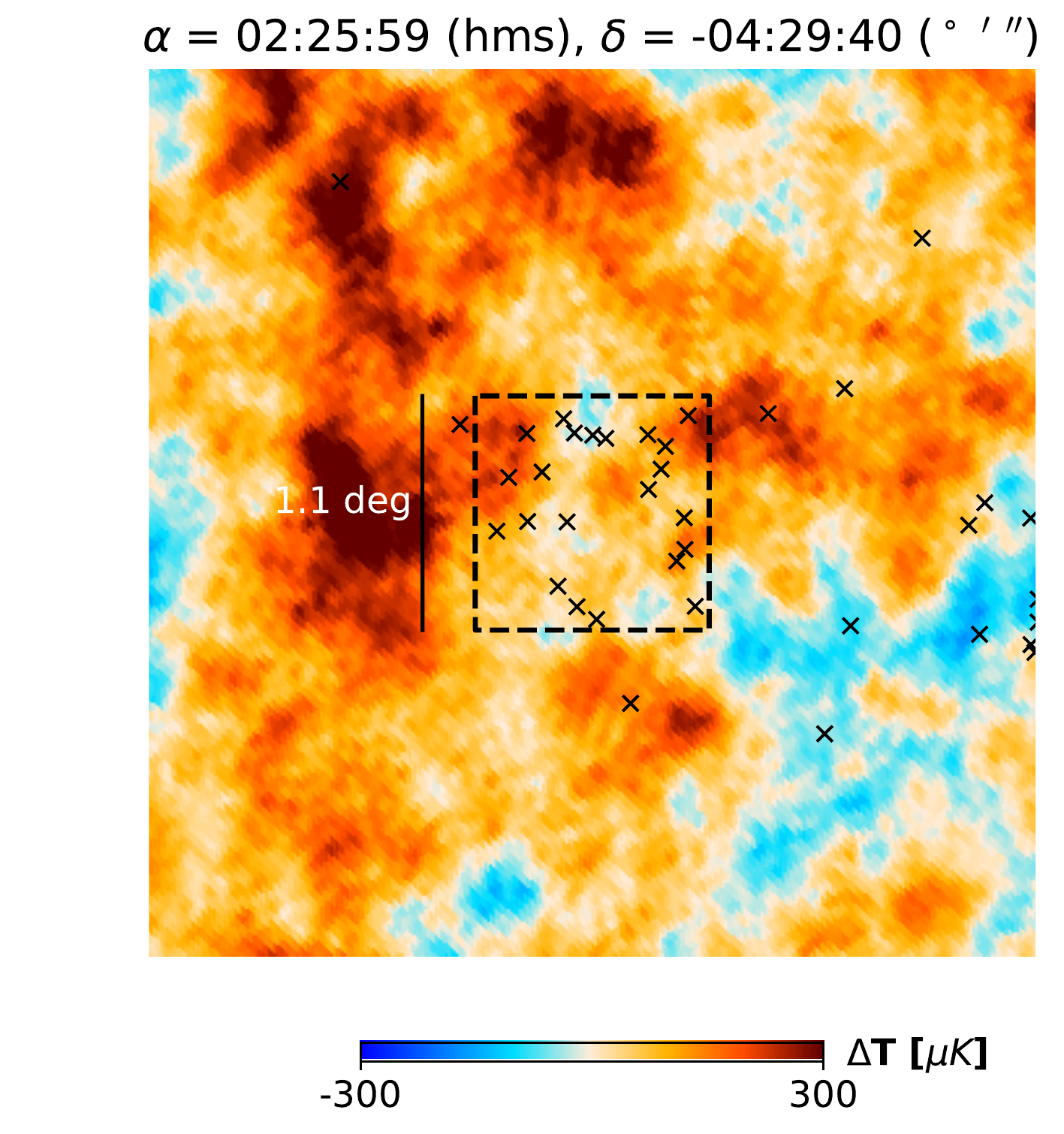} \label{subField4}}
	\subfigure[Field 5, centred on ESSENCE wcc]{\includegraphics[width=0.3\textwidth]{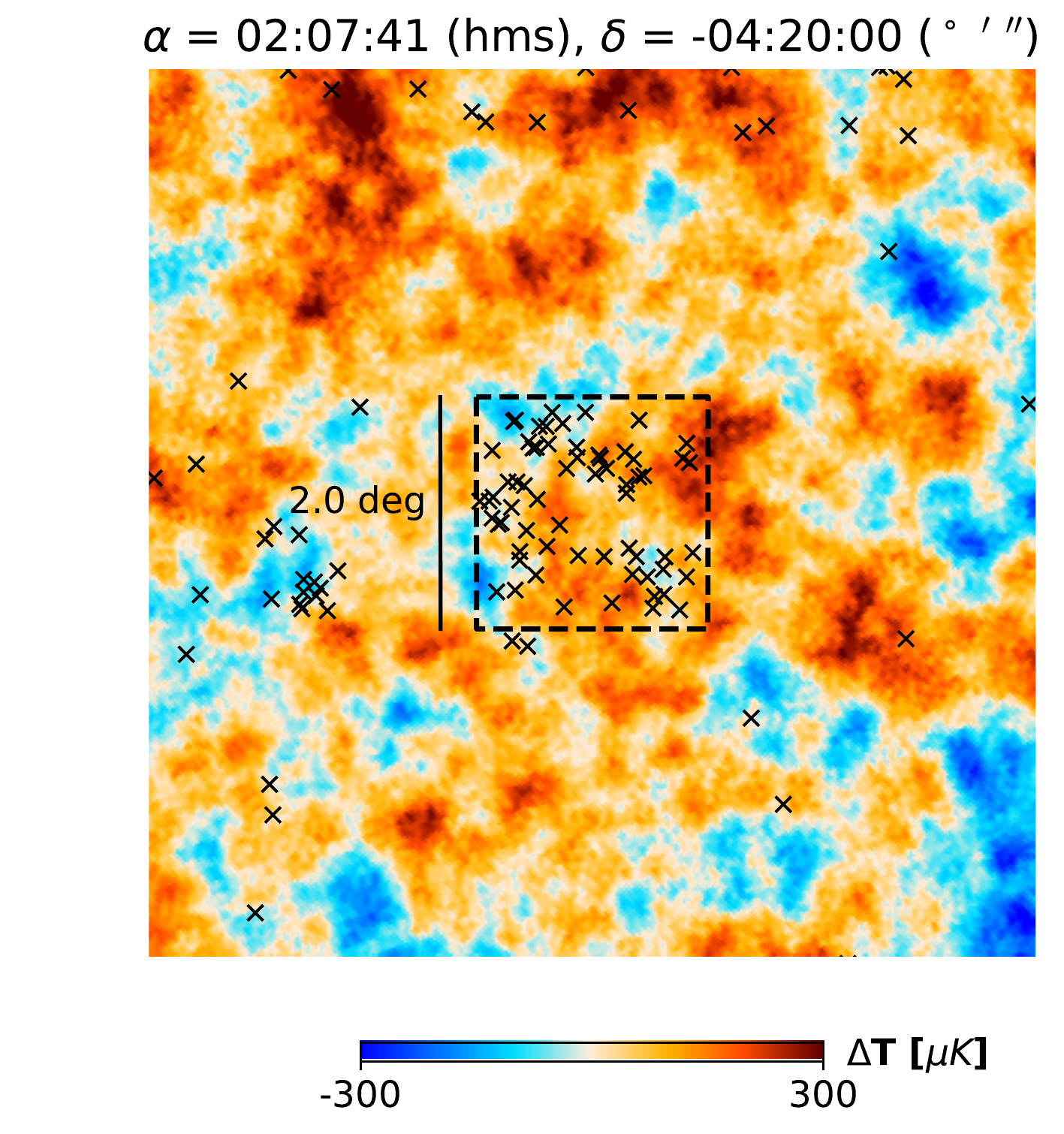} \label{subField5}}
	\subfigure[Field 6, centred on SNLS D4]{\includegraphics[width=0.3\textwidth]{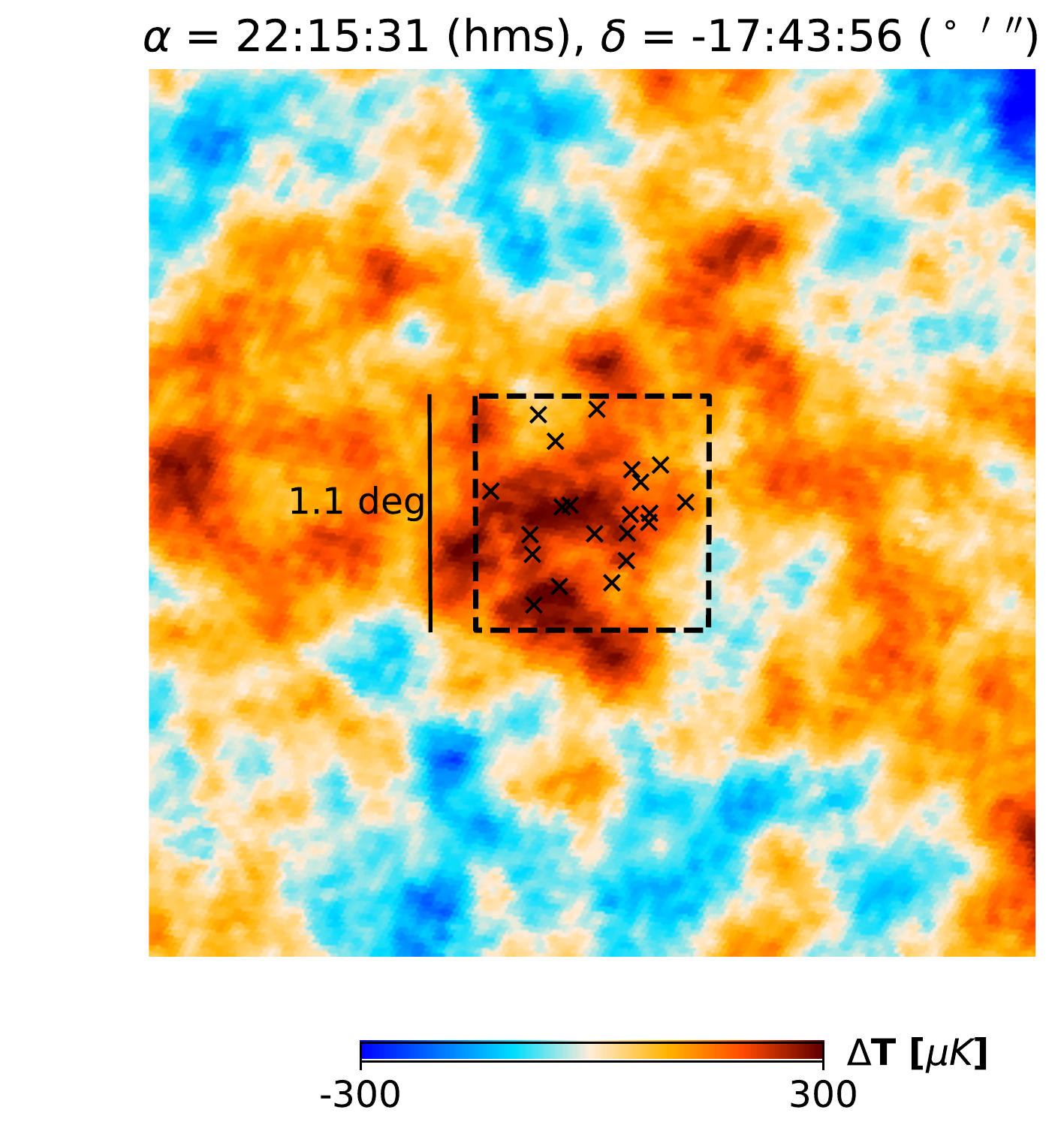} \label{subField6}}
	\\
	\subfigure[Field 7, centred on CDF-S]{\includegraphics[width=0.3\textwidth]{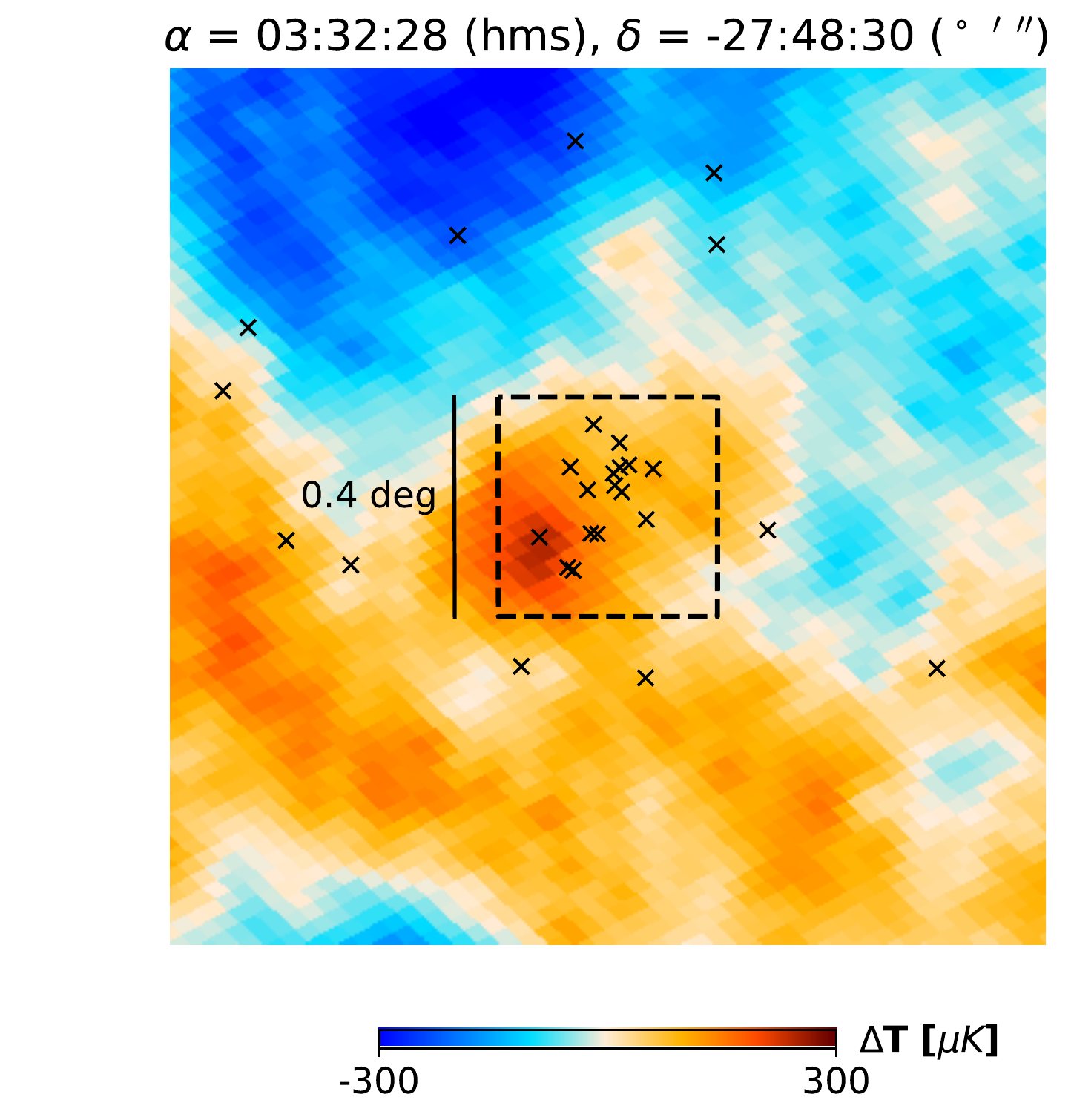} \label{subField7}}
	\caption{Gnomic projections of the \textit{Planck} 2015 SMICA CMB map in the vicinity of fields 1-7 (colour on-line). The location of SAI sample SNe are plotted with crosses ($\times$). The dashed squares are the boundaries of fields 1-7, centred on the corresponding deep survey fields at the specified coordinates ($\alpha$ and $\delta$, J2000).}
	\label{figFields1to7}
\end{figure*}

\section{Variances and smoothing} \label{appVariances}

\begin{figure} 
	\centering
	\subfigure[\label{subHMHDunweighted}Unweighted $\overline{T}$]{
    \makebox[0.8\columnwidth][c]{
    \includegraphics[width=0.43\columnwidth]{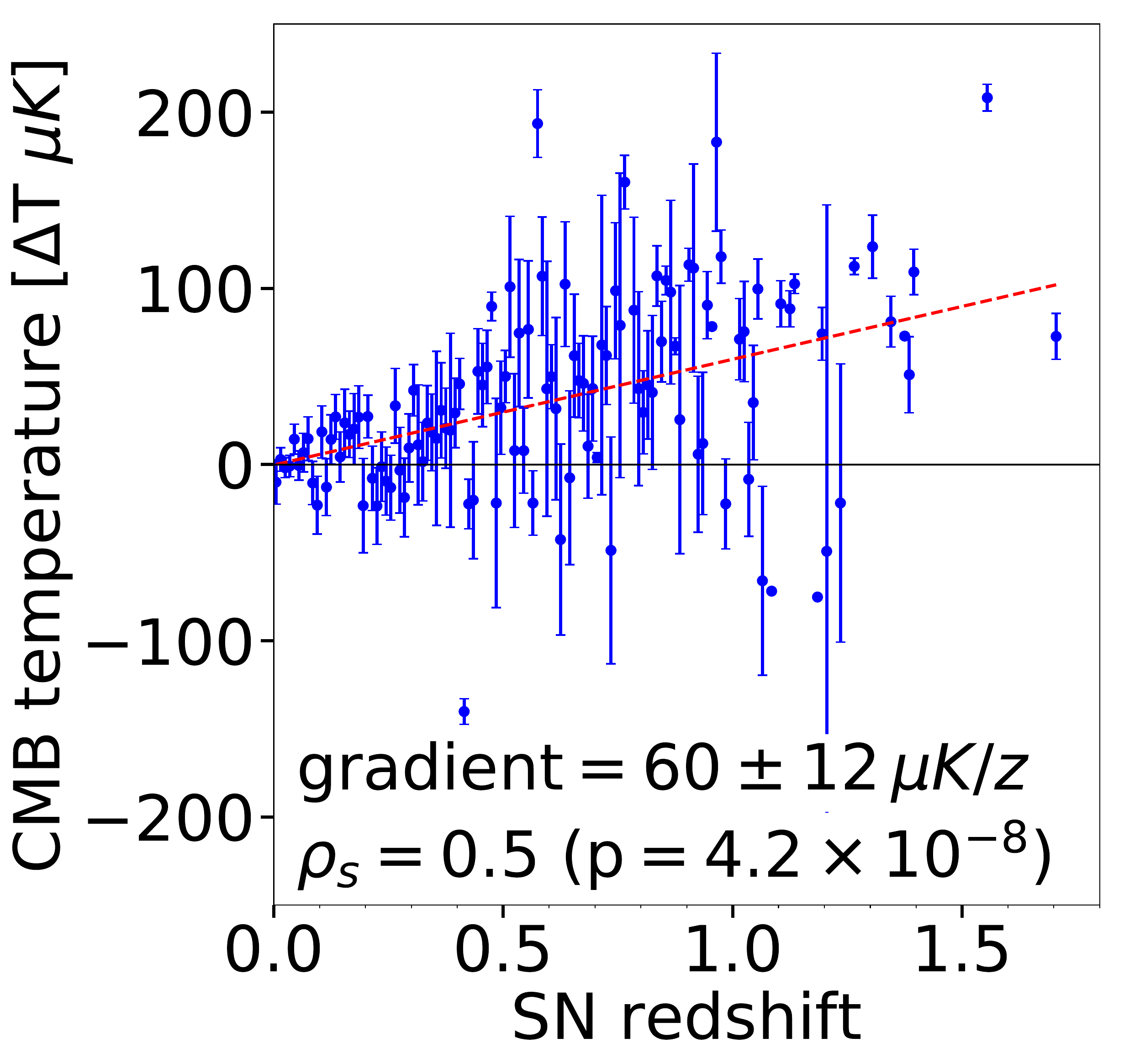}}}
	\\
	\subfigure[\label{subHMHDweightedUnsmoothed}Weighted $\overline{T}$, HMHD variance unsmoothed]{
    \makebox[0.8\columnwidth][c]{
    \includegraphics[width=0.43\columnwidth]{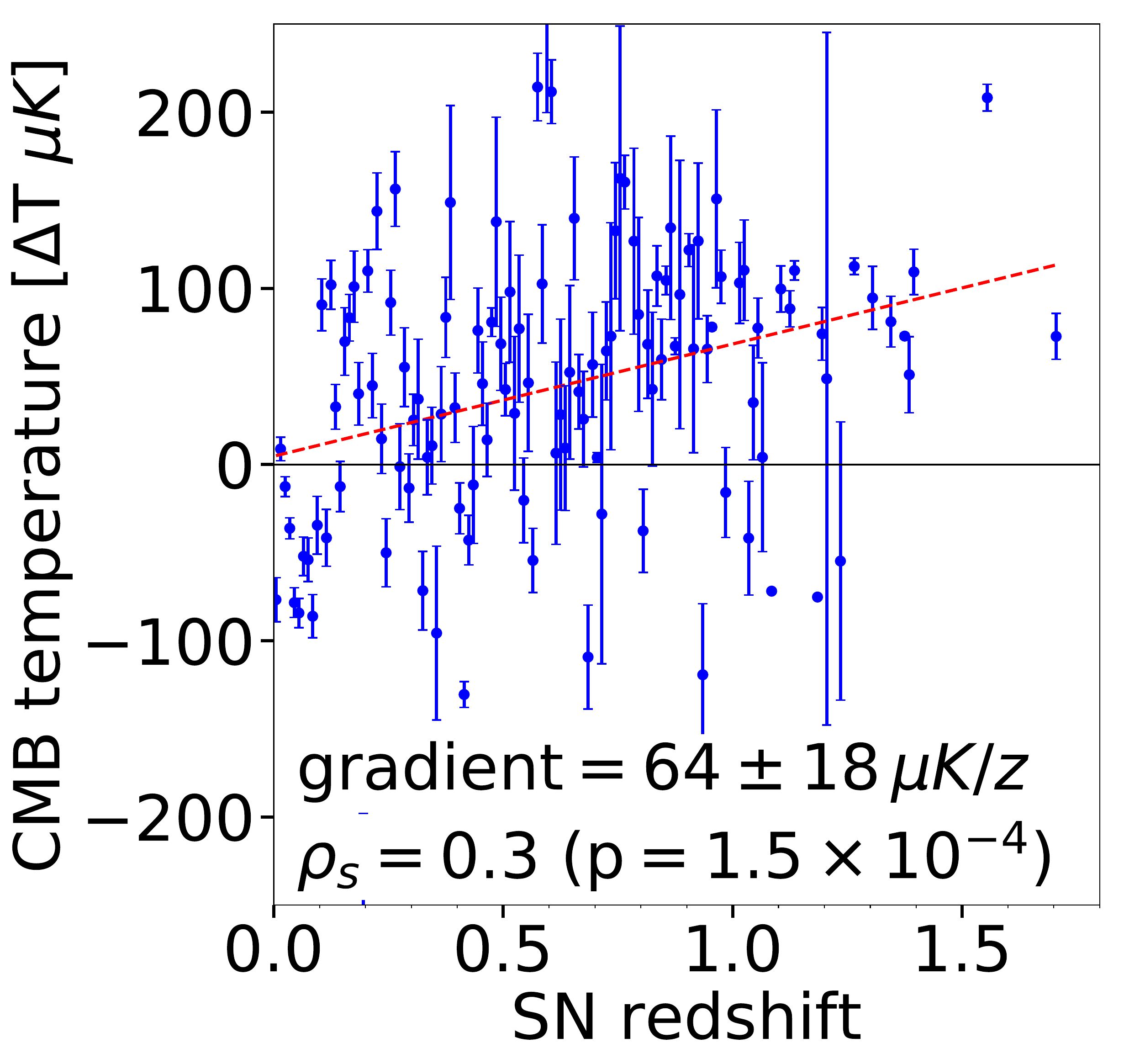}}}
	\\
	\subfigure[\label{subHMHDweightedSmoothed5arcmin}Weighted $\overline{T}$, HMHD variance smoothed to $5'$]{
    \makebox[0.8\columnwidth][c]{
    \includegraphics[width=0.43\columnwidth]{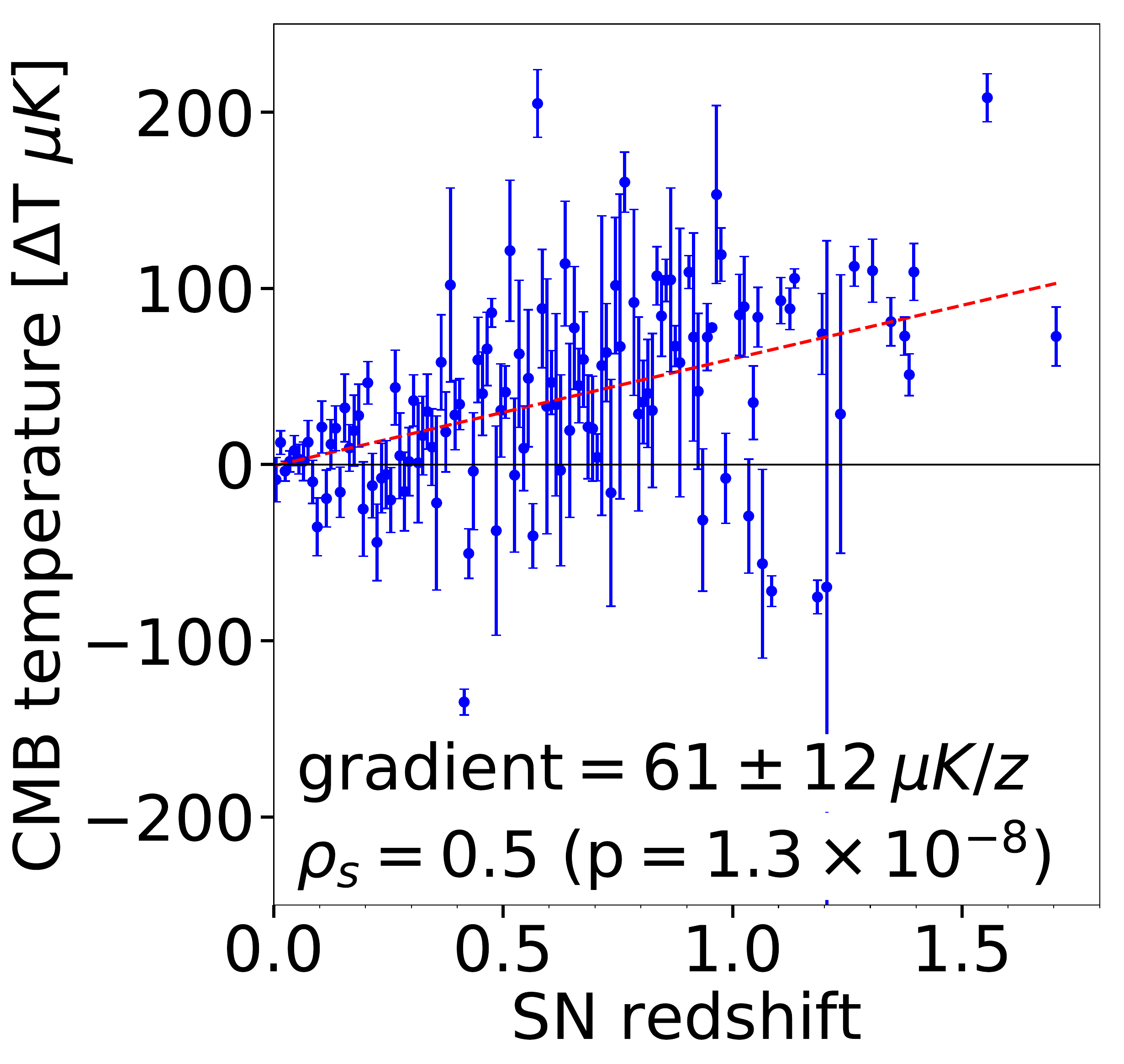}}}
	\\
	\subfigure[\label{subHMHDweightedSmoothedHalfDeg}Weighted $\overline{T}$, HMHD variance smoothed to 0.5$^\circ$]{
    \makebox[0.8\columnwidth][c]{
    \includegraphics[width=0.43\columnwidth]{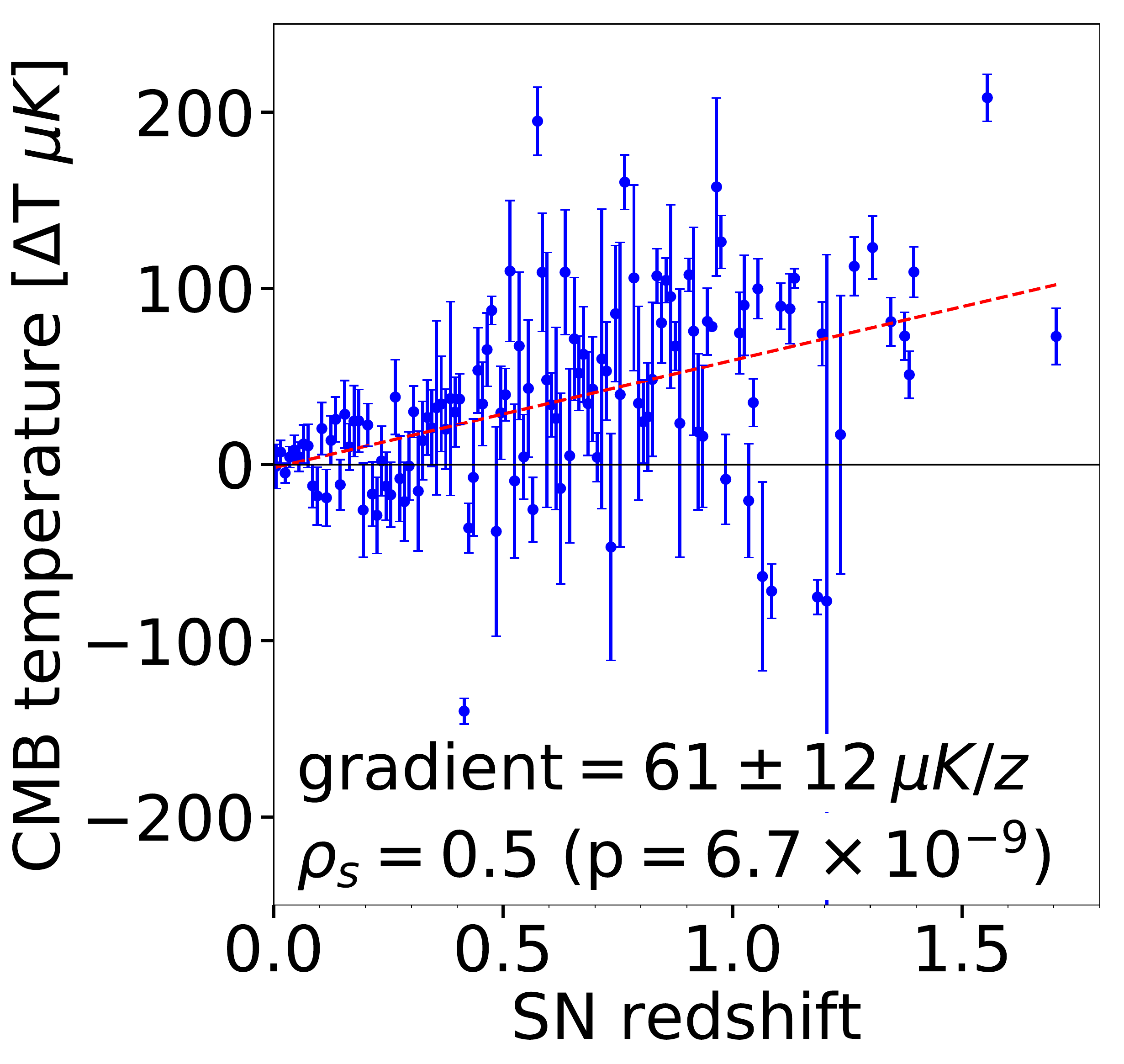}}}
	\\
	\subfigure[\label{subHMHDweightedSmoothed5deg}Weighted $\overline{T}$, HMHD variance smoothed to 5$^\circ$]{
    \makebox[0.8\columnwidth][c]{
    \includegraphics[width=0.43\columnwidth]{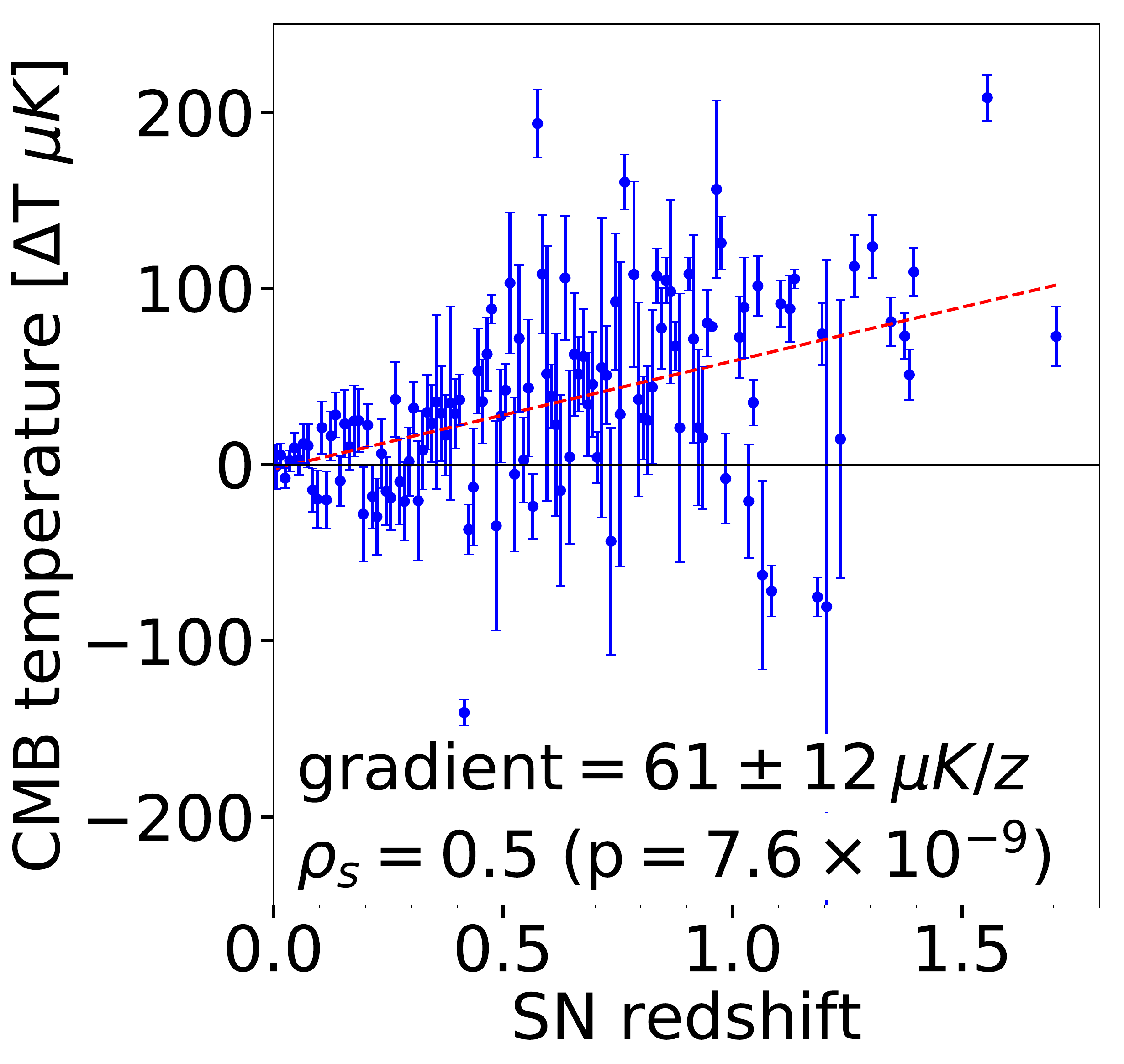}}}
	\caption{Plot of CMB temperature at SNe locations versus SNe redshift (method as section~\ref{secDataMethod}, plot as Fig.~\ref{figBinnedScatter}). \textit{Planck} 2015 CMB data used are: CMB temperature from R2.01 SMICA CMB temperature intensity; variance estimate from R2.01 SMICA half-mission half-difference (HMHD) maps. Sub-figures are \protect\subref{subHMHDunweighted} unweighted $\overline{T}$, \protect\subref{subHMHDweightedUnsmoothed} weighted $\overline{T}$ with unsmoothed HMHD variance, \protect\subref{subHMHDweightedSmoothed5arcmin}, \protect\subref{subHMHDweightedSmoothedHalfDeg}, and \protect\subref{subHMHDweightedSmoothed5deg} weighted $\overline{T}$ with HMHD variance smoothed to $5'$, 0.5$^\circ$, and 5$^\circ$ FWHM respectively. Weight~$= 1/\sigma^2$.}
	\label{figHMDH}
\end{figure}

\begin{figure} 
	\centering
	\subfigure[\label{subHRHDunweighted}Unweighted $\overline{T}$]{
    \makebox[0.8\columnwidth][c]{
    \includegraphics[width=0.43\columnwidth]{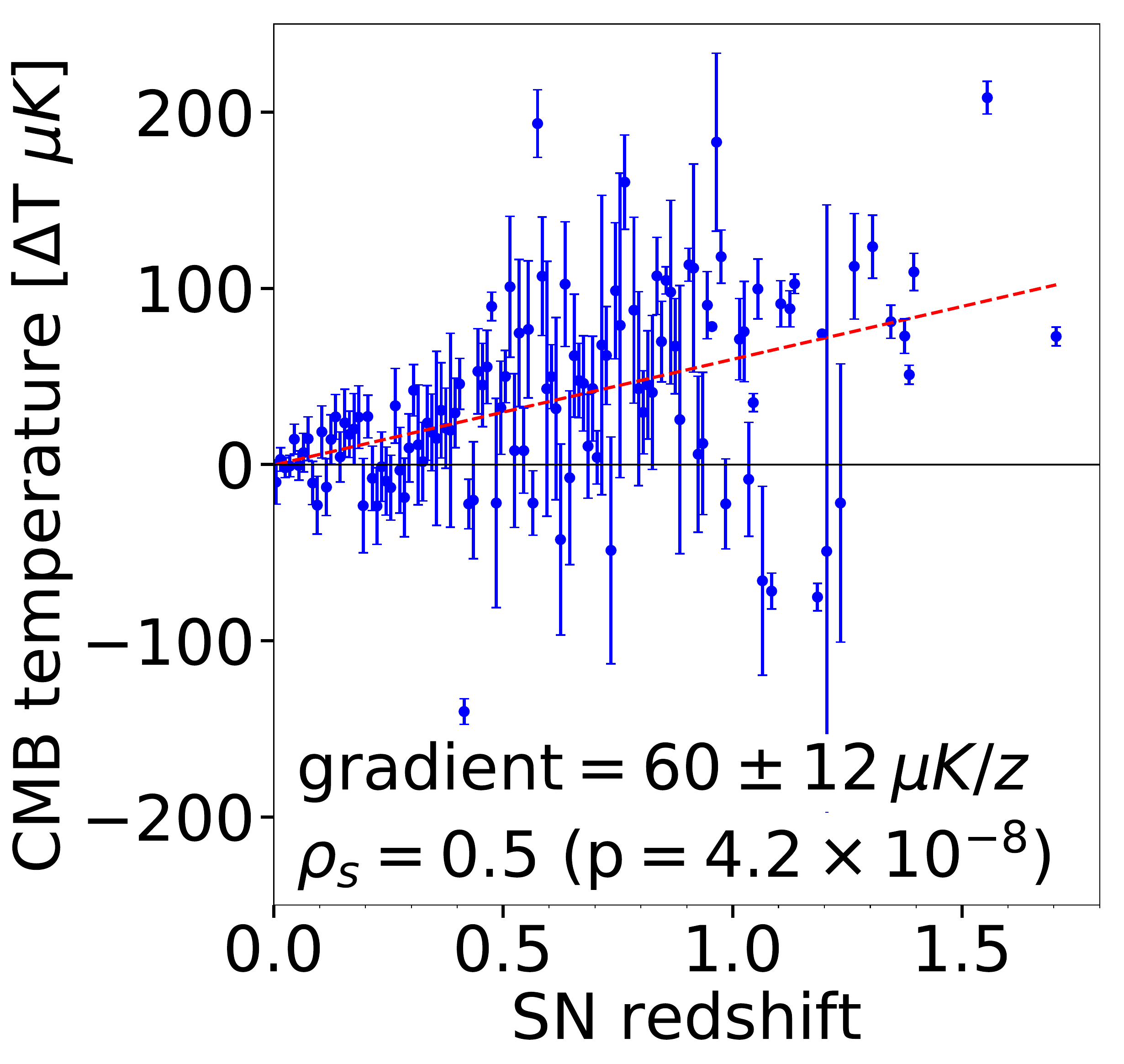}}}
	\\
	\subfigure[\label{subHRHDweightedUnsmoothed}Weighted $\overline{T}$, HRHD variance unsmoothed]{
    \makebox[0.8\columnwidth][c]{
    \includegraphics[width=0.43\columnwidth]{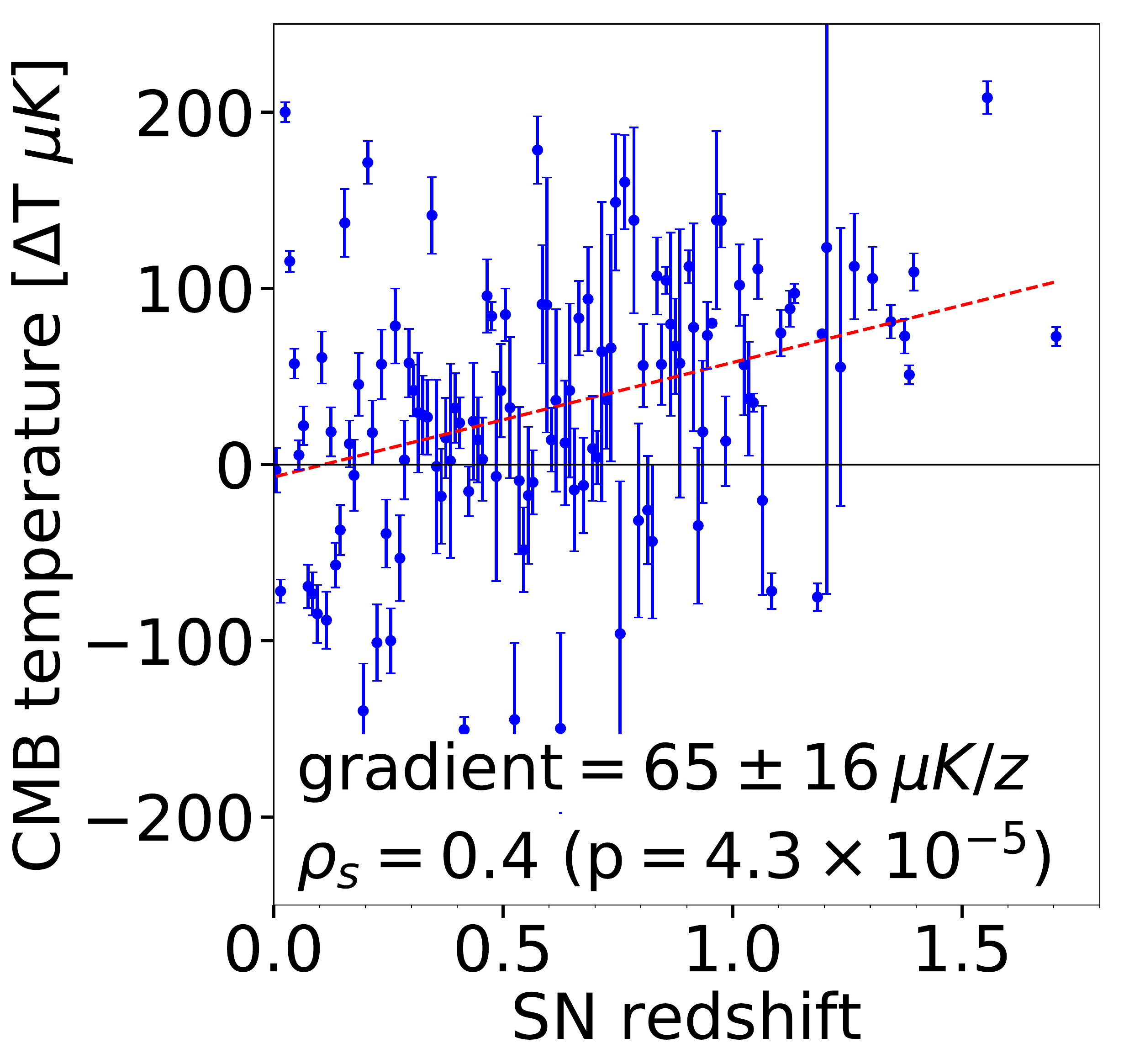}}}
	\\
	\subfigure[\label{subHRHDweightedSmoothed5arcmin}Weighted $\overline{T}$, HRHD variance smoothed to $5'$]{
    \makebox[0.8\columnwidth][c]{
    \includegraphics[width=0.43\columnwidth]{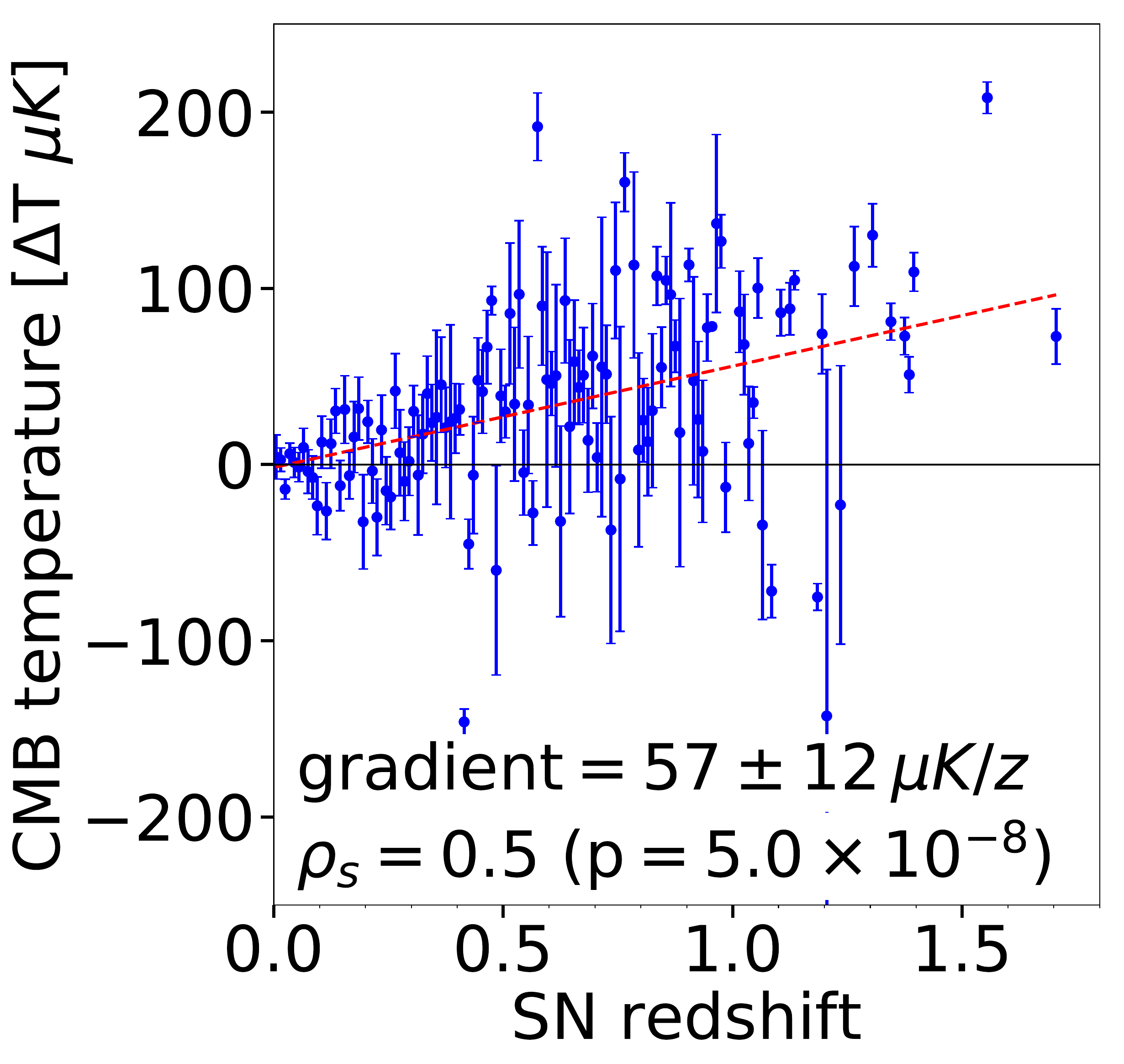}}}
	\\
	\subfigure[\label{subHRHDweightedSmoothedHalfDeg}Weighted $\overline{T}$, HRHD variance smoothed to 0.5$^\circ$]{
    \makebox[0.8\columnwidth][c]{
    \includegraphics[width=0.43\columnwidth]{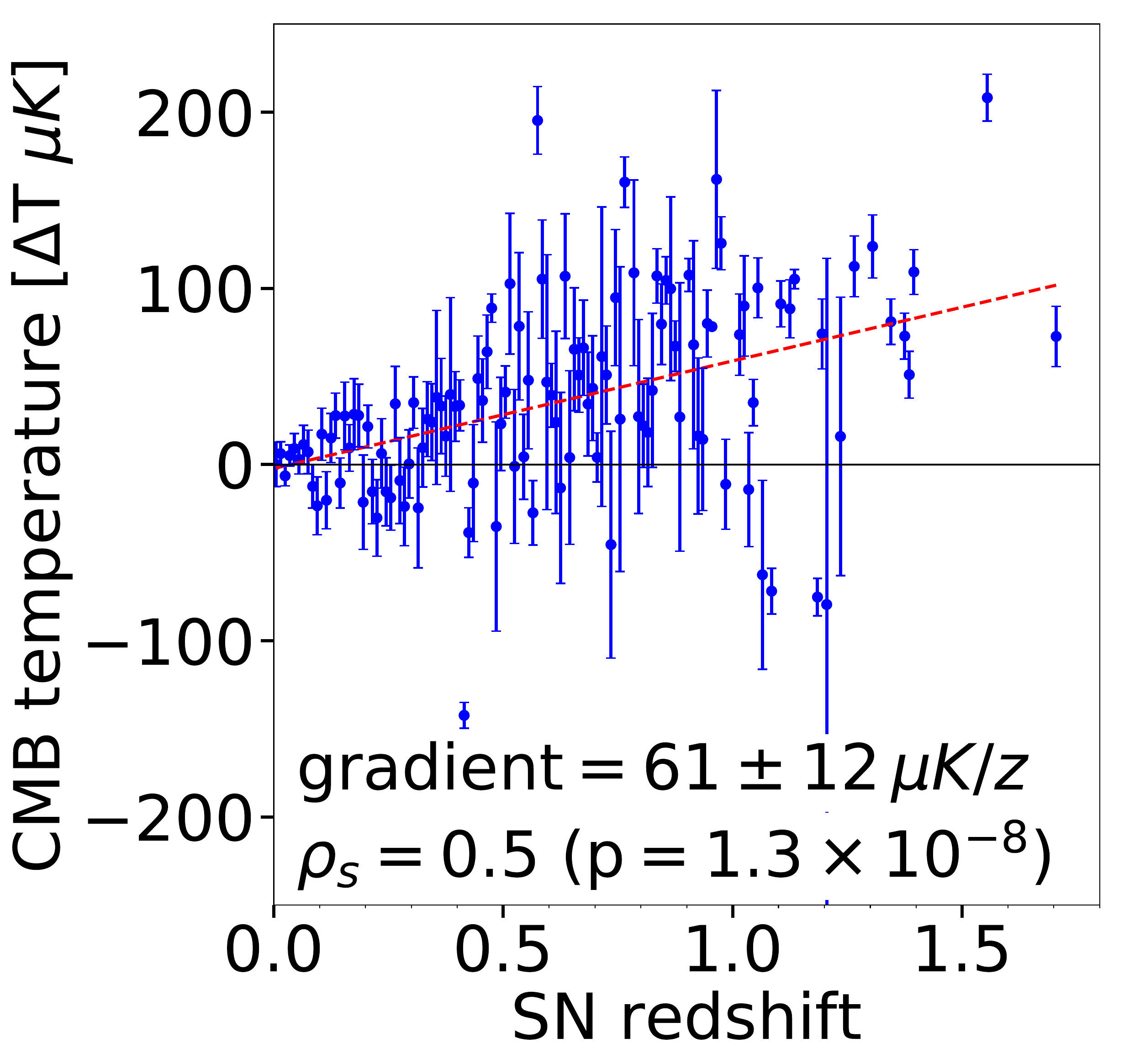}}}
	\\
	\subfigure[\label{subHRHDweightedSmoothed5deg}Weighted $\overline{T}$, HRHD variance smoothed to 5$^\circ$]{
    \makebox[0.8\columnwidth][c]{
    \includegraphics[width=0.43\columnwidth]{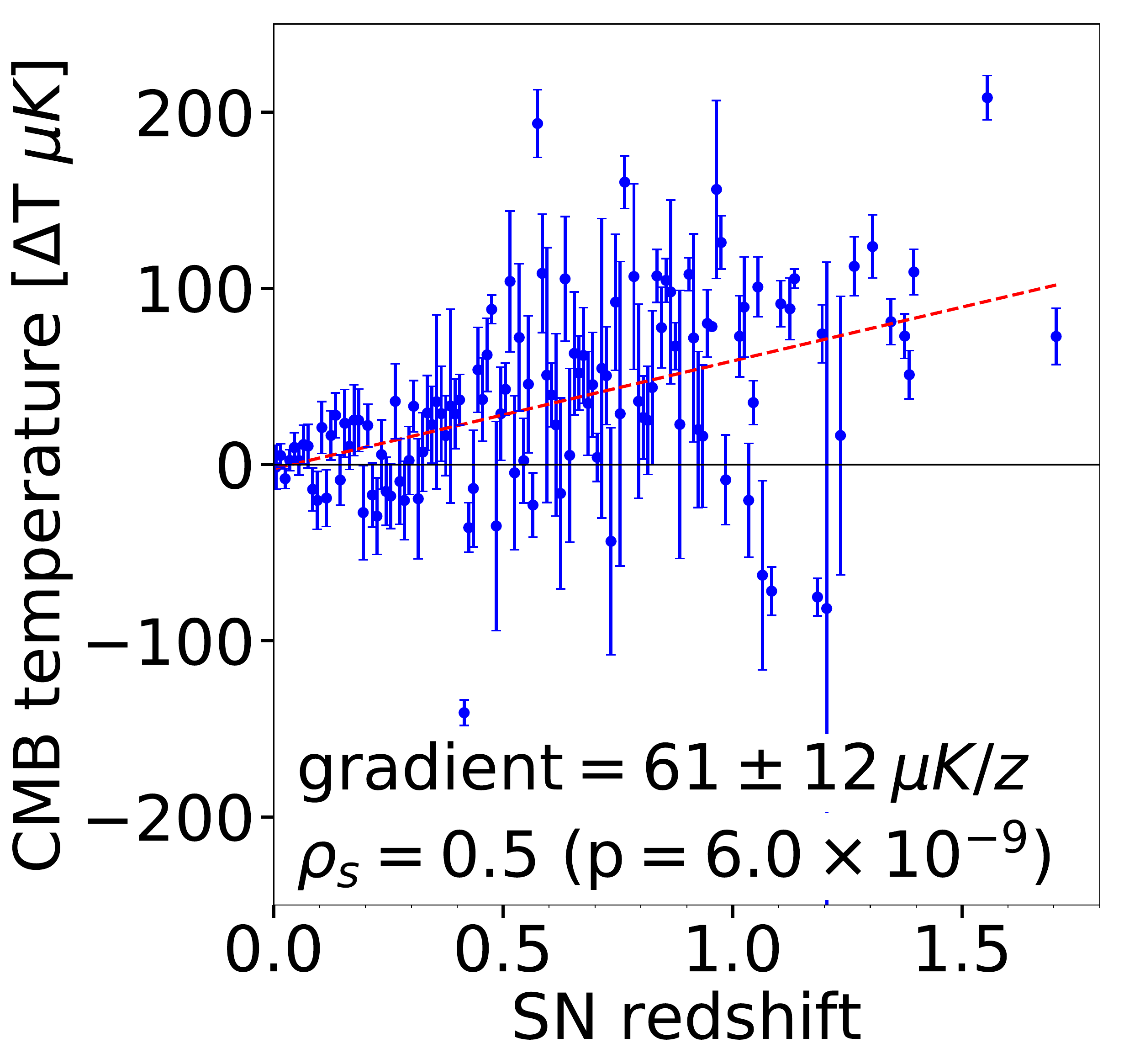}}}
	\caption{Same as Fig. \ref{figHMDH} but for variance estimated using \textit{Planck} 2015 R2.01 SMICA half-ring half-difference (HRHD) maps.}
	\label{figHRHD}
\end{figure}

\begin{figure} 
	\centering
	\subfigure[\label{sub143weightedUnsmoothed}Weighted $\overline{T}$, 143GHz covariance unsmoothed]{
    \makebox[0.8\columnwidth][c]{
    \includegraphics[width=0.43\columnwidth]{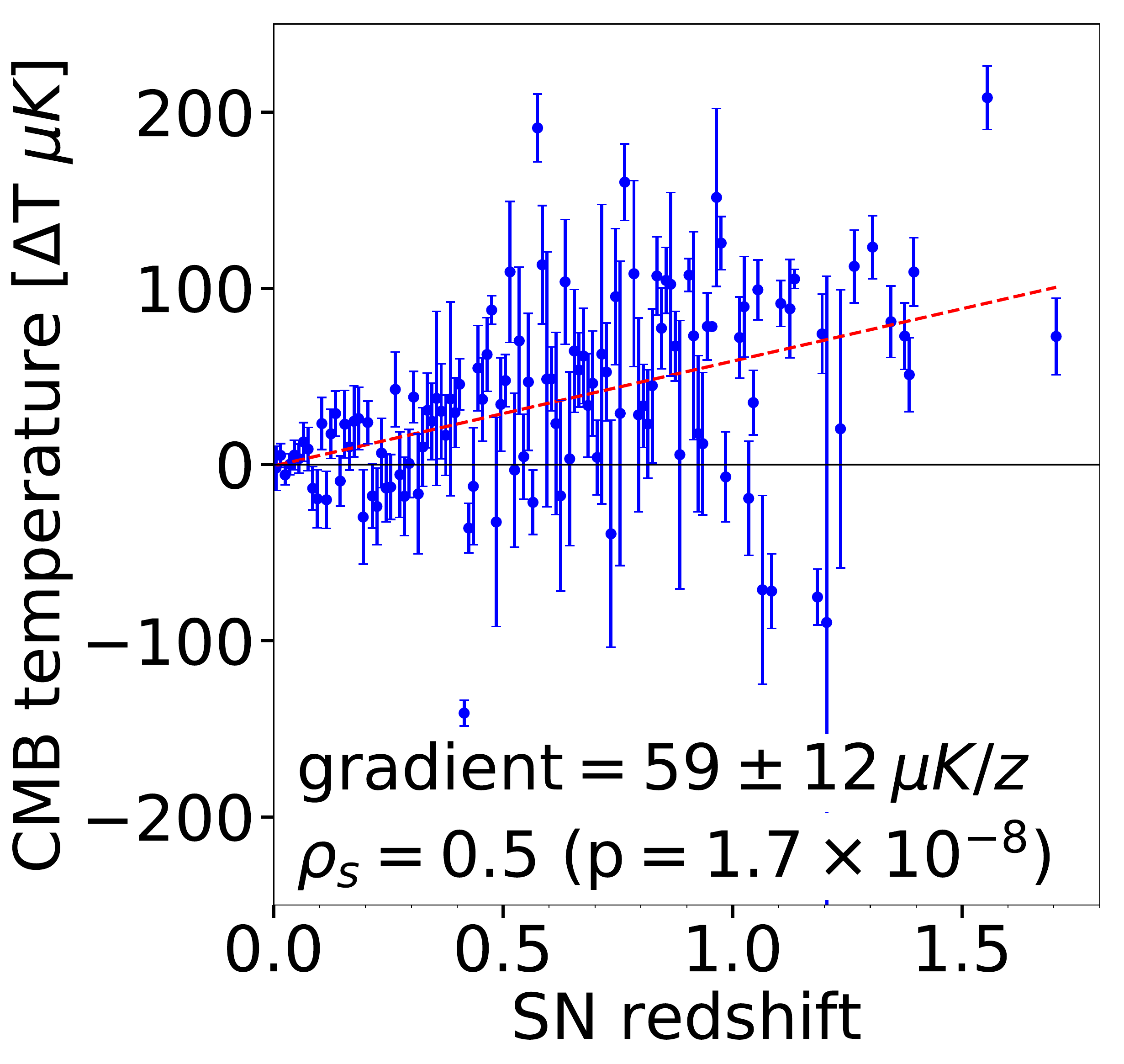}}}
	\\
	\subfigure[\label{sub217weightedUnsmoothed}Weighted $\overline{T}$, 217GHz covariance unsmoothed]{
    \makebox[0.8\columnwidth][c]{
    \includegraphics[width=0.43\columnwidth]{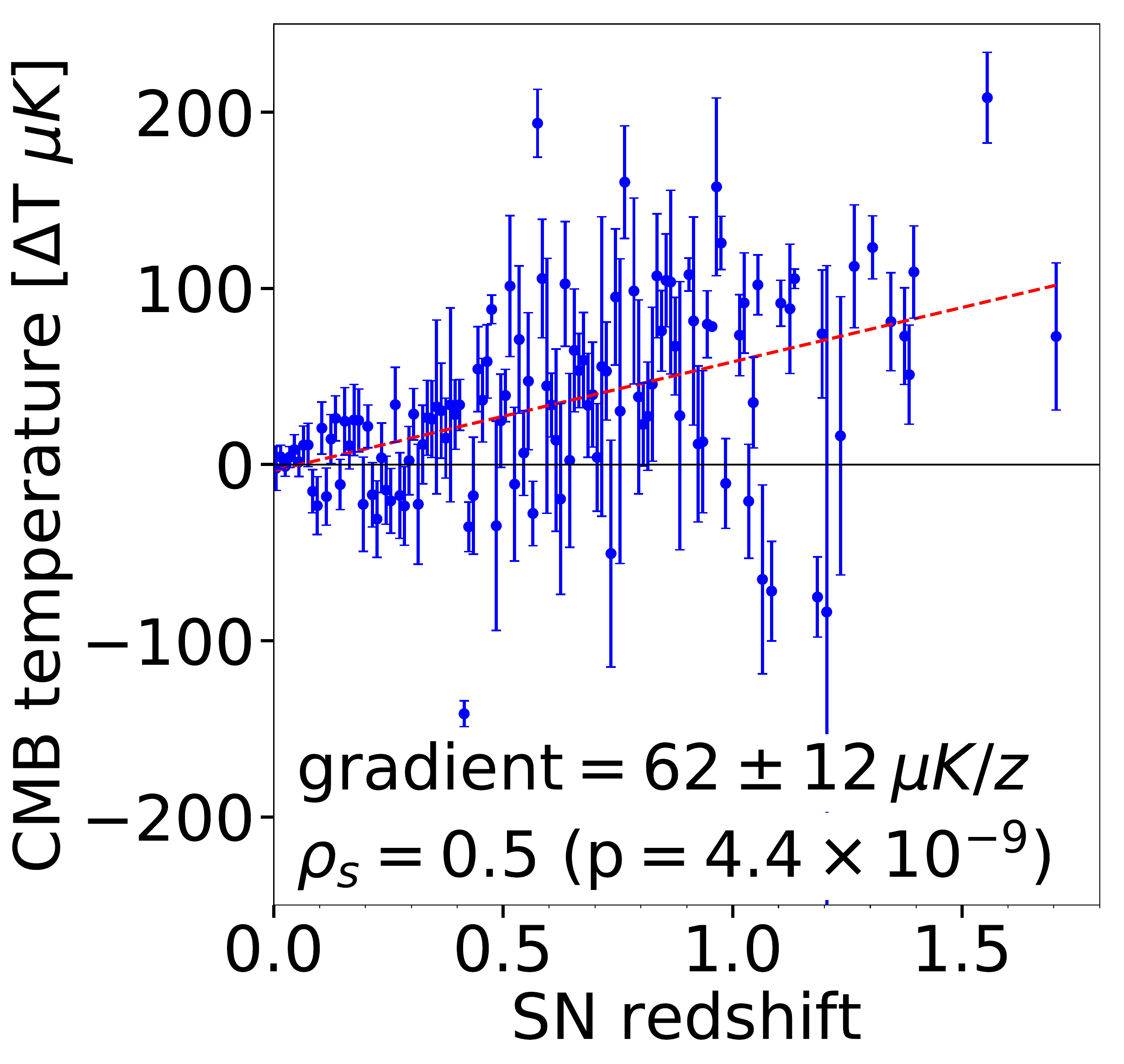}}}
	\caption{Same as Fig. \ref{figHMDH} \protect\subref{subHMHDweightedUnsmoothed} but for variance estimated using intensity covariance in \textit{Planck} 2015 R2.02 143GHz and 217GHz frequency maps. Sub-figures are \protect\subref{sub143weightedUnsmoothed} weighted $\overline{T}$ with unsmoothed 143GHz variance, and \subref{sub217weightedUnsmoothed} weighted $\overline{T}$ with unsmoothed 217GHz variance. Weight~$= 1/\sigma^2$.}
	\label{fig143217}
\end{figure}

\begin{figure} 
	\centering
	\subfigure[\label{sub2013unweighted}Unweighted $\overline{T}$]{
    \makebox[0.8\columnwidth][c]{
    \includegraphics[width=0.43\columnwidth]{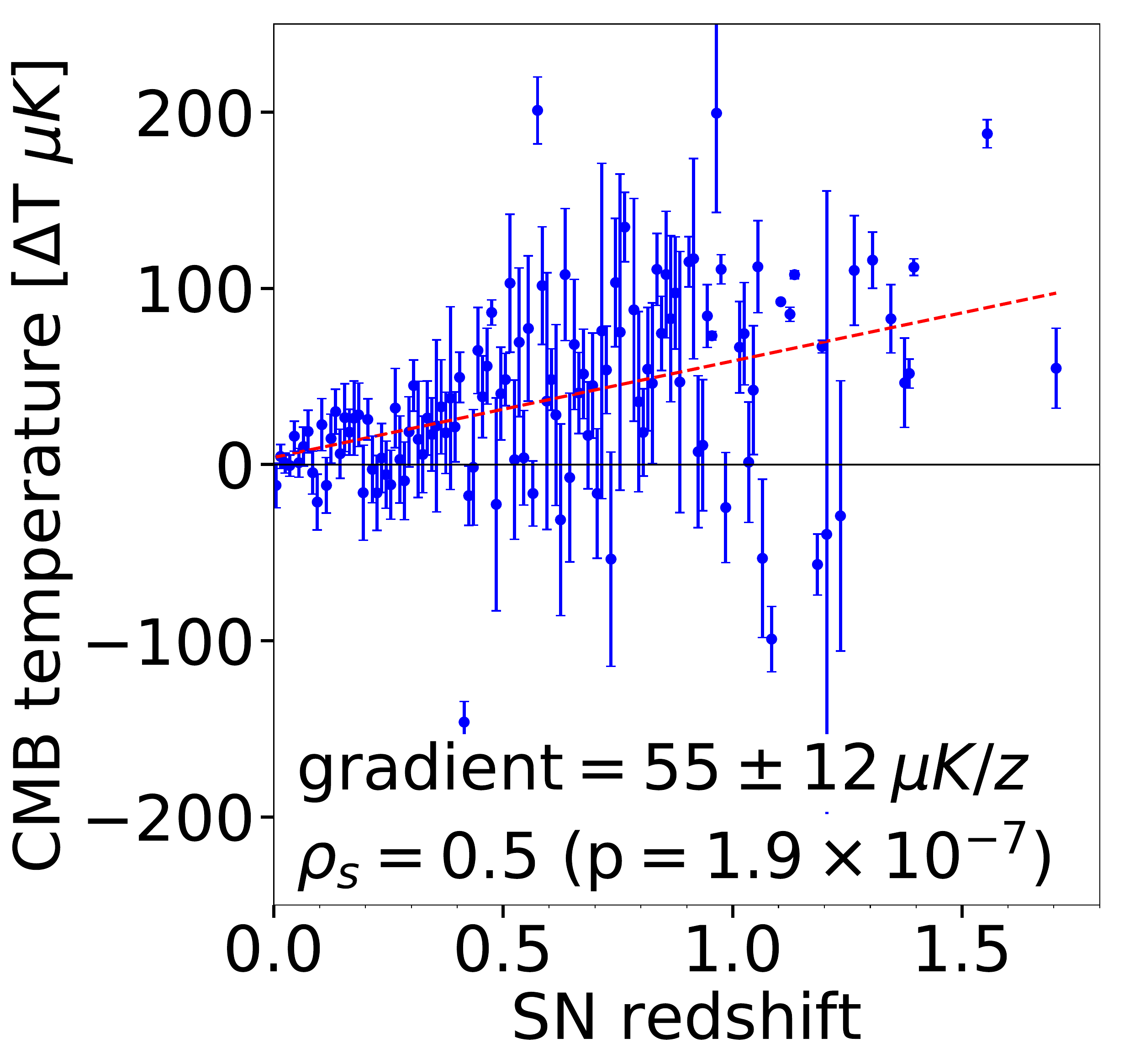}}}
	\\
	\subfigure[\label{sub2013weightedUnsmoothed}Weighted $\overline{T}$, R1.20 variance unsmoothed]{
    \makebox[0.8\columnwidth][c]{
    \includegraphics[width=0.43\columnwidth]{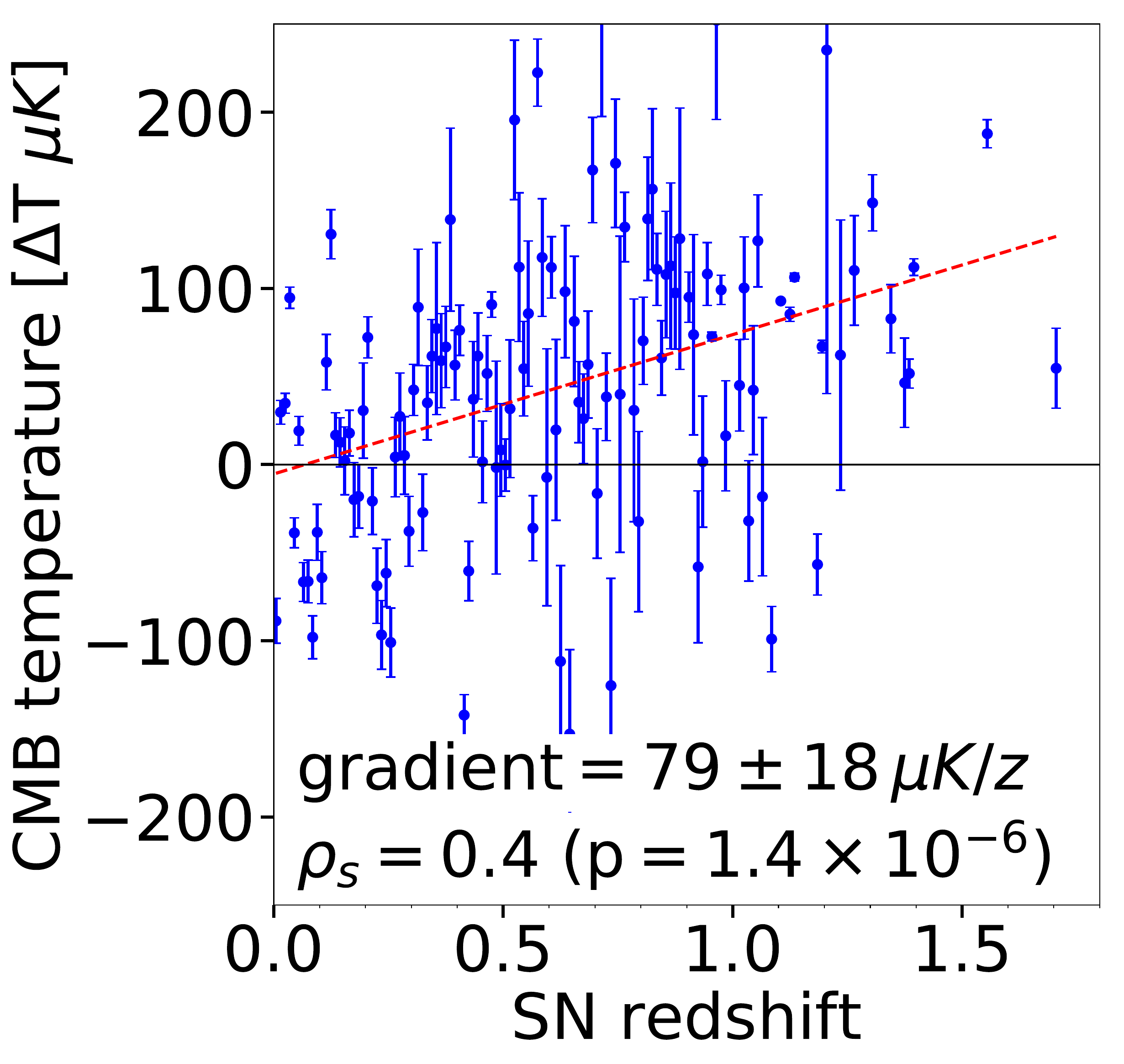}}}
	\\
	\subfigure[\label{sub2013weightedSmoothed5arcmin}Weighted $\overline{T}$, R1.20 variance smoothed to $5'$]{
    \makebox[0.8\columnwidth][c]{
    \includegraphics[width=0.43\columnwidth]{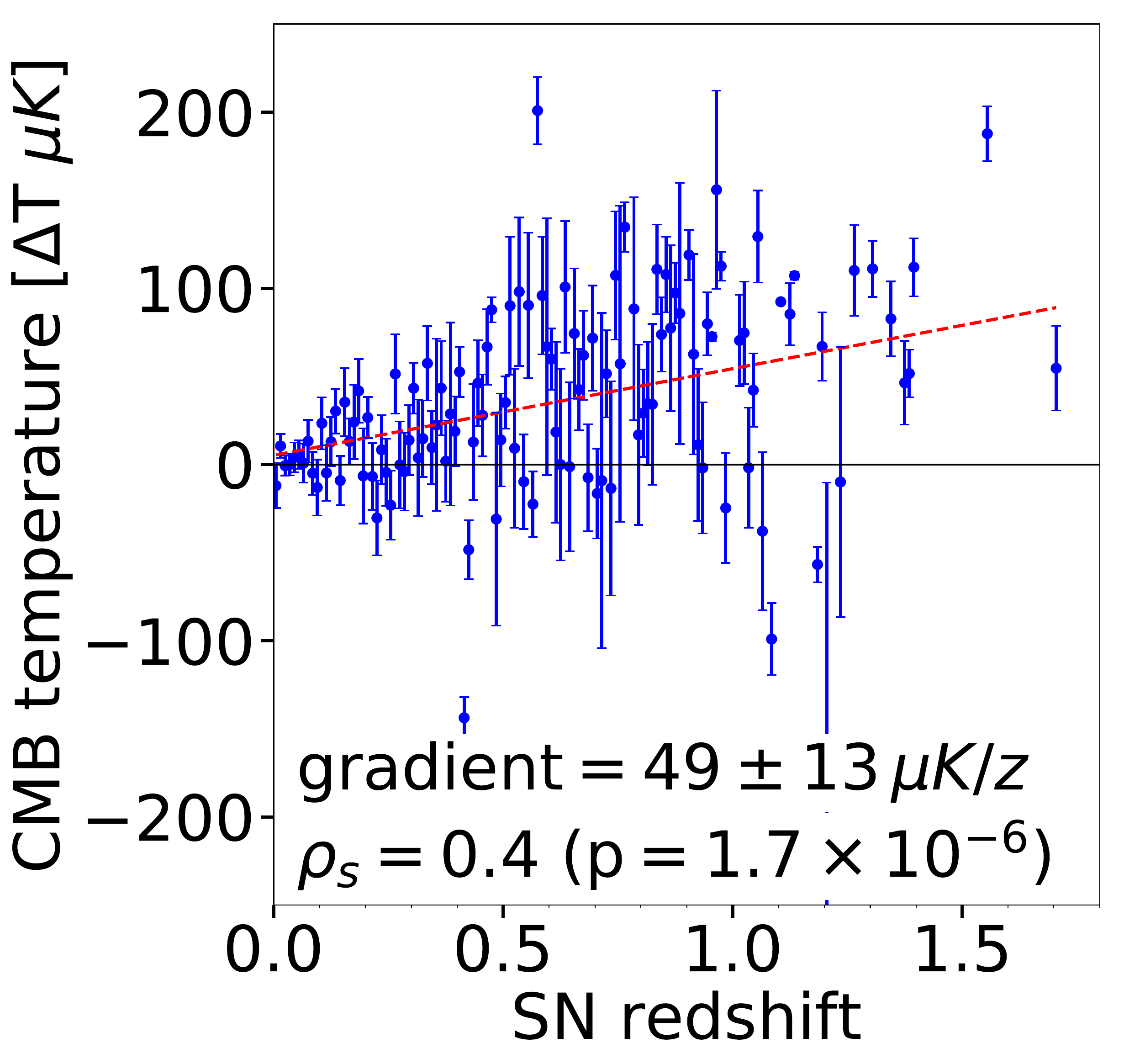}}}
	\\
	\subfigure[\label{sub2013weightedSmoothedHalfDeg}Weighted $\overline{T}$, R1.20 variance smoothed to 0.5$^\circ$]{
    \makebox[0.8\columnwidth][c]{
    \includegraphics[width=0.43\columnwidth]{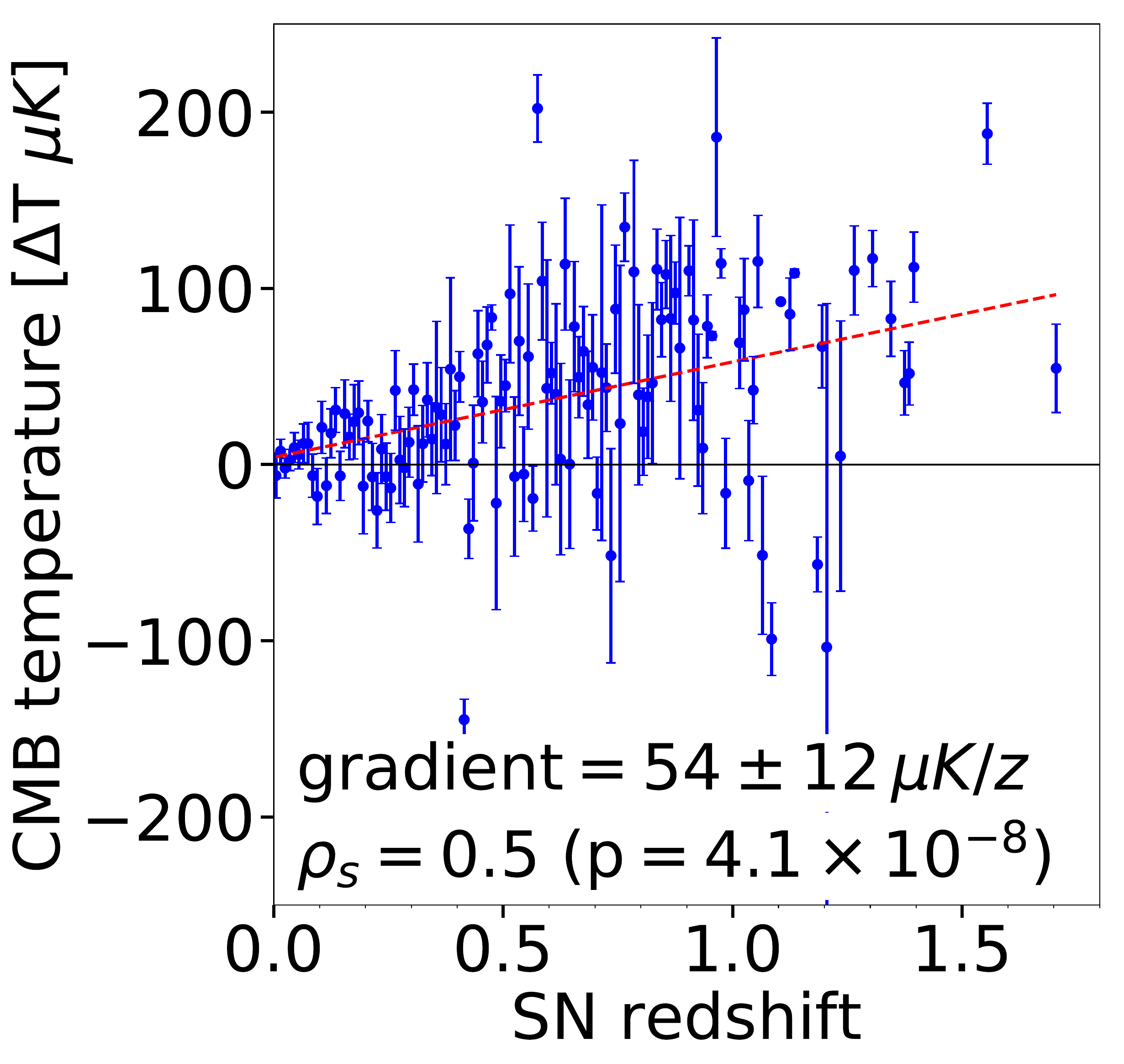}}}
	\\
	\subfigure[\label{sub2013weightedSmoothed5deg}Weighted $\overline{T}$, R1.20 variance smoothed to 5$^\circ$]{
    \makebox[0.8\columnwidth][c]{
    \includegraphics[width=0.43\columnwidth]{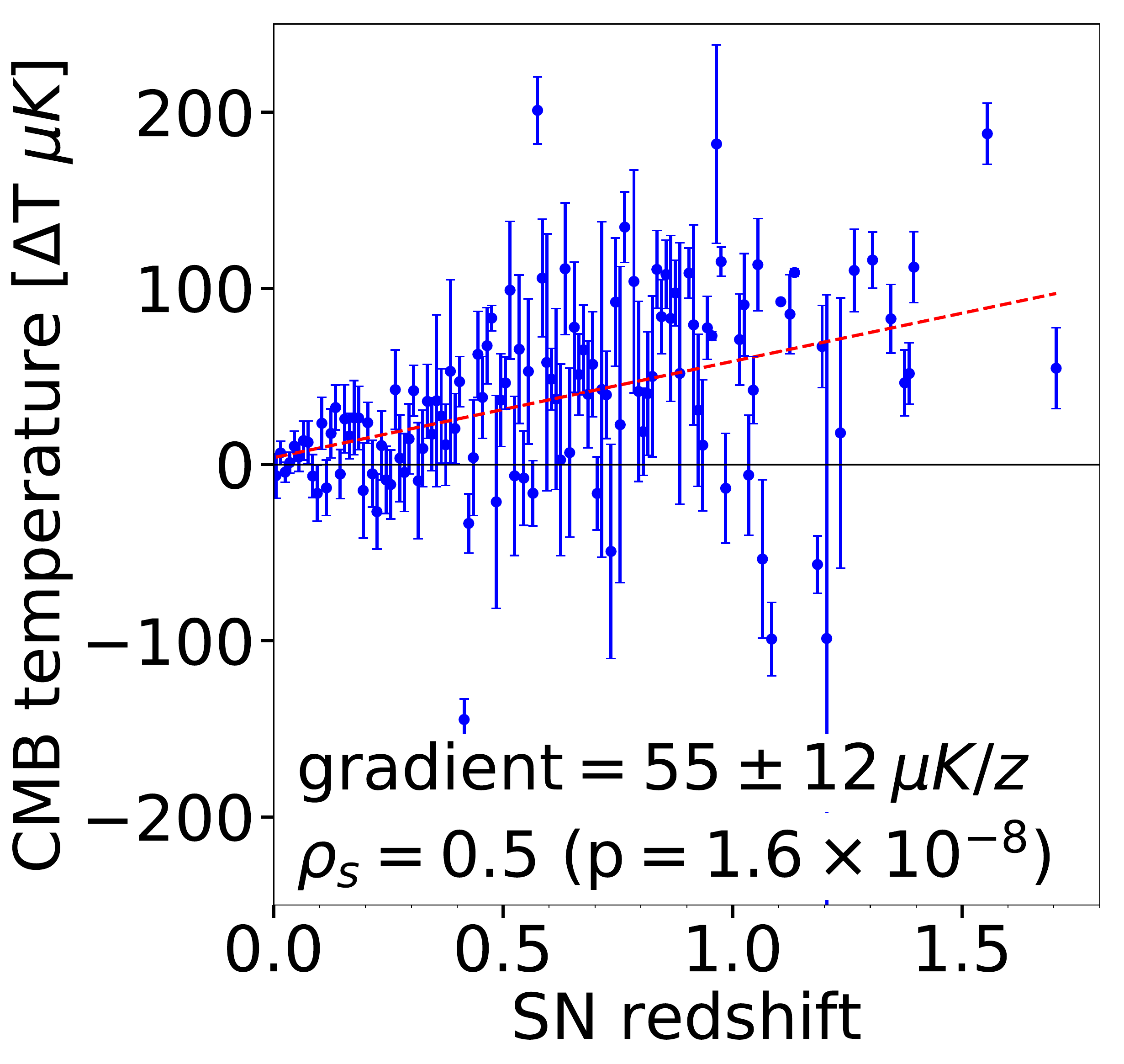}}}
	\caption{Same as Fig. \ref{figHMDH} but for variance estimated using \textit{Planck} 2013 R1.20 SMICA map noise. Note that \textit{Planck} included pixel noise values in the R1.20 SMICA map, unlike R2.01.}
	\label{fig2013}
\end{figure}

\section{Alternative Planck maps} \label{appAltPlanckMaps}
\clearpage

\setlength{\tabcolsep}{1pt} 
\begin{table*} 
	\centering
	\caption{Same as Table \ref{tabHistograms} CMB temperature columns but for temperature measured, and variance estimated, using each of the \textit{Planck} 2015 Commander, NILC, SEVEM, and SMICA maps.}
	\label{tabHistogramsAltPlanckMaps}
	\begin{tabular}{lrclcrclcrclcrclcrclcrclcrclcrclc}
		\hline
        \multirow{3}{*}{Field} & \multicolumn{31}{c}{CMB temperature ($\mu K$)} \\
        & \multicolumn{7}{c}{Commander} && \multicolumn{7}{c}{NILC} && \multicolumn{7}{c}{SEVEM} && \multicolumn{7}{c}{SMICA} \\
		& \multicolumn{3}{c}{SNe} && \multicolumn{3}{c}{pixels} && \multicolumn{3}{c}{SNe} && \multicolumn{3}{c}{pixels} && \multicolumn{3}{c}{SNe} && \multicolumn{3}{c}{pixels} && \multicolumn{3}{c}{SNe} && \multicolumn{3}{c}{pixels}\\
		\hline
		Field 1 & 32.4 & $\pm$ & 9.4 &~~~& 52.3 & $\pm$ & 1.7 &~~~& 30.5 & $\pm$ & 9.1 &~~~& 49.0 & $\pm$ & 1.7 &~~~& 35.7 & $\pm$ & 8.9 &~~~& 51.9 & $\pm$ & 1.7 &~~~& 29.1 & $\pm$ & 9.1 &~~~& 47.5 & $\pm$ & 1.7 \\
        Field 2 & 94.8 & $\pm$ & 2.8 &~~~& 95.2 & $\pm$ & 1.6 &~~~& 90.7 & $\pm$ & 3.1 &~~~& 92.9 & $\pm$ & 1.6 &~~~& 97.0 & $\pm$ & 3.1 &~~~& 96.1 & $\pm$ & 1.7 &~~~& 92.9 & $\pm$ & 3.3 && 94.3 & $\pm$ & 1.7 \\
        Field 3 & 133.0 & $\pm$ & 9.4 &~~~& 138.9 & $\pm$ & 1.0 &~~~& 132.1 & $\pm$ & 9.2 &~~~& 139.0 & $\pm$ & 1.0 &~~~& 132.2 & $\pm$ & 9.3 &~~~& 140.5 & $\pm$ & 1.0 &~~~& 130.8 & $\pm$ & 9.2 && 137.5 & $\pm$ & 1.0 \\
        Field 4 & 91.0 & $\pm$ & 14.7 &~~~& 85.7 & $\pm$ & 1.7 &~~~& 94.2 & $\pm$ & 14.5 &~~~& 88.2 & $\pm$ & 1.7 &~~~& 87.4 & $\pm$ & 13.8 &~~~& 85.6 & $\pm$ & 1.7 &~~~& 85.3 & $\pm$ & 14.3 && 81.7 & $\pm$ & 1.7 \\
        Field 5 & 51.8 & $\pm$ & 8.9 &~~~& 70.2 & $\pm$ & 1.2 &~~~& 52.1 & $\pm$ & 8.6 &~~~& 71.5 & $\pm$ & 1.2 &~~~& 51.8 & $\pm$ & 8.7 &~~~& 70.2 & $\pm$ & 1.2 &~~~& 49.7 & $\pm$ & 8.5 && 68.2 & $\pm$ & 1.2 \\
        Field 6 & 197.3 & $\pm$ & 14.5 &~~~& 169.5 & $\pm$ & 1.9 &~~~& 195.7 & $\pm$ & 14.1 &~~~& 167.4 & $\pm$ & 1.9 &~~~& 197.0 & $\pm$ & 13.8 &~~~& 167.1 & $\pm$ & 1.9 &~~~& 199.3 & $\pm$ &13.6 && 168.9 & $\pm$ & 1.9 \\
        Field 7 & 117.5 & $\pm$ & 12.3 &~~~& 100.5 & $\pm$ & 3.9 &~~~& 121.0 & $\pm$ & 12.4 &~~~& 100.8 & $\pm$ & 4.1 &~~~& 121.9 & $\pm$ & 11.9 &~~~& 103.9 & $\pm$ & 4.0 &~~~& 120.5 & $\pm$ & 12.0 && 100.8 & $\pm$ & 4.0 \\
        Fields 1-7 & 89.4 & $\pm$ & 5.0 &~~~& 103.6 & $\pm$ & 0.7 &~~~& 88.4 & $\pm$ & 5.0 &~~~& 103.7 & $\pm$ & 0.7 &~~~& 90.1 & $\pm$ & 4.9 &~~~& 103.8 & $\pm$ & 4.0 &~~~& 87.4 & $\pm$ & 5.0 && 101.5 & $\pm$ & 0.7 \\
        Stripe 82 & 4.8 & $\pm$ & 4.0 &~~~& 17.9 & $\pm$ & 0.2 &~~~& 6.1 & $\pm$ & 4.0 &~~~& 19.5 & $\pm$ & 0.2 &~~~& 5.8 & $\pm$ & 4.0 &~~~& 18.6 & $\pm$ & 0.2 &~~~& 3.8 & $\pm$ & 4.0 && 17.2 & $\pm$ & 0.2 \\
        Whole sample & 10.8 & $\pm$ & 2.0 &~~~& 3.2 & $\pm$ & 0.0 &~~~& 10.4 & $\pm$ & 2.0 &~~~& 2.2 & $\pm$ & 0.0 &~~~& 11.8 & $\pm$ & 2.0 &~~~& 3.0 & $\pm$ & 0.0 &~~~& 10.3 & $\pm$ & 2.0 && 3.1 & $\pm$ & 0.0 \\
		\hline
	\end{tabular}
\end{table*}
\setlength{\tabcolsep}{6pt}

\clearpage

\begin{figure} 
	\centering
	\subfigure[Whole SNe sample; Commander CMB $\overline{T} \pm \sigma_{\overline{T}}$]{\includegraphics[width=0.8\columnwidth]{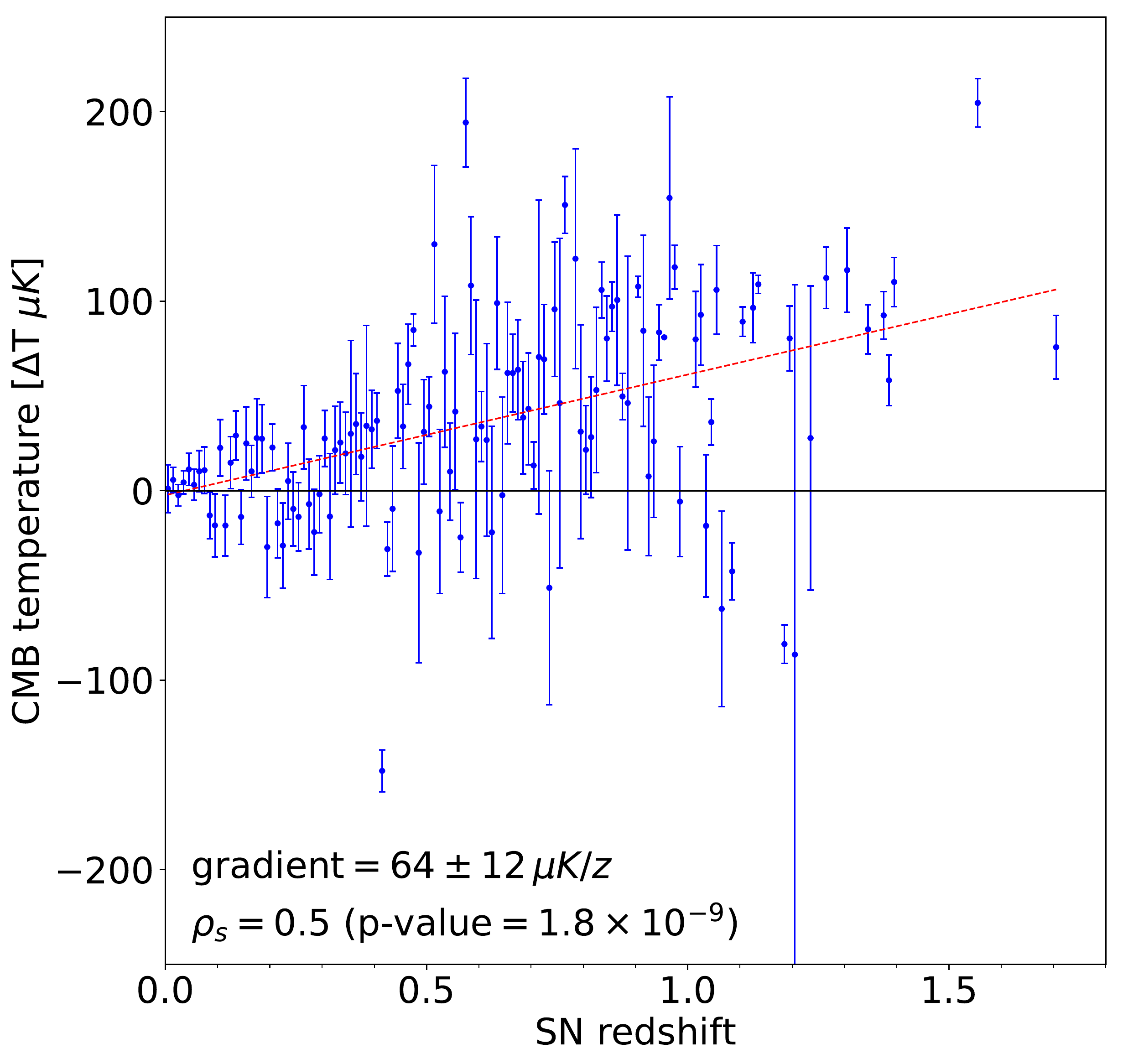} \label{subBinnedScatterFields1to7AllCommander}}
	\\
	\subfigure[Fields 1-7 SNe sample; Commander CMB $\overline{T} \pm \sigma_{\overline{T}}$]{\includegraphics[width=0.8\columnwidth]{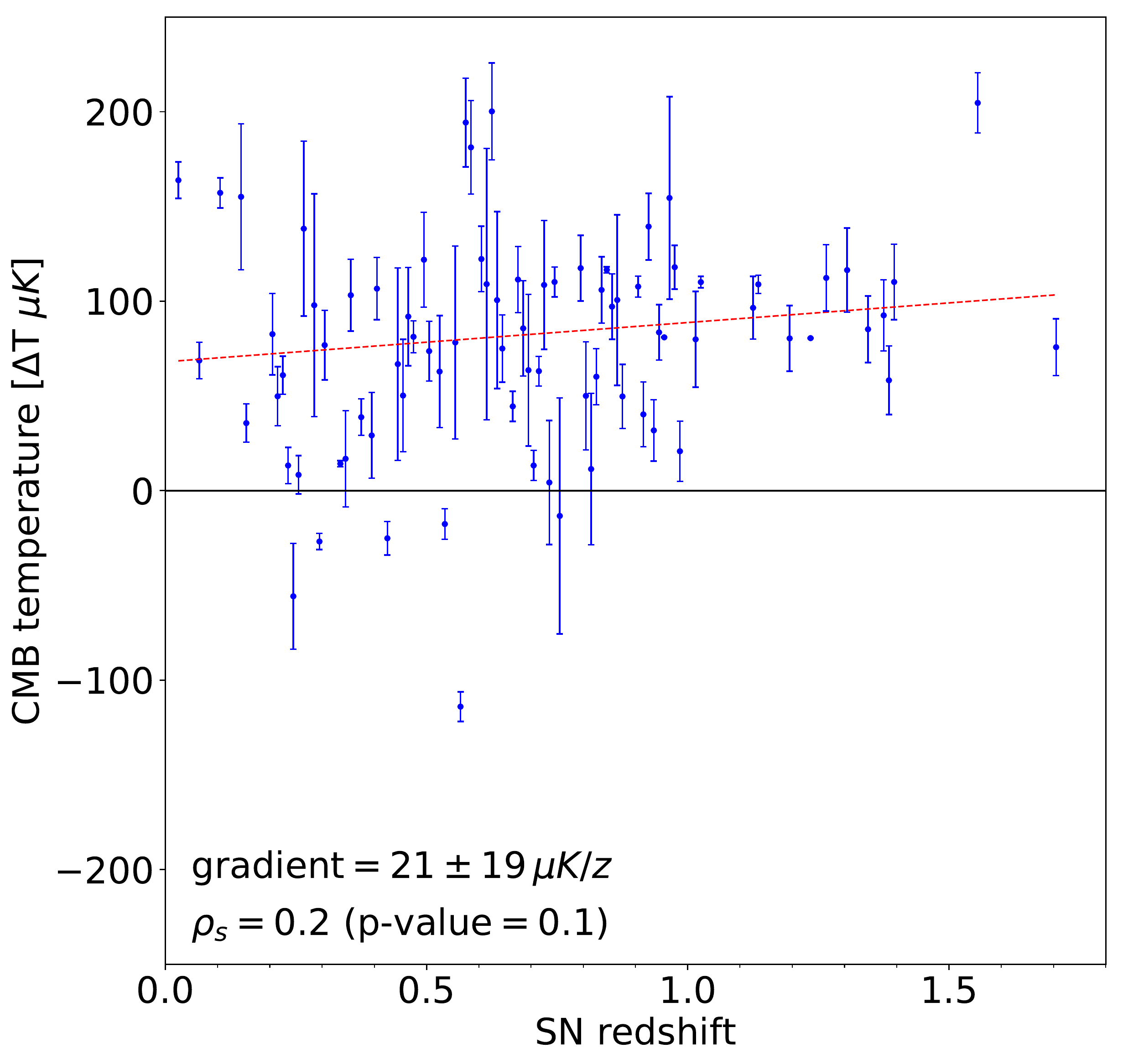} \label{subBinnedScatterFields1to7FieldsCommander}}
	\\
	\subfigure[Remainder SNe sample; Commander CMB $\overline{T} \pm \sigma_{\overline{T}}$]{\includegraphics[width=0.8\columnwidth]{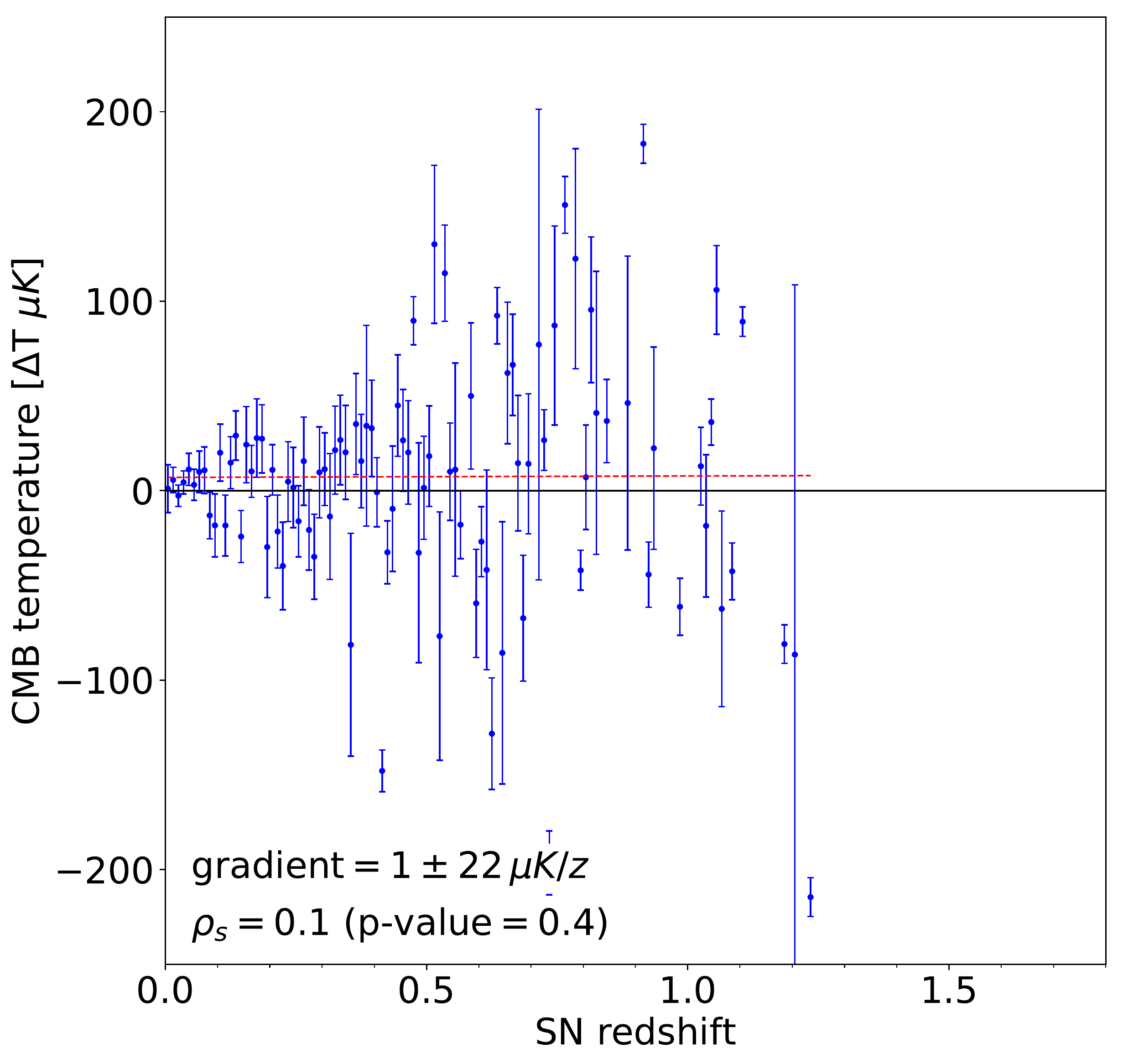} \label{subBinnedScatterFields1to7RemainderCommander}}
	\caption{Same as Fig. \ref{figBinnedScatterFields1to7} but for temperature measured, and variance estimated, using \textit{Planck} 2015 Commander maps instead of \textit{Planck} 2015 SMICA maps.}
	\label{figBinnedScatterFields1to7Commander}
\end{figure}

\begin{figure} 
	\centering
	\subfigure[Whole SNe sample; NILC CMB $\overline{T} \pm \sigma_{\overline{T}}$]{\includegraphics[width=0.8\columnwidth]{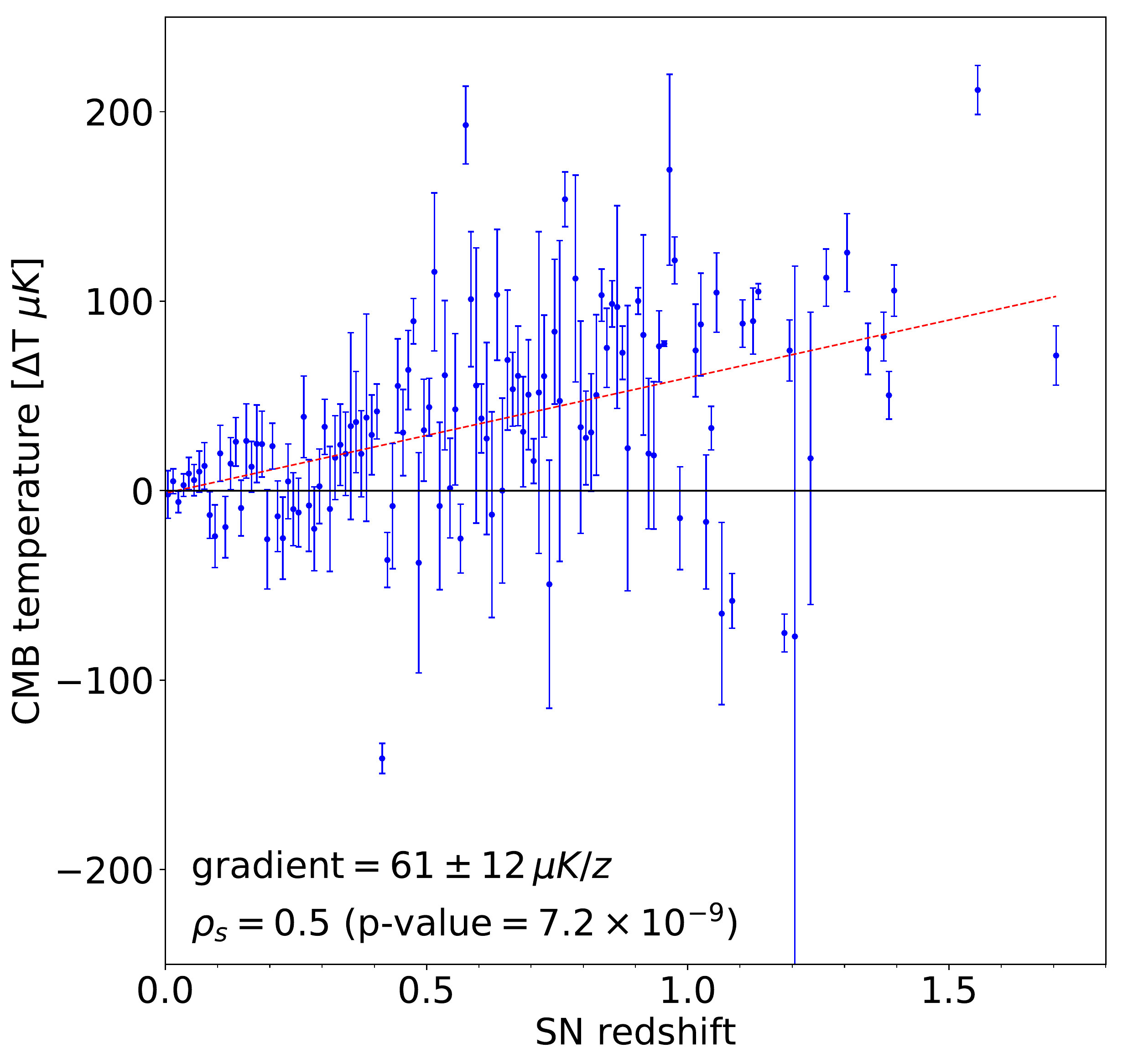} \label{subBinnedScatterFields1to7Allnilc}}
	\\
	\subfigure[Fields 1-7 SNe sample; NILC CMB $\overline{T} \pm \sigma_{\overline{T}}$]{\includegraphics[width=0.8\columnwidth]{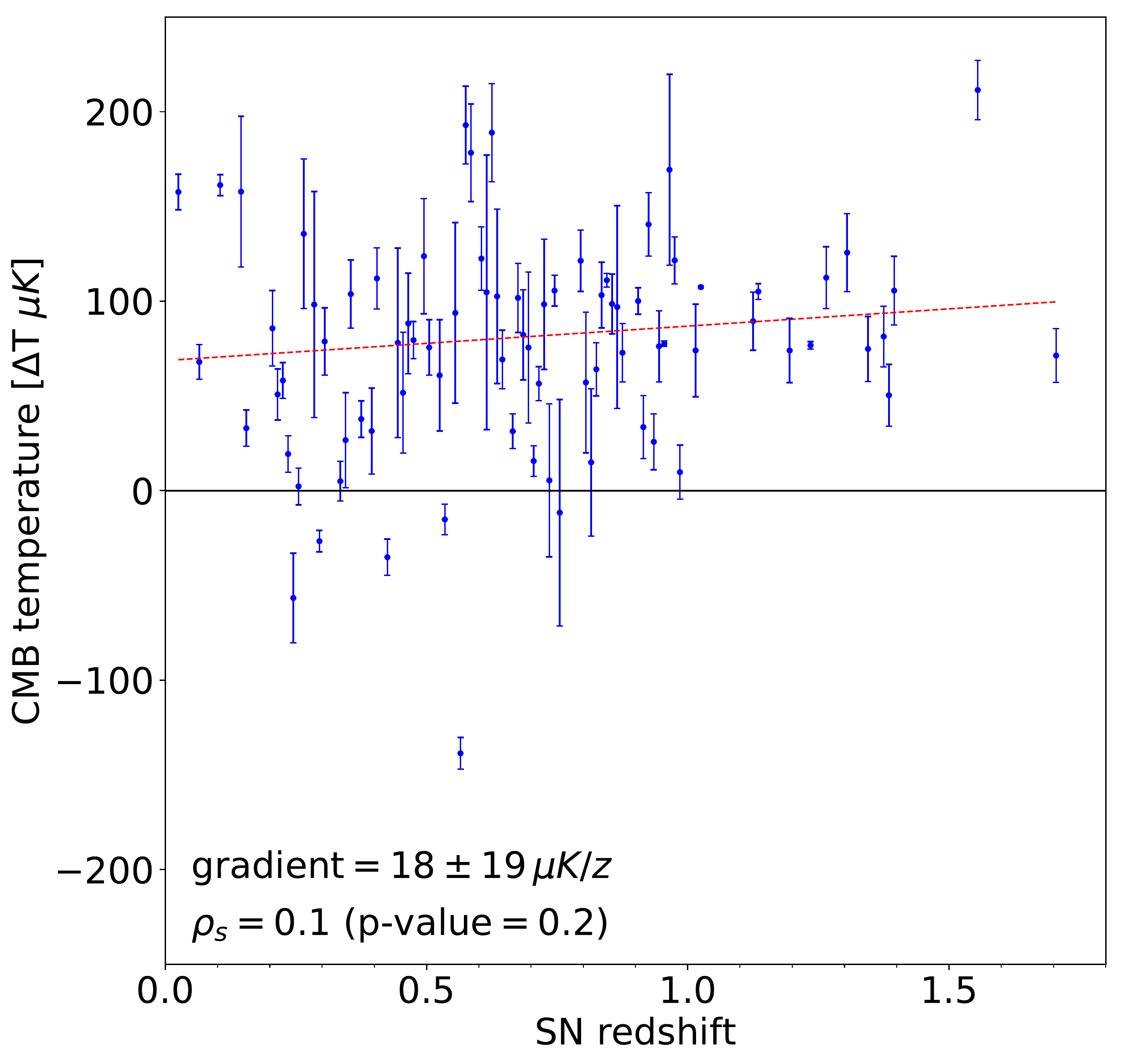} \label{subBinnedScatterFields1to7Fieldsnilc}}
	\\
	\subfigure[Remainder SNe sample; NILC CMB $\overline{T} \pm \sigma_{\overline{T}}$]{\includegraphics[width=0.8\columnwidth]{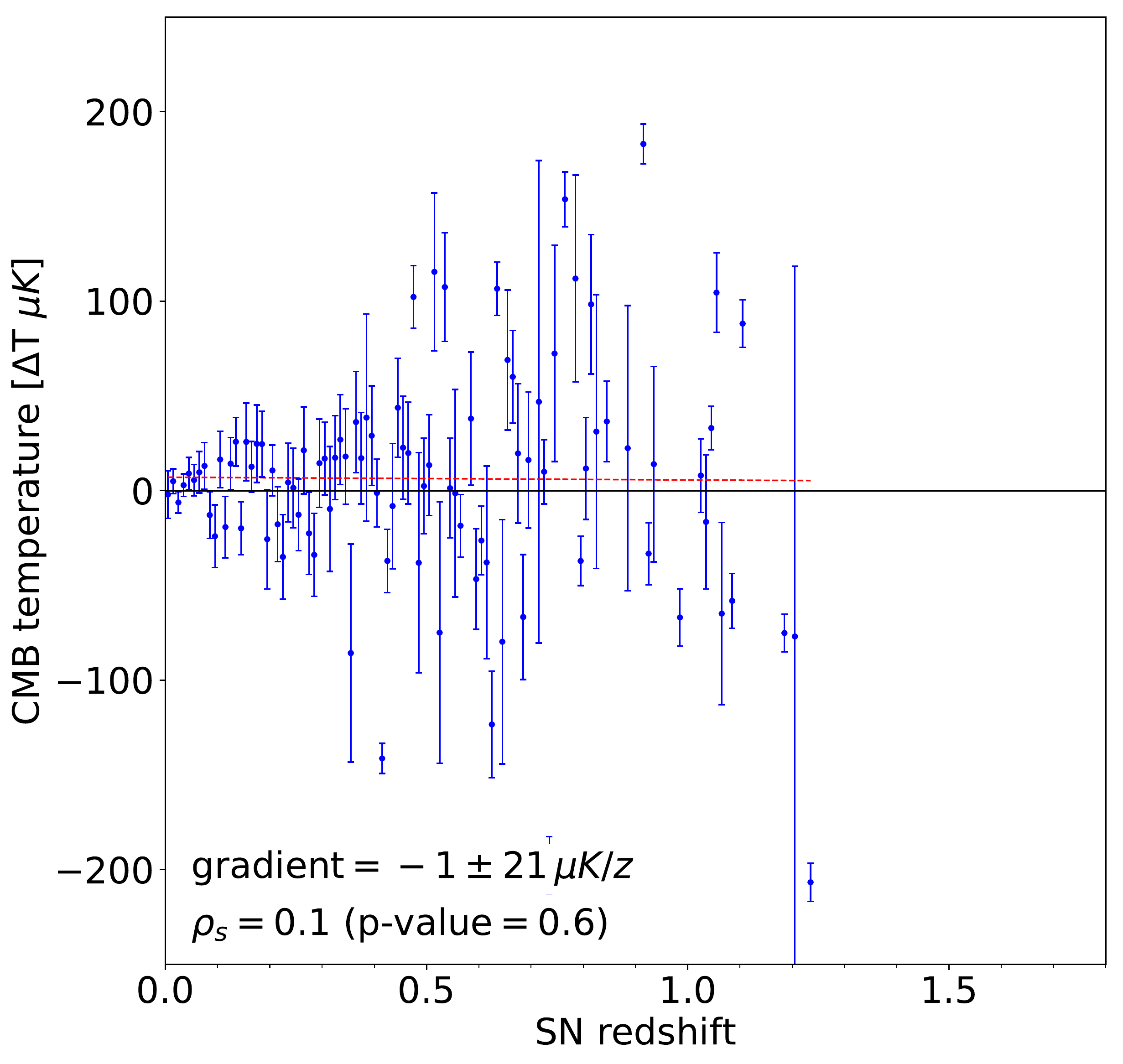} \label{subBinnedScatterFields1to7Remaindernilc}}
	\caption{Same as Fig. \ref{figBinnedScatterFields1to7} but for temperature measured, and variance estimated, using \textit{Planck} 2015 NILC maps instead of \textit{Planck} 2015 SMICA maps.}
	\label{figBinnedScatterFields1to7nilc}
\end{figure}

\begin{figure} 
	\centering
	\subfigure[Whole SNe sample; SEVEM CMB $\overline{T} \pm \sigma_{\overline{T}}$]{\includegraphics[width=0.8\columnwidth]{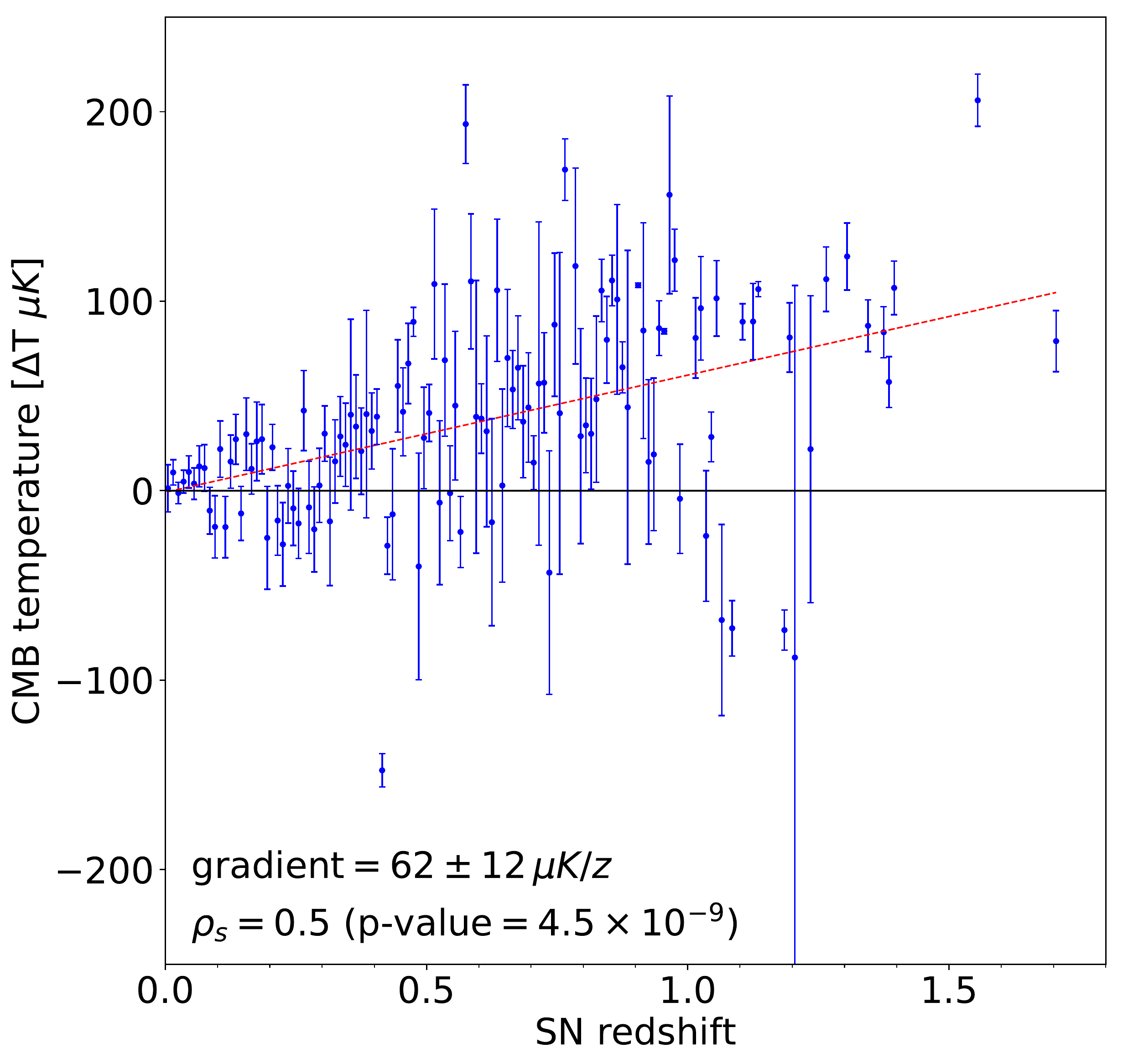} \label{subBinnedScatterFields1to7Allsevem}}
	\\
	\subfigure[Fields 1-7 SNe sample; SEVEM CMB $\overline{T} \pm \sigma_{\overline{T}}$]{\includegraphics[width=0.8\columnwidth]{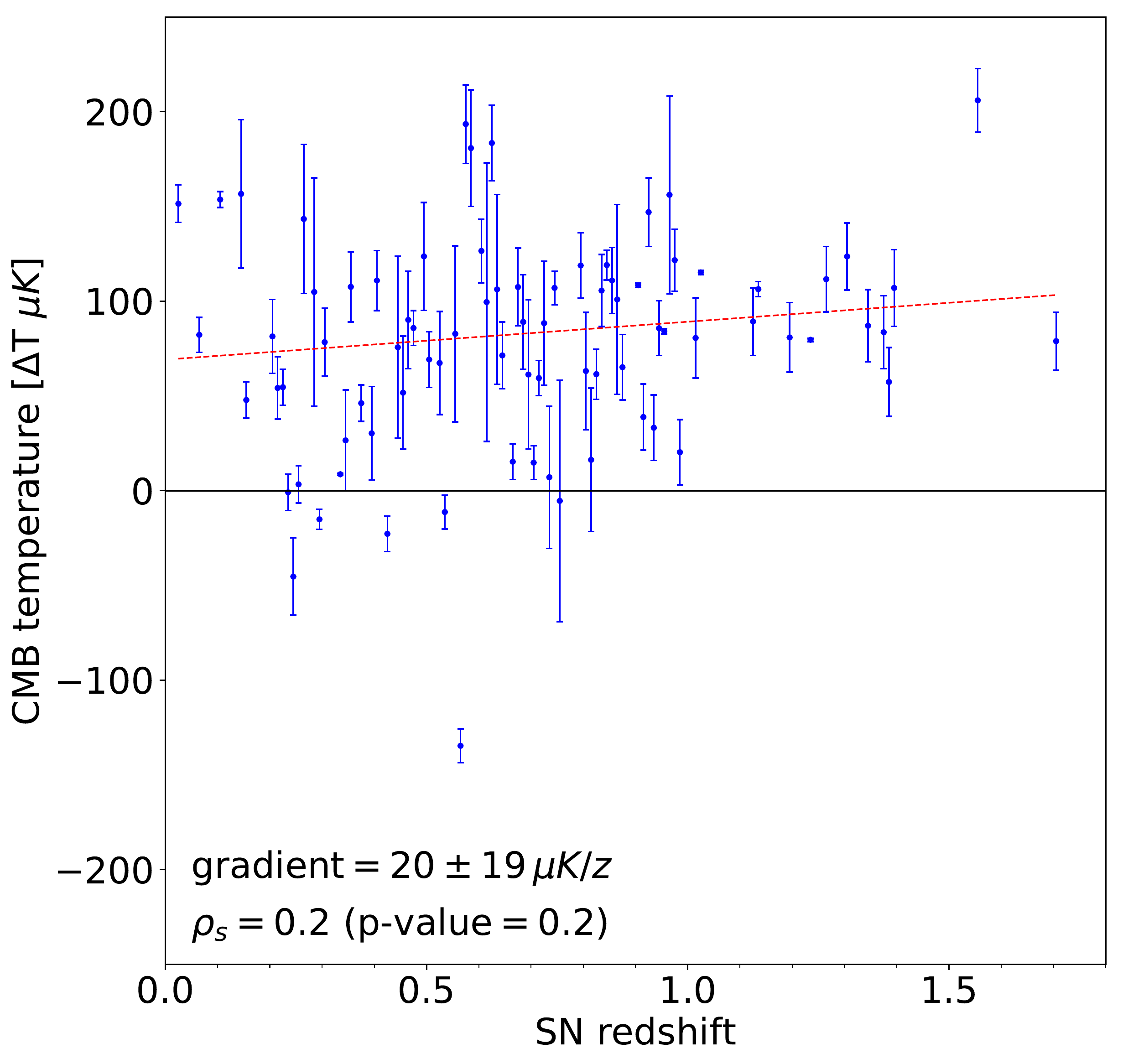} \label{subBinnedScatterFields1to7Fieldssevem}}
	\\
	\subfigure[Remainder SNe sample; SEVEM CMB $\overline{T} \pm \sigma_{\overline{T}}$]{\includegraphics[width=0.8\columnwidth]{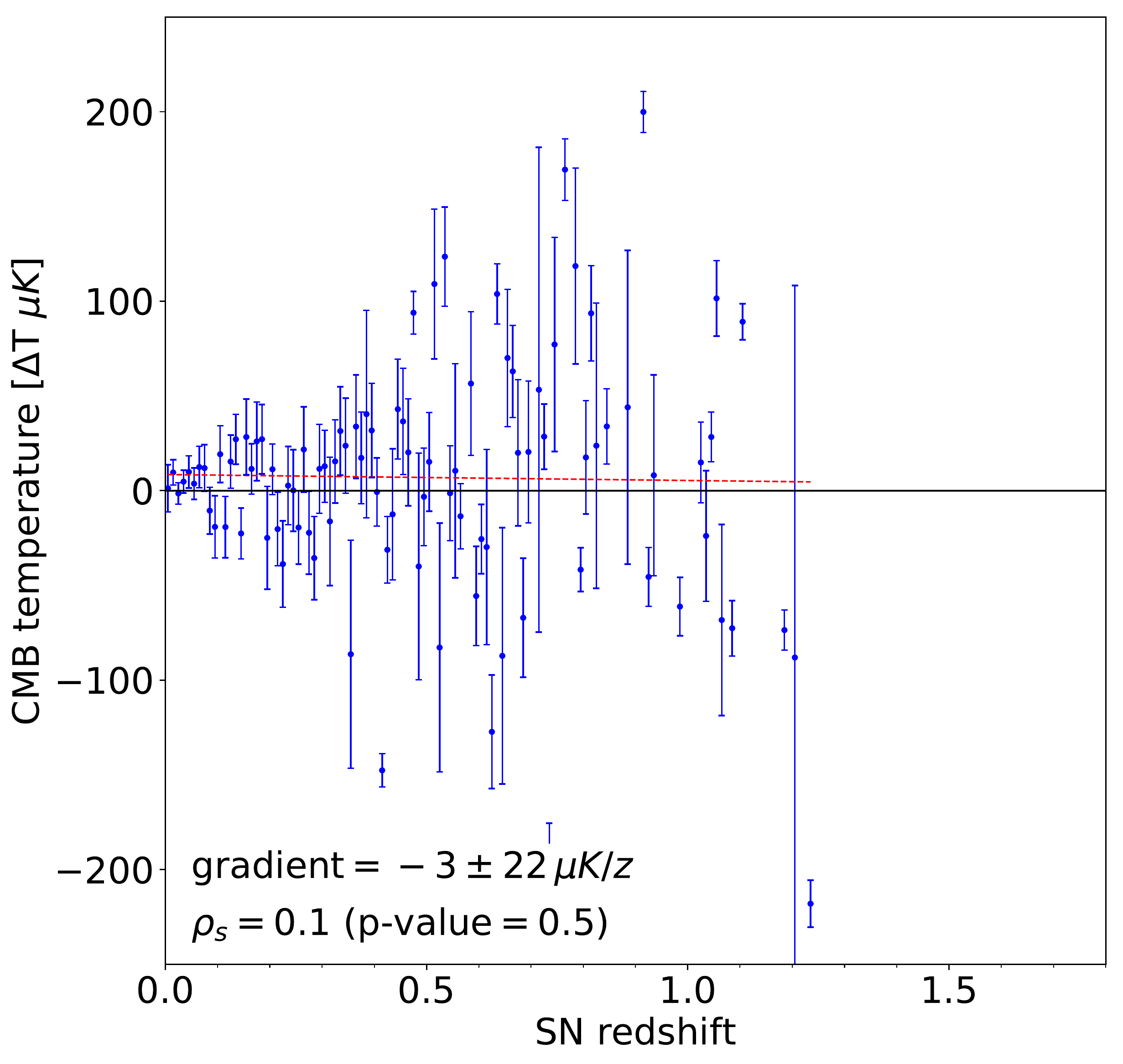} \label{subBinnedScatterFields1to7Remaindersevem}}
	\caption{Same as Fig. \ref{figBinnedScatterFields1to7} but for temperature measured, and variance estimated, using \textit{Planck} 2015 SEVEM maps instead of \textit{Planck} 2015 SMICA maps.}
	\label{figBinnedScatterFields1to7sevem}
\end{figure}

\clearpage


\bsp	
\label{lastpage}
\end{document}